# Neutron Interferometry Using a Single Modulated Phase Grating

I. Hidrovo[1], J.Dey[1], H. Meyer[1], D. S. Hussey[2], N. N. Klimov[2] , L. G. Butler[4], K. Ham[3], W. Newhauser[1],

[1]Department of Physics and Astronomy, Louisiana State University, Baton Rouge, LA, 70803, USA
[2]NIST Center for Neutron Research, 100 Bureau Drive, Gaithersburg, MD 20899, USA
[3]Center for Advanced Microstructures and Devices, Louisiana State University, Baton Rouge, LA, 70806, USA
[4]Department of Chemistry, Louisiana State University, Baton Rouge, LA, 70803, USA
The author to whom correspondence may be addressed: Joyoni Dey, PhD, deyj@lsu.edu

## Abstract

Neutron grating interferometry provides information on phase and small-angle scatter in addition to attenuation. Previously, phase grating moiré interferometers (PGMI) with two or three phase gratings have been developed. These phase-grating systems use the moiré far-field technique to avoid the need for high-aspect absorption gratings used in Talbot-Lau interferometers (TLI) which reduce the neutron flux reaching the detector. We first demonstrate through theory and simulations a novel phase grating interferometer system for cold neutrons that requires a single modulated phase grating (MPG) for phase-contrast imaging, as opposed to the two or three phase gratings in previously employed PGMI systems. The theory shows the dual modulation of MPG with a large period and a smaller carrier pitch P, results in large fringes at the detector. The theory was compared to full Sommerfeld-Rayleigh Diffraction integral simulator. Then we proceeded to compare the MPG system to experiments in the literature that use a two-phase-grating-based PGMI with best-case visibility of around 39%. The simulations of the MPG system show improved visibility in comparison to that two-phase-grating-based PGMI. An MPG with a modulation period of 300 µm, pitch of 2 µm, and grating heights with a phase modulation of $(\pi, 0)$, illuminated by a monochromatic beam, produces a visibility of 94.2% with a comparable source-to-detector distance (SDD) as the two-phase-grating-based PGMI. Phase sensitivity, another important performance metric of the grating interferometer was compared to values available in the literature, viz. the conventional TLI with phase sensitivity of $4.5 \times 10^3$ for a SDD of 3.5 m and a beam wavelength of 0.44 nm. For a range of modulation periods, the MPG system provides comparable or greater theoretical maximum phase sensitivity of $4.1 \times 10^3$ to $10.0 \times 10^3$ for SDDs of up to 3.5 m. This proposed MPG system appears capable of providing high-performance PGMI that obviates the need for the alignment of 2 phase gratings.

# 1. INTRODUCTION

Neutrons are a useful probing tool in measuring material properties and imaging bulk materials due to their dual particle and wave nature, the latter described quantum-mechanically by de Broglie wave packets showing interference phenomena [1]. As neutron waves pass through matter, they undergo phase-shifts due to interactions with local, spatially dependent potentials; the most common being neutron-nucleus interactions [2-3]. Small-angle neutron scattering (SANS) of neutrons, due to nuclear or magnetic interaction potential variations in the sample, also locally degrade the coherence of a well-defined neutron wave front [4-5]. Neutron beams are also attenuated by nuclear reactions and incoherent scattering. Thus, neutron grating interferometry can image variations in phase change (differential phase contrast image), small-angle scattering (dark-field image), and attenuation (transmission image) [1,5].

Currently, there are two grating interferometry methods at the forefront of neutron phase imaging: the Talbot-Lau interferometer (TLI) [6-7] and the phase-grating moiré interferometer (PGMI) [8-11]. The TLI can operate in the full field of a cold neutron beam (contrary to typical Mach-Zehnder interferometers) and has flexible chromatic coherence requirements ($\Delta\lambda/\lambda$). However, the absorption gratings reduce the neutron flux reaching the detector by about a factor 4 for similar source conditions [9-10]. The PGMI also operates in the full-field of a neutron beam, but since it produces directly resolvable interference fringes in the far-field, only a source grating is required, thus reducing the intensity by about a factor of 2. In contrast to the TLI, the PGMI has relaxed grating fabrication requirements, had a broader wavelength acceptance, and permitted control of the fringe period by varying the separation of the phase modulating gratings [10].

The purpose of this study is to investigate by simulation a novel PGMI system for cold neutrons that requires only a single modulated phase grating (MPG). The phase grating has a rectangular modulation in spatial width with a period of W and a finer "carrier" pitch P. An advantage of using a single grating is reducing grating misalignment issues commonly found in multiple-grating systems, which can lead to fringe visibility or contrast loss [9-11].

A non-interferometric grating system which uses only a single attenuation grating was reported by Strobl et al. [12] which has the advantages of its fringe visibilities being independent of the wavelength (achromatic) and its ability to extend the range of autocorrelation lengths probed in materials (particularly in nanometer scale) for dark-field imaging (DFI). However, the system as demonstrated has a SDD of 7.26 m with the grating-to-detector distances of only 50 mm to 300 mm and fringe visibilities of 10% to 70%, limiting the phase sensitivity with its small grating-to-detector distances. Fringe visibility also falls off with higher grating-to-detector distance. This system's visibilities are highly dependent on the geometric blurring due to its pinhole source with the collimation ratio L/D (source-to-grating distance / pinhole size). Thus, the fringe visibilities are susceptible to dropping sharply if the source-to-grating distance is too small or the pinhole size is too large.

The MPG concept was originally investigated in simulations for X-ray by our group [13,14]. In [14] we showed two different components of the grating interfere to create a larger fringe pattern on the detector, directly observable without the analyzer.

In another development, the Modulated Phase Grating was demonstrated in experiments by [15]. The mathematical treatise in [15] was unclear and to the best of our understanding appear to consider the Fresnel kernel to operate directly on the intensity instead of amplitude. The Fresnel

Kernel approximation applies to the amplitude [16-18,8] and then the intensity may be found similar to [8].

In this paper we derived the theory behind a rectangular Modulated Phase Grating from first principles, using Fresnel zone approximation (which is also valid for far-field systems) to show that the intensity pattern on the detector is indeed of directly observable period of magnification times large modulation period W. We also estimate distance conditions for maximum visibility for a given W and P.

We also developed an analytical Somerfield Rayleigh simulator with which we investigated the advantages of performance, visibility, and sensitivity of the MPG system over the two-phase-grating-based PGMI, which we will hereinafter refer to as the "standard-dual-grating" system. Before thoroughly investigating the performance of the MPG, we compared some results of our simulation method with the experiments previously conducted with the standard-dual-grating system and reported by Pushin et al. [9].

## 2. METHODS

### 2.1 Analytical Simulations for Neutron Interferometry

We have constructed an efficient wave-propagation simulator for neutron PGMI applications that can accept various designs (e.g., modulated phase gratings [14] or standard gratings). The code, named "N-SRDI", was written in the C++ programming language and can simulate systems such as those shown in Figures 1-2 using the Sommerfeld-Rayleigh diffraction integrals (SRDI) [16]. These predict the observed complex-valued field amplitude $A(y)$ from a wave that has diffracted from a grating or aperture and which originated from a source wave function $U(P_s)$. The neutron field amplitude at a detector for a single-phase-grating system such as the MPG (Figure 1) can be expressed as

$$A(y) = \frac{1}{j\lambda} \int U(P_s) \; \frac{e^{jkr_1}}{r_1} \; T_1(y_1) \cos(\theta_1) dy_1 \tag{1}$$

and the intensity is

$$I(y) = |A(y)|^2 \tag{2}$$

where $U(P_s)$ is the neutron source wave function, $T_1(y_1)$ is the neutron transmission function though the MPG in a plane perpendicular to the neutron beam propagation direction along the $z$ axis, $k = \frac{2\pi}{\lambda}$ is the wave number, $\lambda$ is the wavelength, $r_1$ is the distance between the grating point $y_1$ to the detector point $y$ given by $r_1 = \sqrt{\left(D_{gd}\right)^2 + (y - y_1)^2}$, and $\theta$ is the angle between $\vec{r_1}$ and the normal of the MPG. The transmission function is given by

$$T_1(y_1) = A_1(y_1)e^{j\phi(y_1)} \tag{3}$$

where $A_1(y_1)$ is the amplitude transmission of the neutron beam through the grating due to attenuation and $\phi(y_1)$ is the phase shift determined by the spatial heights of the MPG.

The SRDI shown in Eq. (1) is a representation of the Huygens-Fresnel principle since the observed complex-valued field amplitude at the detector is a superposition of diverging spherical waves $e^{jkr_1}/r_1$ originating from secondary sources located at each point $y_1$ along the phase grating [16]. In all simulations the neutron source wave function $U(P_s)$ is assumed to be a point source at point $y_0$; however, the fringe patterns from a line source are later acquired with a convolution method. Therefore, Eq. (1) was simplified to

$$A(y) = \frac{D_{gd}}{j\lambda} \int \frac{e^{jk(r_0 + r_1)}}{r_0 r_1^2} \; T_1(y_1) dy_1 \tag{4}$$

where $r_0$ is the distance between the point source at point $y_0$ and the point $y_1$ on the MPG given by $r_0 = \sqrt{(D_{sg})^2 + (y_1 - y_0)^2}$ and the $cos\,\theta$ (between $r_1$ and the $z$ axis) has been replaced by $D_{gd}/r_1$, explaining the $r_1^2$ term inside the SRDI show in Eq. (4).

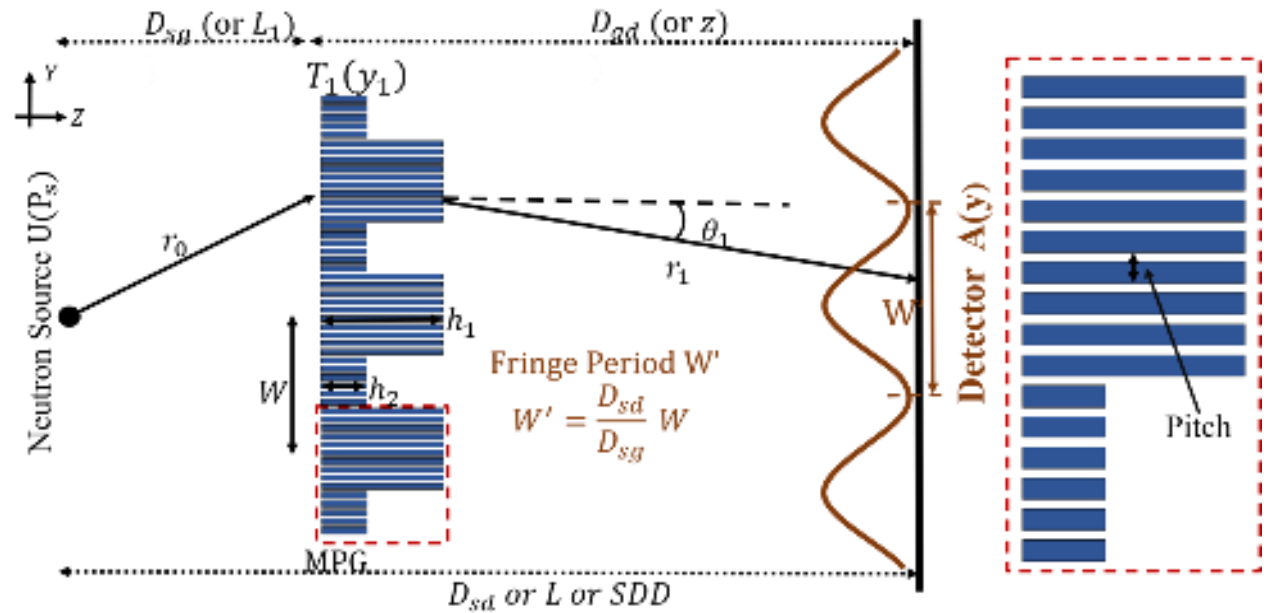

Figure 1. Schematic diagram of the simulated MPG system (not to scale). In our simulations the neutron source is a point source emitting a spherical wavefront along the $z$ axis (fringe patterns from a line source are later acquired with a convolution method). The source-to-MPG distance $D_{sg}$ was chosen to be 0.5 m or 1.0 m (depending on the pitch), and the MPG-to-detector distance $D_{gd}$ can range from 1.0 m to 5.0 m. The grating has a rectangular functional form that modulates in spatial height with a slow-varying period of W and a smaller "carrier" pitch P. The mathematical form of $T(y_1)$ is given in the theory section. The duty cycle of the grating is 50%. The different heights ($h_1, h_2$) of the grating along the z axis correspond to phase shifts for example $(\pi, \pi/4)$ or $(\pi, 0)$ for neutrons of wavelength 0.44 nm. The grating simulated is an ideal phase grating so no attenuation of the beam is considered, i.e. $A_1 = 1$. The fringe period at the detector $W'$ is determined by the geometric magnification $D_{sd}/D_{sg}$ from the point source [14].

To verify the accuracy of the N-SRDI code we also simulated standard-dual-grating systems for neutron PGMI applications and compared our results to experimental data from Pushin et al. [9]. Since two separate gratings are involved for this interferometer system, we use two

SRDIs to calculate the observed complex-valued field amplitude $A(y)$ at the detector. Using the Sommerfeld-Rayleigh formulation for diffraction, the neutron field amplitude at a detector for a standard-dual-grating system (Figure 2) was simplified, using the same logic as before, to

$$A(y) = -\frac{DL_2}{\lambda^2} \iint \frac{e^{jk(r_0+r_1+r_2)}}{r_0 r_1^2 r_2^2} T_1(y_1) T_2(y_2) dy_1 dy_2 \qquad (5)$$

In all simulations (MPG or standard-dual-grating systems), we obtained results with a point source (that is $U(P_s)$ was a point source at point $y_o$ emitting spherical wave $e^{ikr_0}/r_0$) and then used a convolution method on the fringe pattern of a point source to simulate the line-source, as will be described next.

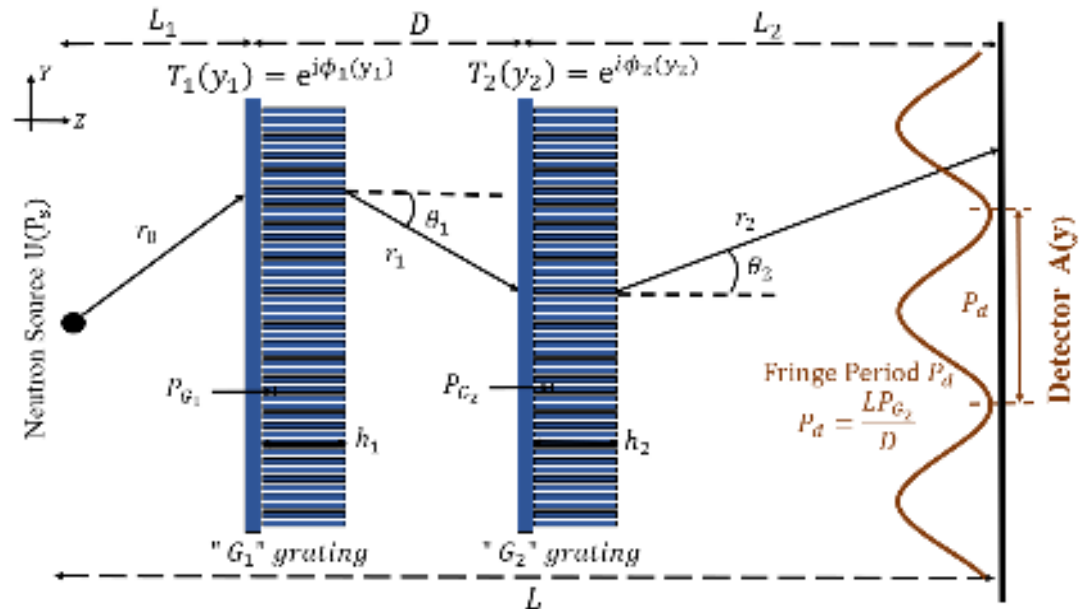

Figure 2. Schematic diagram of the simulated standard-dual-grating system (not to scale). Simulations were done to compare to fringe visibility measurements done by Pushin et al. [9]. In our simulations the neutron source is a point source emitting a spherical wavefront along the z axis (fringe patterns from a line source are later acquired with a convolution method). $L_1 = 1.2$ m and $L_2 = 1.79$ m. The inter-grating spacing D

ranges from 7 mm to 16 mm. The gratings have standard binary form with no modulation in spatial height. The periods of the gratings are $P_{G_1} = P_{G_2} = 2.4$ μm. The duty cycles of the gratings are 50%. The heights of the $G_1$ and $G_2$ gratings are $h_1$ and $h_2$, respectively. They both correspond to a phase shift of $.27\pi$ for neutrons of wavelength 0.44 nm. The gratings simulated are ideal phase grating so there is no attenuation of the beam for either grating, i.e. $A_1 = A_2 = 1$. The fringe period at the detector $P_d$ is given by the ratio $LP_{G_2}/D$ as given in Ref. [8].

## 2.2 Fringe Pattern Blurring due to Slit Width and Pixel Size via Convolution Method

*Slit Width:* In the first step of the simulation, the neutron source was assumed to be a point source emitting a spherical wave in the direction of the *z*-axis towards the imaging detector. A line source, such as that in Ref. [9], can be thought of as an array of point sources that are mutually incoherent. As shown in Figure 3, each point along a line source creates a fringe pattern at the detector that is spatially displaced with respect to the fringe pattern from the central point source. Their superposition causes the observed composite fringe pattern to be washed out, reducing the overall visibility of the fringe pattern. Consequently, a line source will have lower fringe visibility compared to a single, monochromatic point source.

Instead of repeating costly SDRI computations by taking a series of point sources along with the extent of a line source (Figure 3), we modeled the fringe pattern of a line source by first simulating the fringe-pattern due to a center-point source and then *convolving* it with a rectangular window function of the same size as the line source. This has the equivalent "wash-out" effect on the fringe intensity pattern at the detector from simulating an array of point sources that make up a line source.

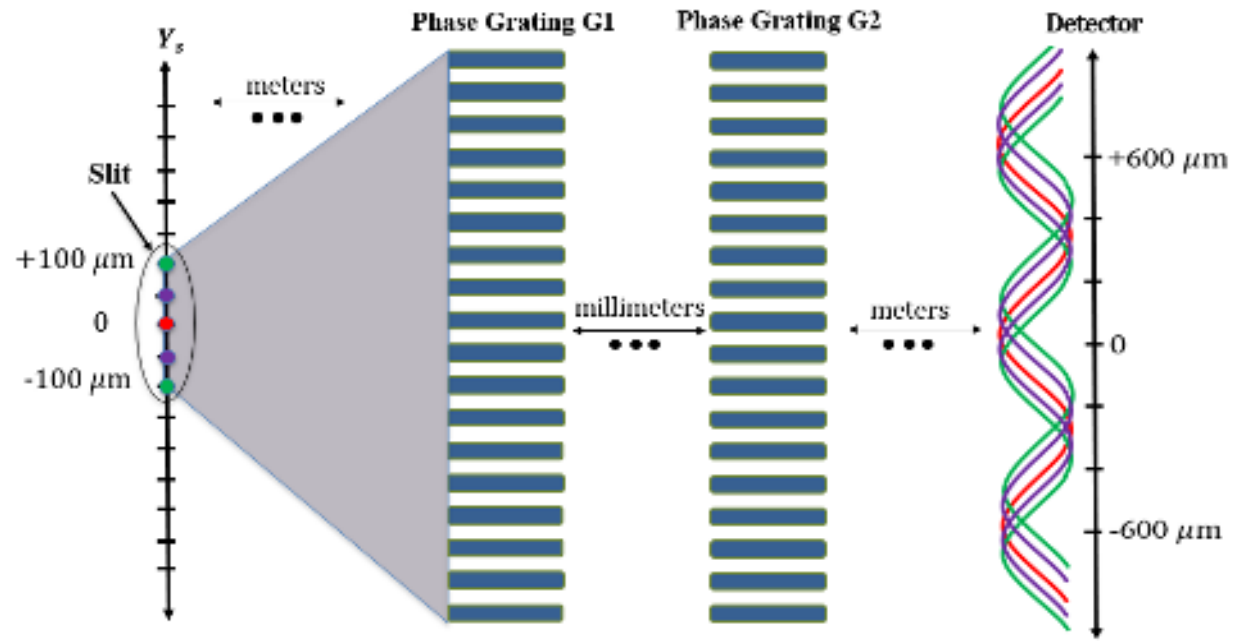

Figure 3. Schematic illustration demonstrating visibility loss of an approximated line source (not to scale). The red fringe pattern on the detector is due to the central point source, and the green fringe patterns are due to the endpoints of the approximated line source. Superposition of these fringe patterns will result in visibility loss due to the green fringe patterns being shifted away from the red fringe pattern. With more intermediate point sources added to approximate the line source, the visibility is improved since there are more fringe patterns that are closer in phase to the red fringe pattern. However, the visibility of this approximated line source will always be lower than that of the fringe pattern of a single point source because of the wash-out effect many slightly shifted fringe patterns create.

*Note on simulation of source grating G0:* The G0 simulations are similar to simulations of a slit. When a source grating is used, instead of a single slit we have several slits adjacent to each other in the $Y_s$-axis of Figure 3 with center separation multiples of $P_0$. The $P_0$ (or source period $P_s$ in [8]) is chosen such that each slit sends a net fringe pattern co-registered with the fringes from the other slits, resulting in a simple scale factor of intensity. Therefore, simulating a single slit versus source grating simulation is identical in procedure, except for a large-scale factor increasing the intensity at the detector. The calculations of $P_0$ and slit opening are available in the literature [20,14]. In the Discussion section we calculated the pitch $P_0$ of G0 and open-ratio requirement for the MPG system for an example geometry we investigated.

*Pixel-size convolution and subsampling:* Furthermore, to account for the fringe blurring due to the pixel resolution and to have the correct number of sampled data points given a certain pixel size, two steps are taken: (1) another convolution of the fringe pattern at the detector by a rectangular window function that is the same size as the pixel (2) subsampling of the fringe intensity pattern at pixel-size increments since the intensity pattern at the detector is originally sampled at 0.1 μm. This is followed by an interpolation to reconstruct the signal at higher sampling rates.

We simulated a case from the experimental setups from Ref. [9] to assess our convolution method that used a standard-dual-grating system with a monochromatic beam ($\lambda = 0.44$ nm) exiting a 200 μm slit source with the system parameters of $L_1 = 1.2$ m, $D = 12$ mm, and $L_2 = 1.78$ m (Figure 2). The identical phase gratings had pitch $P_{G_1} = P_{G_2} = 2.4$ μm and a $0.27\pi$ phase shift for 0.44 nm wavelength neutrons. Both gratings had duty cycles of 50%. The detector pixel size resolution was reported to be ~100 μm. In simulations the grating function is sampled at every 2 nm and the intensity pattern at the detector is sampled at every 0.1 μm. We used the two independent convolutions previously mentioned on the fringe pattern to account for blurring due to the slit width and pixel size. The fringe pattern is subsampled at 100 μm increments and intensity values are interpolated between the sampled values by a factor of 1000 using MATLAB's *interp* function to return to a 0.1 μm sampling rate.

We compared our convolution method against two other methods: (1) the expected closed-form visibility function of slit size given in Ref. [9]: $V = V_o|sinc(\pi s/P_s)|$, where $V_o$ is the optimal fringe visibility from one point source, s is the width of the slit, and $P_s$ is the source period. The fringe visibility for a 200 µm source using our convolution method was 32.9% which closely matched the expected close form 32.7% visibility (blue and orange bars in Figure 4). (2) The brute force method of simulating many individual point sources on the slit, illustrated in Figure 3 was also compared. The yellow bar in Figure 4 represents the visibility for 21 points evenly space about the 200 µm width of the slit, yielding a 32.7% visibility. As more intermediate point sources were uniformly added as represented by the red bar (41 total point sources), the visibility remained at the expected visibility value of 32.7%. Thus, we verified that a central point source simulation followed by a convolution with a rectangular window function can adequately model the fringe pattern of a slit source, reducing computation cost considerably by avoiding brute force computations of several point sources.

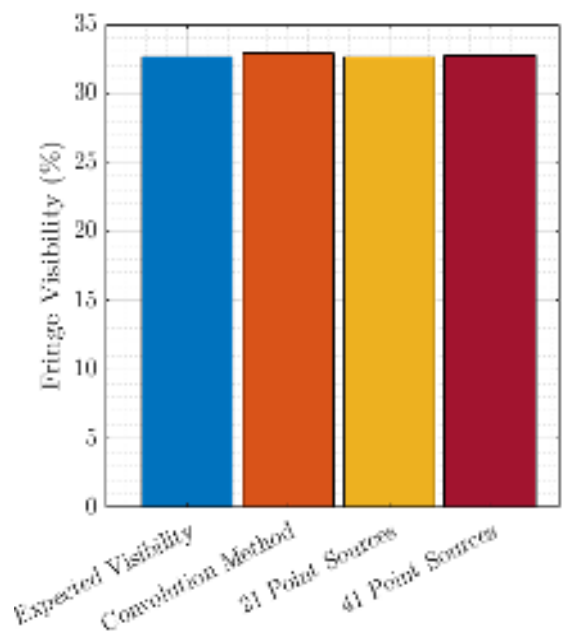

Figure 4. Verification of convolution method visibility for a 200 µm slit source in a monochromatic beam. The fringe visibility using the convolution method is 32.9% closely matched the expected 32.7% visibility, theoretically calculated with Eq. 10 given in Ref. [9]: $V = V_o|sinc(\pi s/P_s)|$ where $V_o$ is the optimal fringe visibility given by one point source (39.6%), $s$ is the slit size (200 µm), and $P_s$ is the source period (598 µm). Another method tested was modeling the slit as a series of evenly spaced point sources. The visibility converged after 21 points, with 21 and 41 points each yielding the expected visibility of 32.7%.

## 2.3 Evaluation of N-SRDI Simulations for a Standard-Dual-Grating System, Comparisons with Simulations and Experiments

We evaluated the N-SRDI obtained fringe visibilities V against experimental results with standard-dual-grating systems for the neutron PGMI in Ref. [9] using the same methods detailed in the previous section. When the gratings were close D = 7 or 8mm, $\cos(\Theta_1)$ in Figure 2 is assumed close to 1 to avoid numerical problems. This is akin to a Fresnel-zone approximation in this region. We also evaluated whether the simulated fringe periods agreed with the theoretically predicted values [8]. The evaluations are shown in Results under the corresponding heading.

*Comparison of Visibility:* The simulation parameters to match experiments are detailed in the previous section. The grating-to-grating separation distance *D* was varied from 7 mm to 16 mm, while keeping the *L* (or $D_{sd}$) and $L_1$ (or $D_{sg}$) fixed (Figure 2). The $L_2$ (or $D_{gd}$) was adjusted accordingly. Our convolution method previously described was applied to the fringe patterns, accounting for the 200 μm slit and 100 μm pixel used in the experiments. Again, evaluations are compiled in the Results section.

It is important to note that the slit size or pitch of a source grating $G_0$ have a direct one-to-one effect on the fringe periods at the detector without a pinhole geometric magnification when $P_{G1}=P_{G2}$. Thus, the source period (or the pitch of $G_0$) is directly equal to the fringe period as shown in Ref [8] or [9].

This not true for the single MPG. The pinhole factor Ref [14,20] is present for the slit-opening as tested in the next section with simulations and theory.

## 2.4 Theory of Modulated Phase Grating (MPG) and comparison with N-SRDI MPG Simulations

The idea behind Modulated Phase Grating was originally demonstrated by our group in simulations using the Somerfield-Rayleigh Simulations in Ref [14]. The Modulated Phase Grating has an envelope function with two grating phase heights, with a large period W (order of 100 μm) and a small modulation of pitch P (order of few µm), Figure 1. The grating has two clear components with different phase heights. As shown in [14], these two components interact *in amplitude* to produce *intensity* patterns on the detector which have periods with magnified version of W.

In what follows we show the (a) effect of slit size on MPG system fringe patterns, (b) mathematical theory behind the MPG (with details in Appendix A) and finally the (c) Z-condition for maximum and minimum visibility.

*(a) Effect of slit size on MPG system interference pattern*

Consider a geometry where the center source to MPG distance is $L_1 = 100$cm and MPG to detector distance is $L_2=200$cm, ($L_2/L_1 =2$). Figure 5 shows the N-SRDI simulated intensity pattern (Eq.. 1 and 2) for this case with a point source at the center, and two others shifted at +100µm and -100µm (shift perpendicular to propagation direction).   As shown, these create three beam-patterns shifted from each other by +/-200 µm. This shows that for a 200 µm slit (with extremities at +/-100µm), one expects the patterns shifted over a net 400 µm range (+/-200 µm). This can be explained by the pinhole magnification factor of the slit, $L_2/ L_1 =2$. In general, if the actual slit size is $L_s$ µm, the convolution kernel to be used is $K_s= L_s (L_2/ L_1)$.

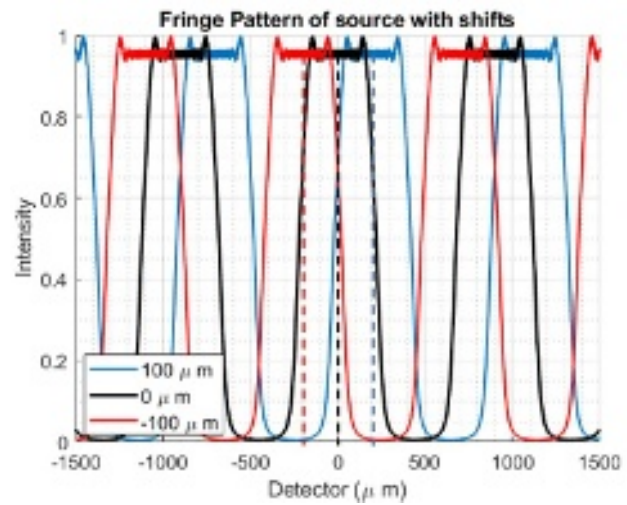

Figure 5. N-SRDI simulations for point sources at the center (black, 0 µm) and at +/-100 µm (blue and red respectively). The dashed vertical lines show that red and blue patterns are separated by a total 400 µm, which is explained by point source separation of 200 µm multiplied by a pinhole-factor of 2 for the geometry setting.

This result is mathematically proved in Appendix B (after the main theory in Appendix A, highlighted next).

*(b) Mathematical Theory behind the MPG*

In this section and in the Appendix A, the field equation for MPG is derived using the Fresnel approximation of the Somerfield-Rayleigh Equation.

We consider a 1D detector, $y$-axis as the axis along the detector, $y_1$-axis as the axis of the grating and z-axis as the propagation axis. First, we consider the Grating Function such as in Figure 1. This can be written as

$$T(y_1) = \left[ \left\{ g(y_1) \sum_{n=-\infty}^{\infty} \delta(y_1 - nP) \right\} + \sum_{n=-\infty}^{\infty} \delta\left(y_1 - \frac{nP}{2} - P/2\right) \right] \otimes rect\left(\frac{y_1}{P/2}\right) \qquad (6)$$

Note that second term inside the square brackets, $1 \times \sum_{n=-\infty}^{\infty} \delta\left(y_1 - \frac{nP}{2} - P/2\right)$ is to account for the fact that in between the spikes of the comb-train $\sum_{-\infty}^{+\infty} \delta(y_1 - np)$ sampling the basic envelope function (which is given below), there is a transmission factor of 1 (and not zero). The convolution by the small half-pitch width *rect* function is to take into account the small post-size ($P/2$) of the grating, (assuming a duty cycle of 50%) [21]. The equation is similar to a standard phase grating as shown in Wilde and Hesselink [17], except with the envelope function is allowed to be more complex than a single constant phase transmission factor of $exp(i\varphi)$.

For the RECT MPG discussed in this paper, the envelope function is given by, $g(y_1) = \left\{ exp(i\phi_1) \, rect\left(\frac{y_1}{W/2}\right) + exp(i\phi_2) \, rect\left(\frac{y_1 - W/2}{W/2}\right) \right\} \otimes \left\{ \frac{1}{W} \sum_{n=-\infty}^{\infty} \delta(y_1 - nW) \right\}$, the $\phi_1$ and $\phi_2$ being the phase heights in regions $h_1$ and $h_2$, each of width W/2 for the MPG of interest. The first curly bracket shows one period of the envelope which is repeated via the comp-convolution in the second curly bracket.

Since P/2 is effectively the opening of the grating, the coherence requirement should be of the order of P (not W). One of the periodicities of $T(y_1)$ is W from the envelope function. But this alone would necessitate high coherence requirement. But since the function is also multiplied by a carrier "sampling" period P, we expect the $T(y_1)$ also creates multiple periodic frequencies in amplitude, a combination of harmonics of $\frac{1}{W}$ and $\frac{1}{P}$ and not just $\frac{1}{W}$. We explicitly derive this in Appendix A.

Note, the Fresnel approximation holds where the incident and generated Huygens's spherical waves in Eq. 4 can be approximated as paraboloid [16]. In the Appendix A, the field amplitude is derived explicitly in terms of the Fourier coefficients of the grating transmission factor $T(y_1)$ from first principles outlined in Goodman [16].

The field amplitude, $U(y,z)$, for the case of the RECT MPG in Eq. 6 is derived in the Appendix A. The key findings are represented here. The amplitude field may be divided into (W,P)-dependent and a P-dependent parts.

$$U(y,z) = A \exp(iB) \frac{1}{L_1 + z} \left[ \sum_{n=-\infty}^{+\infty} \sum_{m=-\infty}^{+\infty} g_m \, sinc\left\{\frac{1}{2}\left(\frac{mP}{W}+n\right)\right\} exp\left\{-\frac{j\pi\lambda L_1 z}{L_1+z}\left(\frac{m}{W}+\frac{n}{P}\right)^2\right\} exp\left\{\frac{j2\pi L_1 y}{L_1+z}\left(\frac{m}{W}+\frac{n}{P}\right)\right\} \right.$$

$$\left. + \sum_{n=-\infty}^{+\infty} exp\{-j\pi n\} sinc\left\{\frac{n}{2}\right\} exp\left\{-\frac{j\pi\lambda L_1 z}{(L_1+z)}\left(\frac{n}{P}\right)^2\right\} exp\left\{\frac{j2\pi L_1 y}{(L_1+z)}\left(\frac{n}{P}\right)\right\} \right]$$

$$= A \exp(iB) \frac{1}{L_1+z}[U_1(y,z,W,P) + U_2(y,z,P)] \qquad (7)$$

These terms are reminiscent of [18-19] except here the frequency is related to a combination of harmonics of $\frac{1}{W}$ and $\frac{1}{P}$. More specifically, in the first term $U_1$ the frequencies are related to $\frac{m}{W} + \frac{n}{P}$ and in the second term $U_2$ to $\frac{n}{P}$.

The intensity as both $\frac{m}{W} + \frac{n}{P}$ modulation terms (Appendix A) but since n/P will be hardly visible directly in the detector, it is approximated by the n = 0 harmonic:

$$I(y,z) \approx \left(\frac{A}{L_1+z}\right)^2 \left[ C_{22}(z,0) + \sum_{m=-\infty}^{+\infty} [C_{11}(z,m,0) + C_{12}(z,m,0) + C_{12}^*(z,-m,0)] \, exp\left\{\frac{j2\pi L_1 y}{L_1+z}\left(\frac{m}{W}\right)\right\} \right]$$

$$= \left(\frac{A}{L_1+z}\right)^2 \left[ I_0 + \sum_{m=1}^{+\infty} I_m(z) \, cos\left\{\frac{2\pi L_1 y}{L_1+z}\left(\frac{m}{W}\right) + \theta_m\right\} \right] \tag{8}$$

Where $C_{ij}$ are autocorrelation terms shown in Appendix A. *Therefore, the intensity produces fringe patterns of period $\frac{m}{MW}$ where the magnification $M = \frac{L_1+z}{L_1}$.* In our case of an even grating, $\theta_m = 0$.

Figure 6 shows an example where the $C_{ij}$ (in Eq. 8 above) were evaluated for the MPG with phase heights $(\pi,0)$ and W=300 μm in order to calculate the intensity. The N-SRDI simulations for the same MPG parameters and geometry settings are also plotted in Figure 6, showing an excellent match.

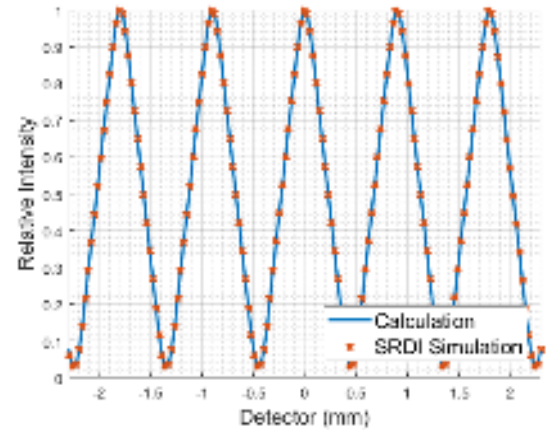

Figure 6. Intensity from Eq. 8 compared to the N-SRDI simulation, for MPG W= 300μm, phase heights $(\pi,0)$ with geometry Dgd (or $L_2$) = 200cm and Dsg (or $L_1$) = 100cm. Pixel size is 100μm and the actual slit size is $L_s$ = 200 μm, with the slit convlution kernel $K_s = L_s$ x $L_2/L_1$ = 400 μm.

*(c) Conditions for Maximum/Minimum Visibility*

As shown in the Appendix A, the visibility of the first harmonic will oscillate in Z with extrema occurring at

$$Z = \frac{L_1 W P k'}{2\lambda L_1 - W P k'} = \frac{L_1}{\frac{2\lambda L_1}{W P k'} - 1} \tag{9}$$

where $k'$ is an integer. For the Z=1 m to 5 m of interest, we found some peaks and troughs as reported in the Table 1, which was corroborated by the N-SRDI simulations. For example, for W=120 $\mu m$, P=1 $\mu m$, the visibility has a local maximum at Z = 2.14 m and local minimum at Z = 4.5 m.

*Note that the maximum Z condition (which is independent of the envelope function Fourier parameters, $g_m$) should be similar for all MPG height pairs. But the envelope function or the height difference in the RECT MPG will affect the visibility.*

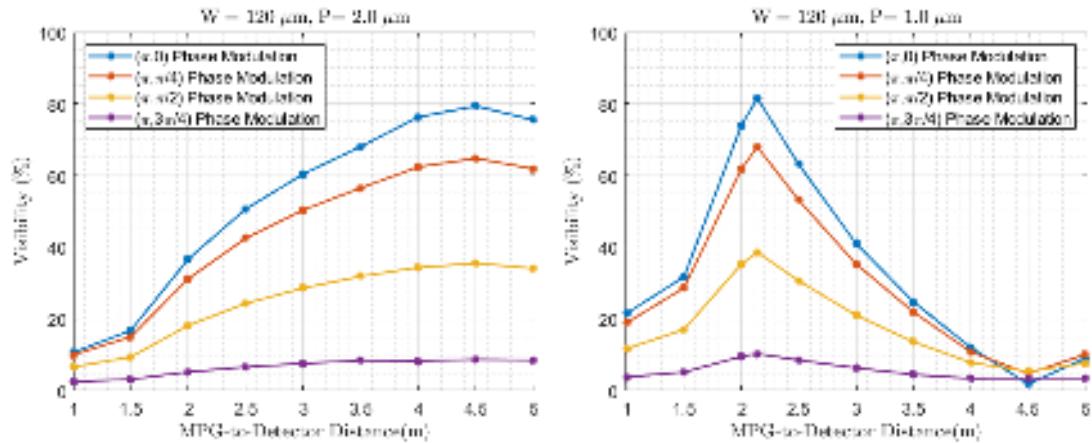

Figure 7. The visibility plots for N-SRDI simulations for different phase height pairs, keeping larger height fixed at $\pi$ and changing the smaller height from 0 to $3\pi/4$. Left and right shows (a) W=120 $\mu$m, P=2 $\mu$m and (b) W=120 $\mu$m, P=1 $\mu$m, respectively. Note maximum or minimum visibility position is independent of the phase heights. The ($\pi$,0) yielded the highest visibility. The maxima and minima correspond to what is predicted by Eq. 9.

To confirm this, we applied the N-SRDI to simulate different MPGs with $h_2$ for the distance ranges of interest here (Dgd = 1 m - 5 m or Dsd = 2 m - 6 m from source) and the wavelength 0.44 nm. We kept the maximum height of grating $h_1$ such that it yields $\pi$-shift and varied the $h_2$ to provide 0, $\pi/4$, $\pi/2$ and $3\pi/4$ phase shifts (for $\lambda$ = 0.44 nm). The results for W=120 $\mu$m, P=2 $\mu$m and P=1 $\mu$m are shown in Figure 7. We therefore confirmed that the Z at maximum visibility is independent of the MPG-type, but the visibility will drop as the $h_2$ is increased. Note for the simulations, the location of peaks was very close whether we applied the slit smoothing or not. In Figure 7 a pixel size of 100 μm and slit of $L_s$ = 75 μm with the magnified convolution kernel $K_s$ = $L_s$ x L2/L1 were applied for each point.

We observed that the peak visibility and minimum visibility matched for the N-SRDI and the values from Eq. 9. In Table 1 we tabulate the distances and the corresponding k-values in the distance range of interest Z from 1 m to 5 m.

**Table 1. Distance for maximum or minimum visibility with Simulation versus Eq. 9**

| W (μm) | P (μm) | Z@Vmax (N-SRDI) (m) | Z@Vmax (Eq. 9) (m) | Z@Vmin (N-SRDI) (m) | Z@Vmin (Eq. 9) (m) |
|--------|--------|---------------------|--------------------|---------------------|--------------------|
| 120 | 2 | 4.5 | 4.5 (for k = 3) | - | - |
| 120 | 1 | 2.14 | 2.14 (for k=5) | 4.5 | 4.5 (for k=6) |

The fringe visibility was highest for $h_2 = 0$ for all MPG-detector distances. This is expected because it would provide the maximum difference from the two components of the grating. This was also the grating which was investigated for X-rays in [15].

Since the fringe visibility was highest for $h_2 = 0$ for all MPG-detector distances, henceforward in this work we used (π,0), as shown in the schematic diagram Figure 8.

***Evaluations of N-SRDI for MPG with (π,0) case***: Fringe visibility, fringe period, and maximum phase sensitivity were evaluated for different modulation periods W, pitches P, slit widths $S_w$, source-to-MPG distances $D_{sg}$, and MPG-to-detector distances $D_{gd}$ as shown in the results Section 3.

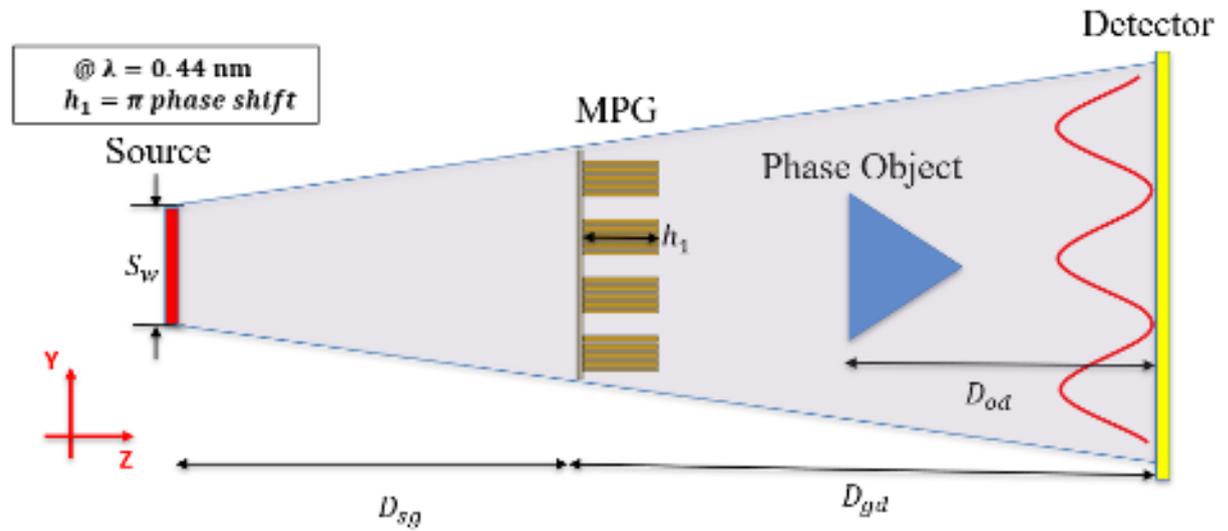

Figure 8. Schematic of the MPG system (not to scale) with a slit source with width $S_w$, MPG with grating heights with phase modulation of $(0, \pi)$, and detector with a phase object at an object-to-detector distance $D_{od}$.

## 2.5 N-SRDI Simulations for a Modulated Phase Grating (MPG) Neutron Interferometer Illuminated by a Polychromatic Beam

We investigated how a polychromatic beam may degrade the visibility for our MPG system. We considered the configuration that produced the best visibility from the previous section. The polychromatic beam was modeled after the one at the NG6 Cold Neutron Imaging Facility (NCI) at the NCNR, which is approximately given by a Maxwell-Boltzmann distribution with a peak wavelength of $\lambda_c = 0.5$ nm [9]. The same MPG is used in both the monochromatic and polychromatic simulations. Since the polychromatic beam can be thought of as the aggregate of incoherent sources, the fringe patterns of each wavelength in the spectrum can be added in intensity according to their weight in the spectrum. Again, the evaluation results are in Results Section.

## 2.6 Single-shot Phase Contrast Recovery with MPG

Phase objects, that is objects that introduce a pure phase shift on the wavefront on the path of the beam were simulated mathematically, and the effect on the interference fringe at the detector was analyzed.

The shift in the interference pattern $\Delta y$ at the detector is related to the refractive angle $\alpha$ imposed by an object on the wave field and therefore the object's differential phase-shift $d\Phi/dy$ as follows [14,25].

$$\Delta y = D_{od} \tan \alpha \approx \frac{\lambda D_{od}}{2\pi} \frac{d\Phi}{dy} \qquad (10)$$

Estimating $\Delta y$, (which itself is a function of y in general), can therefore yield the object's differential phase shift $d\Phi/dy$. Note that $\Delta y$ also depends on system parameters: object-to-detector distance $D_{od}$ and the wavelength $\lambda$, which have to be corrected for a true estimate of the object's differential phase.

Since we have large interference pattern period $W'$ (usually greater than the pixel size), and the detector sampling rate above the Nyquist sampling rate, we use a single-shot recovery using Fourier transforms to demonstrate our system properties, following the method in Ref. [22]. The steps followed are shown below. Before showing the steps, we like to remind that the object's differential phase at the detector space is denoted by $d\Phi/dy$ (this is to be calculated) and the measured $\Delta\phi$ is the actual phase difference calculated via delay $\Delta y$ observed at the detector.

**Step 1.** (a) Take Fast Fourier transform (FFT) of each simulated fringe pattern with and without object. Isolate the first harmonic by windowing the Fourier transforms on one side. A window-width of total width $1/W'$ is chosen symmetrically around the first harmonic peak located at $1/W'$ (i.e., the window extends $\pm 1/(2W')$ around $1/W'$). (b) Perform Inverse Fast Fourier Transform (IFFT) of the one-sided windowed FFTs and take the angle difference of the two complex spatial domain signals. The difference $\Delta\phi$ is related to the spatial shift between the fringe pattern signals $\Delta y$ as

$$\Delta\phi = \phi - \phi_b = \frac{2\pi}{W'}\Delta y \qquad (11)$$

Note: the IFFT of the windowed FFT yields complex exponentials in the spatial domain, ideally the first harmonic signals. These have the frequency $2\pi/W'$. The difference of phase angles $\Delta\phi$ between the blank fringe pattern phase angle $\phi_b$ and the with-object fringe pattern phase $\phi$ is related to the spatial shift $\Delta y$ between these two signals as given by Eq. (11).

**Implementation details:** Note: The N-SRDI code outputs detector data with a detector pixel size of 0.1 μm. We convolve the fringe patterns by independent rectangular window functions to compensate for the fringe blurring due to the slit width and pixel size. Then we subsample the data at a rate equal to the pixel size which is 100 μm.

We interpolate the subsampled pixel-size detector data up by a factor of 1000 by using the MATLAB's *interp* function to resample the function at 0.1 μm before performing the FFT and IFFT (these are the same processing steps used for the standard-dual-grating simulations). The FFT, IFFT, and angle difference of the fringe patterns are all performed in MATLAB. Each phase angle $\phi_b$ and $\phi$ are obtained modulo $\pm\pi$. The resultant phase angle difference has to be unwrapped before integration. Note for some small objects, the $\Delta y$ shift can be sub-pixel. However, since we sample the fringe pattern above the Nyquist rate, (i.e., pixel size $< W'/2$), we can always reconstruct the signals with interpolation and obtain the phase difference. The interpolation used is the default one used by MATLAB's *interp* function, which uses a symmetric special FIR filter that passes the original samples unchanged while minimizes the sum-squared error between the original and the ideal values.

The interpolation step is optional but adds to the robustness of the recovery for cases where the Nyquist edge (frequency) is too close the first harmonic window The subpixel interpolation yields a higher sampling rate and moves the Nyquist edge away from the first harmonic in Fourier domain, which makes it easier to place an automated symmetric window in the Fourier domain to retrieve the first harmonic.

**Step 2 (a)**. To obtain the object's differential phase $d\Phi/dy$ at the detector space (y) from the measured phase difference $\Delta\phi$ we multiply by $(S\lambda)^{-1}$. We show this by combining Eq. (10) and Eq. (11) to obtain

$$\Delta\phi = \frac{\lambda D_{od}}{W'}\frac{d\Phi}{dy} = S\lambda\frac{d\Phi}{dy} \qquad (12)$$

Where is $S = D_{od}/W'$ is the phase sensitivity for the interferometer. Expressed in terms of the object's differential phase profile $d\Phi/dy$:

$$\frac{d\Phi}{dy} = (S\lambda)^{-1}\Delta\phi \qquad (13)$$

Thus, the measured phase difference $\Delta\phi$ must be "corrected" by $1/(\lambda S)$ to obtain $d\Phi/dy$ .

**Step 2 (b)** To obtain the object's differential phase $d\Phi/dy_{ob}$ at the object space ($y_{ob}$), we have to scale (de-magnify) the spatial variable of the detector (y) by $1/M_{obj}$ where $M_{obj} =$ object magnification.

**Step 3.** We integrate $d\Phi/dy_{ob}$ to obtain the object phase profile. We use the MATLAB's *cumtrapz* function for the integration. The scaled grid (object grid) is provided into MATLAB's *cumtrapz* function as the grid argument. Note that we could have also magnified the true object's grid and compared both phase profiles in the detector space (y), but we wanted to avoid any processing with the true object phase profile.

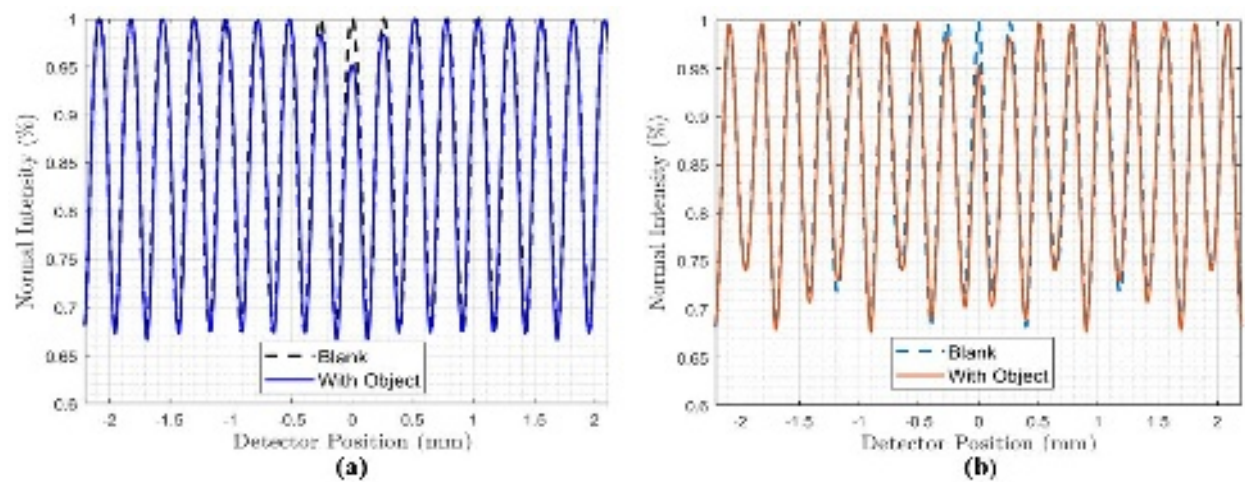

Figure 9. (a) Interference pattern with and without a triangle phase object with a peak phase of $8\pi$ rad using the system parameters W= 50 µm, Pitch = 2.0 µm, $D_{sd} = 5$ m, $D_{sg} = 1$ m, and $D_{od} = 2.5$ m. Intensity values are originally sampled along the detector plane at 0.1 µm increments and are filtered by two independent rectangular window functions for blurring effects due to slit width and pixel size. (b) Shows both fringe patterns subsampled at 100 µm increments and interpolated back to a sampling rate of 0.1 µm. The dark lines on the left and right are locations where the $\Delta y$ shift of the fringe pattern were checked manually.

**Table 2. Measured fringe patterns shifts from left and right windows in Figure 9**

| Fringe Pattern $\Delta y$ | Left Window $\Delta y$ (µm) | Right Window $\Delta y$ (µm) |
|---|---|---|
| Expected | 5.5 | -5.5 |
| Measured from Plot | 5.0±1.4 | -5.1 ±1.3 |

Figure 9 shows an example simulated blank and with-object fringe pattern. The phase object is a triangular object with a maximum phase shift of $8\pi \ rad$, ramping up/down over $\pm 800$ µm spatial extent. It is placed an object-to-detector distance of 2.5 m. In reality, this phase profile could be a silicon wedge sample with a maximum height of 275.6 µm at the center and falling off to zero over $\pm 800$ µm on either side of its peak. Figure 9(a) shows both fringe patterns just after being convolved with two independent rectangular filters to account for the blurring of the slit width of 50 µm and the pixel size of 100 µm. Figure 9(b) shows the fringe patterns after being subsampled at every 100 µm and then interpolating between sampled values to return to a 0.1 µm sampling rate. In the N-SRDI code, the objects are represented simply as mathematical phasors $e^{i\Phi(y_{ob})}$ which are applied to the wave field at the object-to-detector distance, where $\Phi(y_{ob})$ is the object phase profile.

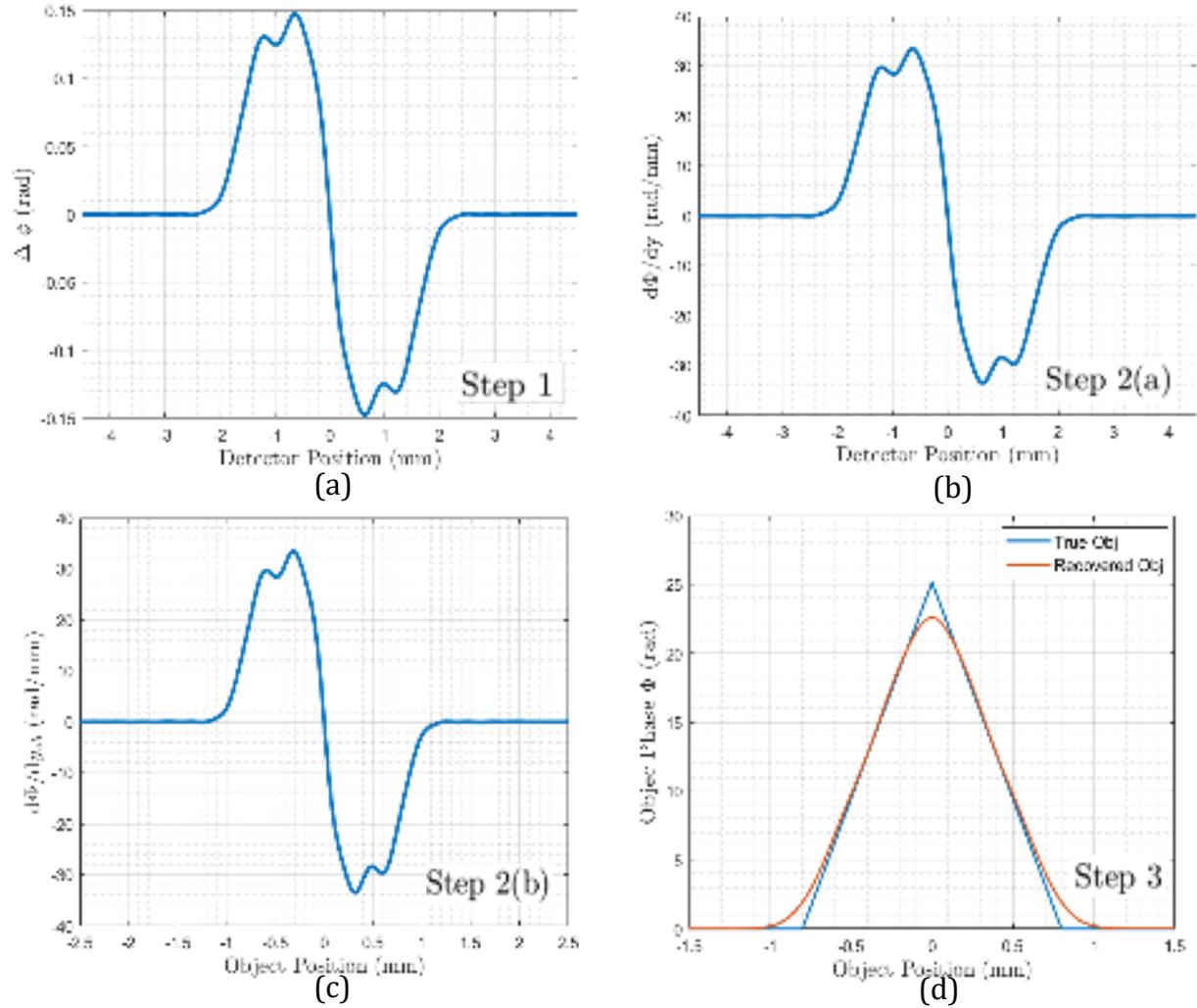

Figure 10 (a) Shows the measured phase difference $\Delta\phi$ at the detector from Figure 9(b) Obtained via the single-shot method explained in Step 1. (b) The measured phase difference $\Delta\phi$ is corrected by a $(S\lambda)^{-1}$ factor to obtain the object differential phase $d\Phi/dy$ at the detector space (y). (c) The grid of $d\Phi/dy$ is scaled by the inverse object magnification factor $D_{so}/D_{sd}$ to obtain the object differential phase $d\Phi/dy_{ob}$ in the object space ($y_{ob}$). (d) Shows the object phase $\Phi$ retrieved from $d\Phi/dy_{ob}$ by integration (Step 3). For this instance, no detector noise was added.

The expected spatial shifts in the fringe pattern due to this triangular phase object are given by Eq. (10). Since the object phase consists of an up and down ramp, the spatial shift $\Delta y$ is a positive constant on the left side of the fringe pattern center (detector position 0 μm) and a negative constant on the right side of the fringe pattern center. As a manual check, we calculated the average $\Delta y$ in the two regions over three cycles each at the 90% normalized intensity line in Figure 9(b). In Table 2 we compare the expected $\Delta y$ and our measurements showing that the values are very close to each other.

Figure 10 shows the **Steps 1-3** recovery with the blank and with-object fringe patterns shown in Figure 10(b). First, we show the measured $\Delta\phi$ after **Step 1** in Figure 10(a). This approximately shows the rectangular pattern expected from the differential phase of the triangular object consisting of two ramps. The differential phase is shown in Figure 10(b) after correction by $(S\lambda)^{-1}$ as in **Step 2**. In Figure 10(c), the detector grid is scaled by the object magnification to switch to the objects grid space ($y_{ob}$) instead of the detector grid space (y). Then Figure 10(d) shows the integrated phase of the object (**Step 3**).

Lastly, Figure 11 elaborates the **Step 1** and **Step 2(b)** of the phase recovery process for the triangular and parabolic phase objects for object-to-detector distances varied from 0.5 m to 3.5 m. Figure 11(a) and Figure 11(c) shows the measured phase $\Delta\phi$ (**Step 1**) at the detector grid. Figure 11(b) and Figure 11(d) shows the object differential phase $d\Phi/dy_{ob}$ (**Step 2(b)**) on the object grid space ($y_{ob}$). They both approximately show the rectangular and linear $d\Phi/dy_{ob}$ expected for triangular and parabolic phase profiles, respectively.

While these concepts were demonstrated in this Methods section without adding detector noise, we added realistic Poisson noise to the detector in the Results section before phase recovery.

In the results Section 3, we show quantitative phase recovery error analysis of triangular, parabolic, and trapezoidal objects with object-to-detector distances of 0.5 m to 3.5 m and maximum object phases of $0.8\pi$ rad and $8\pi$ rad. *This was done after adding realistic Poisson noise to the detector pixels which we explain in details in the results section.*

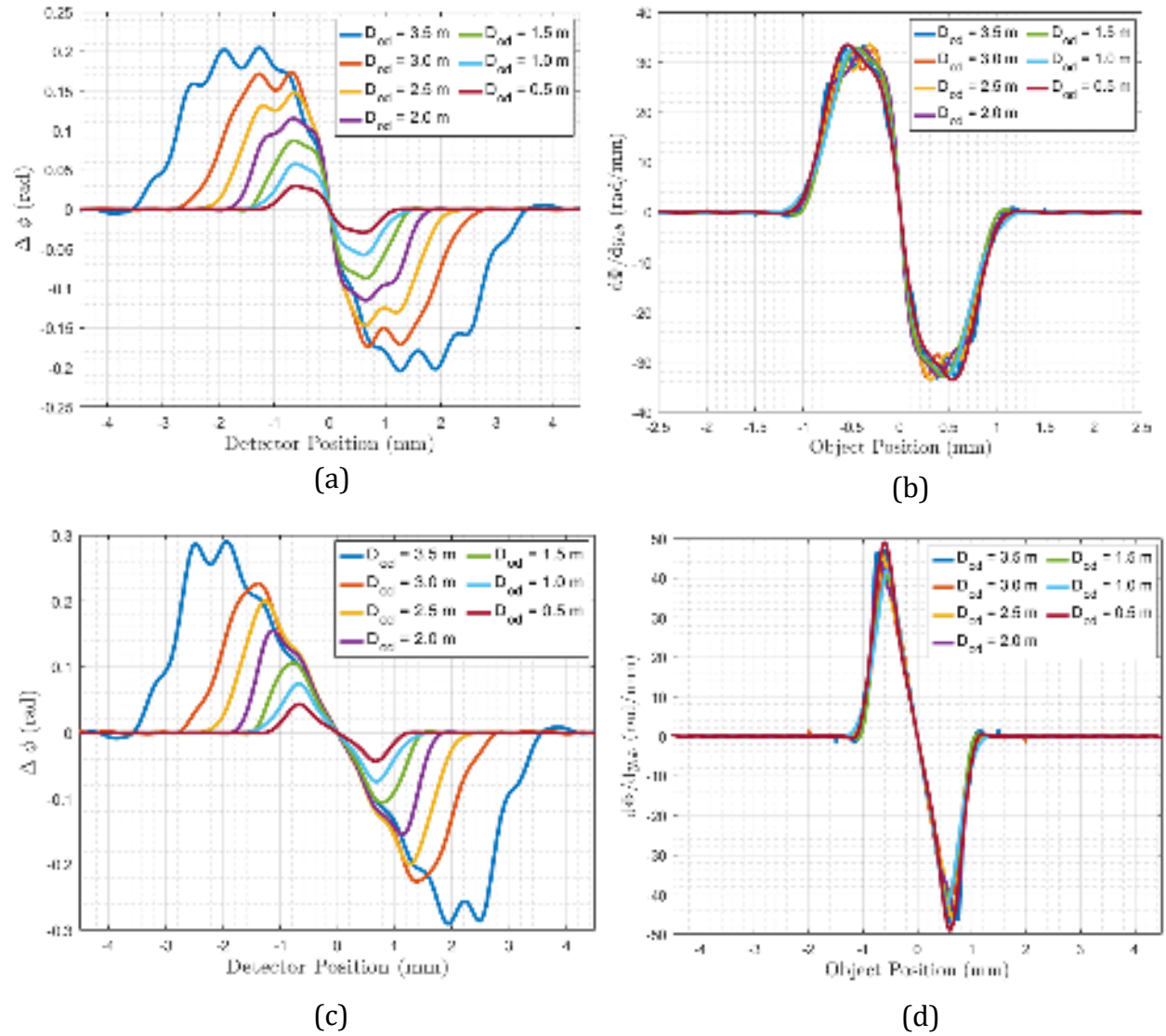

Figure 11. (a) Measured phase difference $\Delta\Phi$ at the detector for object-to-detector distances of $0.5$ m to $3.5$ m for the $8\pi$ $rad$ triangle phase object. (b) Shows the corresponding object differential phase $d\Phi/dy_{ob}$ at the object plane (c) Measured phase shifts $\Delta\Phi$ at the detector for object-to-detector distances of $0.5$ m to $3.5$ m for the $8\pi$ $rad$ parabolic phase object (d) Shows the corresponding object differential phase $d\Phi/dy_{ob}$ at the object plane. $\Delta\phi$ are corrected by a $(S\lambda)^{-1}$ factor to obtain $d\Phi/dy$ and the detector grids are also corrected by the inverse object magnification factor $D_{od}/D_{sd}$ .

## 3. Results

### 3.1 Evaluation of N-SRDI Simulations for a Standard-Dual-Grating System, Comparisons with Theory and Experiments

To test our simulator N-SRDI, we simulated the standard-dual-phase grating system and compared our results to theory and experiments. We varied the grating-to-grating distance from 7 mm to 16 mm in increments of 1 mm and obtained fringe periods and fringe visibilities for the monochromatic experiments performed in Ref. [9] with $\lambda = 0.44$ nm. The fringe periods obtained via the simulations were compared to the theoretical closed-form prediction given by in Eq. (11) in Ref. [8] as shown in Figure 12. We observed an excellent agreement of the simulated fringe periods to the corresponding theoretical values.

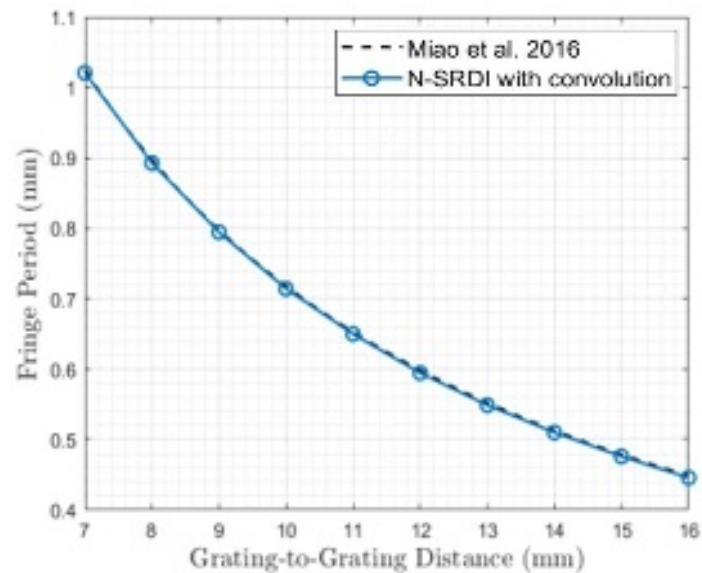

Figure 12. Fringe period from standard-dual-grating simulations compared to theory. N-SRDI simulations were done with the monochromatic configuration parameters used in Ref. [9]. The theoretical fringe period is given by Eq. (11) in Ref. [8].

Figure 13 plots the N-SRDI with convolution fringe visibility results and the experimental fringe visibility results for the monochromatic experiments in Ref. [9]. Not all but every other grating-to-grating distance was simulated and shown below since we varied the grating-to-grating distance in spacings of 1 mm. The general trend was captured by the N-SRDI with convolution simulations with the visibility peaking around 11 mm to 12 mm. The curve also fell within the statistical agreement of experiments (indicated by the error bars) except for the grating-to-grating distances D of 7 mm, 8 mm, and 11 mm.

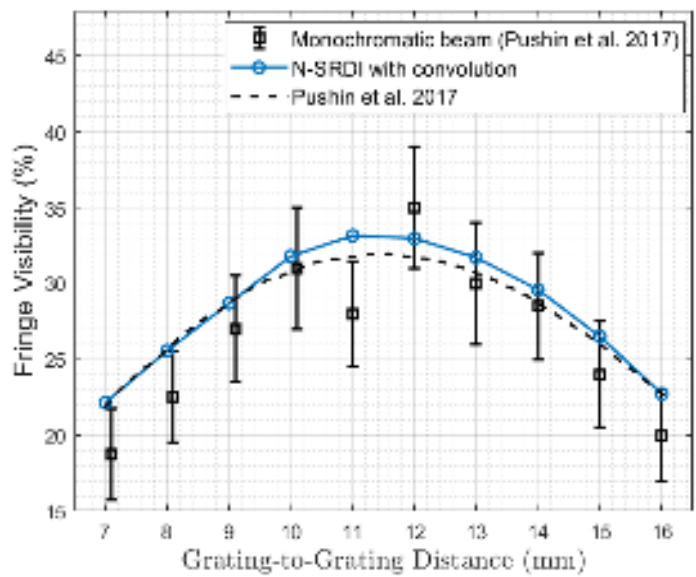

Figure 13. Fringe visibility comparison of N-SRDI with convolution simulations to experiments performed in Ref. [9] The dashed lines show the closed-form expected visibility given in Eq. 12 in Ref. [8] which was plotted in Ref. [9].

## 3.2 N-SRDI Simulations for a Modulated Phase-Grating (MPG) Neutron Interferometer Illuminated by a Monochromatic Beam

We simulated the MPG system with grating heights modulation of $(0, \pi)$, the same slit-source in a monochromatic beam ($\lambda$=0.44 nm) used for the standard-dual-grating system in the previous section, different modulation periods W, and the two different pitches of 2.0 µm and 1 µm. The pitch of P = 2.0 µm is approximately the period of some of the gratings currently available at the National Institute of Standards and Technology Center for Neutron Research (NCNR).

We show an excellent performance for an equivalent geometry as the set up in Ref [9]. This is shown in Figure 14(a) of interference fringe "carpet" for W = 300 µm and P = 2 µm, where the intensity carpet is obtained by placing the detector at different distances from the MPG. The carpet shows a diverging self-image of the modulation pattern. At the source-to-MPG distance of 1 m and MPG-to-detector distance of 2.0 m (SDD of 3 m), we are at a nearly equivalent SDD of 2.99 m used in the setup in Ref. [9] for their monochromatic configuration. A fringe visibility analysis of the whole carpet is shown in Figure 14(b). The normalized fringe pattern with the maximum visibility of 94.2% at the MPG-to-detector distance of 2 m is shown in Figure 14(b-c).

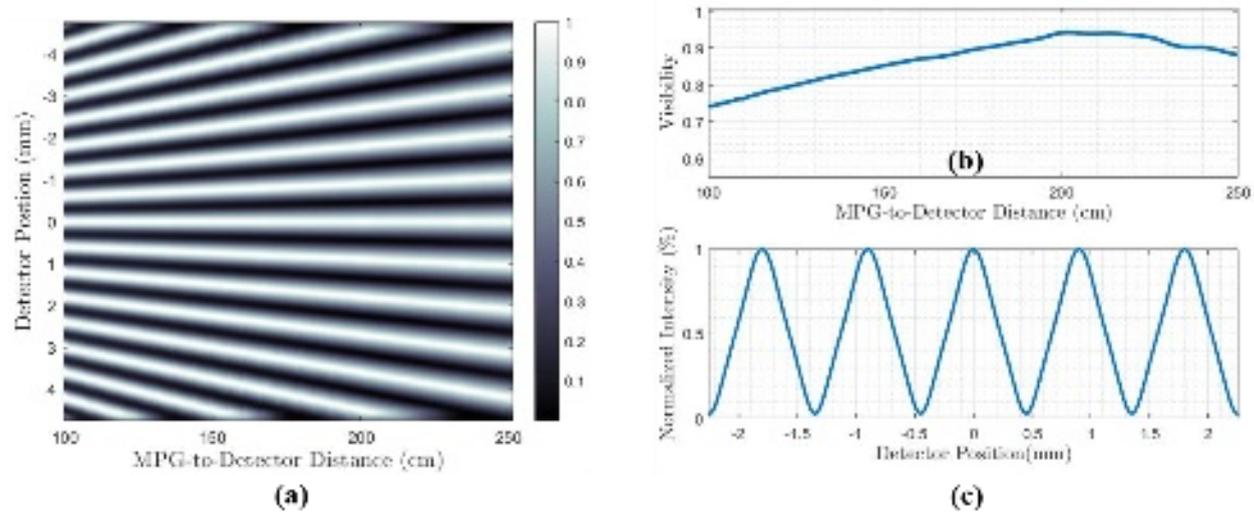

Figure 14. (a) Interference fringe "carpet" generated with N-SRDI and pixel and source convolution, for the source-to-MPG distance of 1 m, W= 300 μm, and pitch P = 2.0 μm. The pixel size is 100 μm and the slit width $L_s$=200 μm (kernel $K_s = L_s \times L_2/L_1$). (b) Fringe visibility analysis of the carper shows the maximum visibility of about 94.2% at the MPG-to-detector distance of about 2 m. (c) The normalized fringe pattern at the MPG-to-detector distance of 2 m, with maximum visibility of about 94.2%.

The example shows that the MPG system may be expected to yield more than twice the visibility compared to the standard-dual-grating system's maximum visibility ~39 % in a nearly equivalent set-up geometry to the one used in Ref. [9]. Note that at this magnification the fringe period is about 900 μm, therefore the slit or pixel size effects are relatively small.

We summarize other cases in Figures 15 and 16 with high visibility. In Figure 16 we also point out which cases the geometry is more compact than the system in Ref [9]. In Figures 15 (a) and (b) we show the visibility and fringe period for different modulation periods W=50-600 μm (all with a pitch of 2.0 µm) at varying MPG-to-detector distances. We make sure to not include cases where the pixel size > $W'/2$, which would lead to insufficient sampling of the fringe pattern (Nyquist theorem) and cause aliasing. We observed several operating points with fringe visibility V > 40% in Figure 15(a-b) for different fringe periods. The ability to control the fringe period is important as the fringe period determines the autocorrelation probing lengths in dark-field imaging (DFI) [24], and it is also inversely proportional to the phase sensitivity [14].

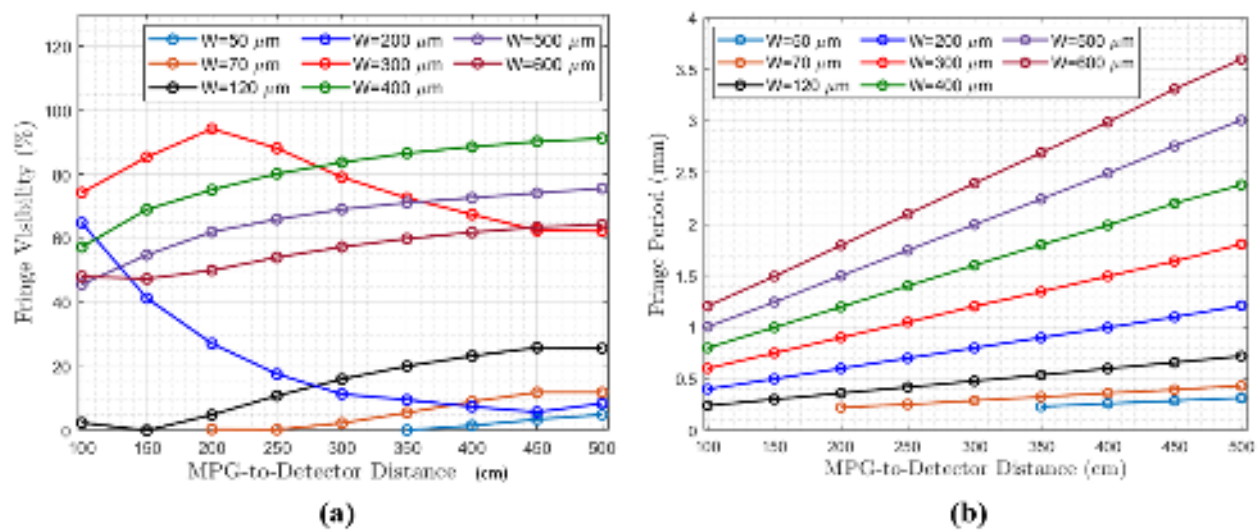

Figure 15. Measured (a) fringe visibility and (b) fringe periods from simulations with varying MPG-to-detector distances of 1.0 m to 5.0 m. The fringe period $W'$ is given by the geometric magnification $D_{sd}/D_{sg}$ of W. The source-to-MPG distance is 1.0 m and the pitch is 2.0 µm. The pixel size is 100 µm and the slit width $L_s$ = 200 µm (kernel $K_s = L_s$ x $L_2/L_1$). Fringe sampling frequency is ensured to be greater than the Nyquist rate (pixel size < $W'/2$).

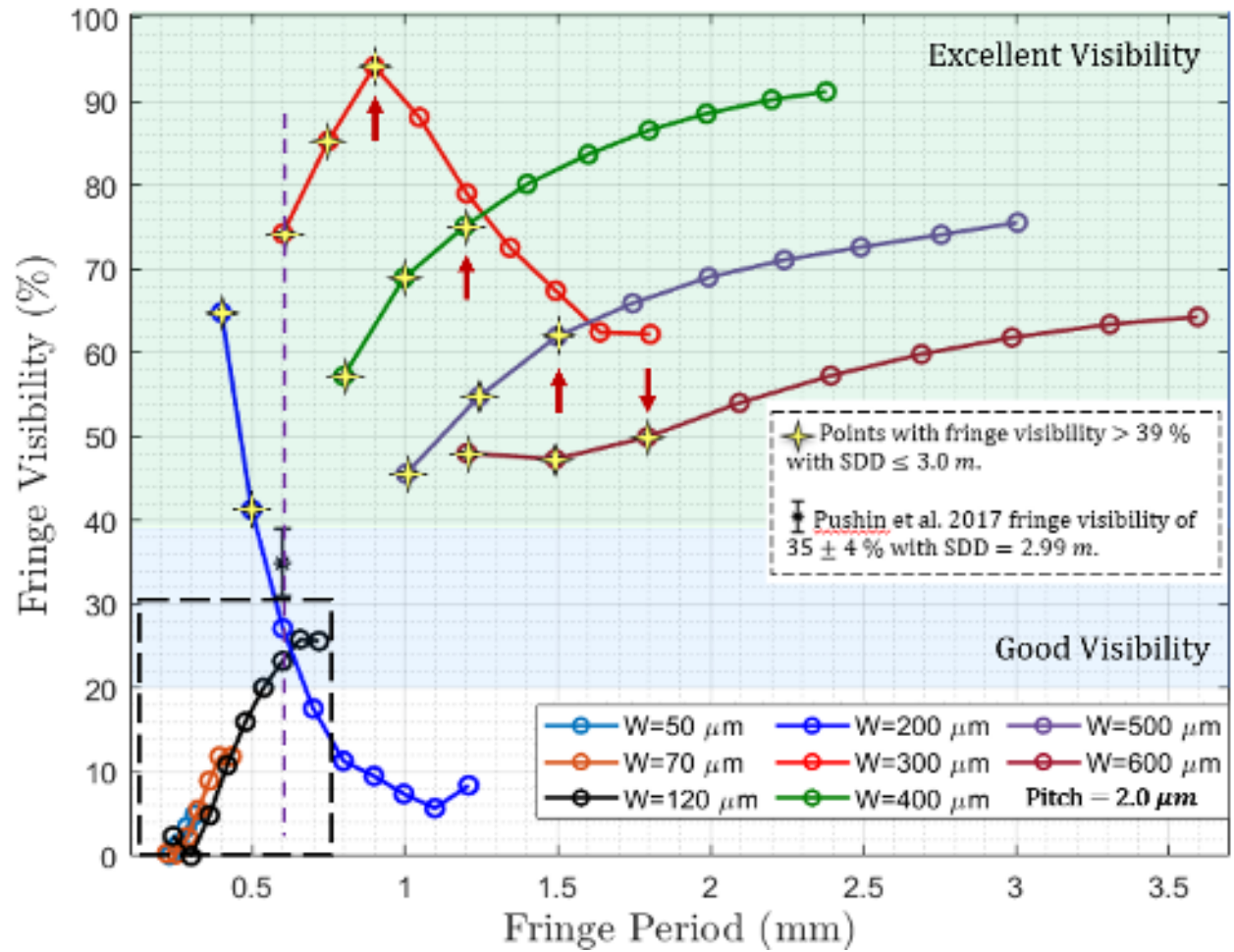

Figure 16. Fringe visibility versus period for a source-to-MPG distance of 1.0 m, varying MPG-to-detector distances from 1.0 m to 5.0 m, and pitch of 2.0 µm. The pixel size is 100 µm and the slit $L_s$=200 µm (kernel $K_s = L_s$ x $L_2/L_1$). The cases with a SDD of 2.99 m ($\approx$ 3.0 m) like the monochromatic experimental configuration in Ref. [9] are shown with the red arrows.

To visualize the information in a meaningful way, we plot the visibility versus fringe-period in Figure 16, with the MPG-to-detector distance implicitly varied. The 9 circles in each curve correspond to the visibilities at 9 different MPG-to-detector distance (100 - 500 cm, with intervals of 50 cm), i.e., the SDDs range from 200 cm to 600 cm. We partitioned into green and light blue zone where the MPG provided high visibility, V > 39%, and acceptable 20-39% respectively. In order to compare to the standard-dual-grating system (in Figure 13), the peak visibility from those experiments is shown as the asterisk with an error bar (falling on the "good visibility" zone). The dashed (purple) vertical line (through the asterisk case) shows that for the same fringe period for the asterisk case (from the monochromatic setup in Ref. [9] with the best-case visibility of 39%), there is the $W = 300\,\mu m$ MPG design that can yield a better visibility of approximately 75% visibility for MPG-to-detector distance of 100 cm (SDD of 200 cm). Other designs such as $W = 200\,\mu m$ yields 40-65% visibility for fringe period 0.5mm or lower.

If one is looking for high visibility $V > 39\%$ with a compact setup geometry, we point out the operating points marked with yellow stars in Figure 16 where the SDD $\leq 300$ cm.

We therefore show several operating points of the MPG system with higher visibility than the highest reported visibility of 39% from the standard-dual-grating system in Ref. [9], with similar, smaller or larger fringe periods and same or more compact geometry.

We investigated a lower the pitch for the lower fringe period region in dashed square in Figure 16. Lowering pitch also enabled the $D_{sg}$ to be smaller (0.5m) and the coherence requirement hold. The smaller fringe period zone is shown in Figure 17 for the $P = 1\,\mu m$ cases. We make sure not to include cases where the pixel size $> W'/2$, which would lead to insufficient sampling of the fringe pattern (Nyquist theorem). Unfortunately, the slit kernel convolution effect dominated at these lower fringe periods and the visibility was poor for $L_s = 200\,\mu m$ with convolution kernel varying as $K_s = L_s \times L_2/L_1$.

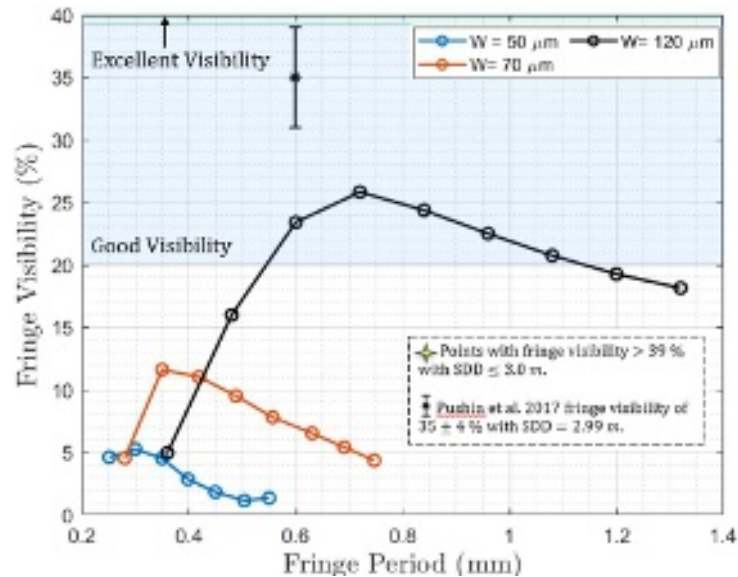

Figure 17. Fringe visibility versus period for $D_{sg} = 0.5$m, $D_{gd} = [1.0$ m - $5.0$ m], and pitch = 1.0 μm. Fringe sampling frequency is ensured to be greater than the Nyquist rate (pixel size < $W'/2$). The pixel size is 100 μm and the slit $L_s$=200 μm (kernel $K_s$ = $L_s$ x $L_2/L_1$). The slit kernel convolution dominated at the lower fringe periods and visibility suffered.

We also investigated if the visibility for the smaller W cases (50 μm, 70 μm, and 120 μm) where the fringe period is smaller, could be improved by using narrower slit widths $S_w$ (albeit knowing that the flux reaching the detector would be reduced proportionately and typically a $G_0$ grating would be required instead of a slit). We kept the $S_w$ (or $L_s$) = 200 μm as a reference. As expected in general, we see in Figure 18 (a-c), using a narrower slit width $S_w$ improved the visibility for the W = 50, 70 and 120 μm for these fringe patterns with smaller periods.

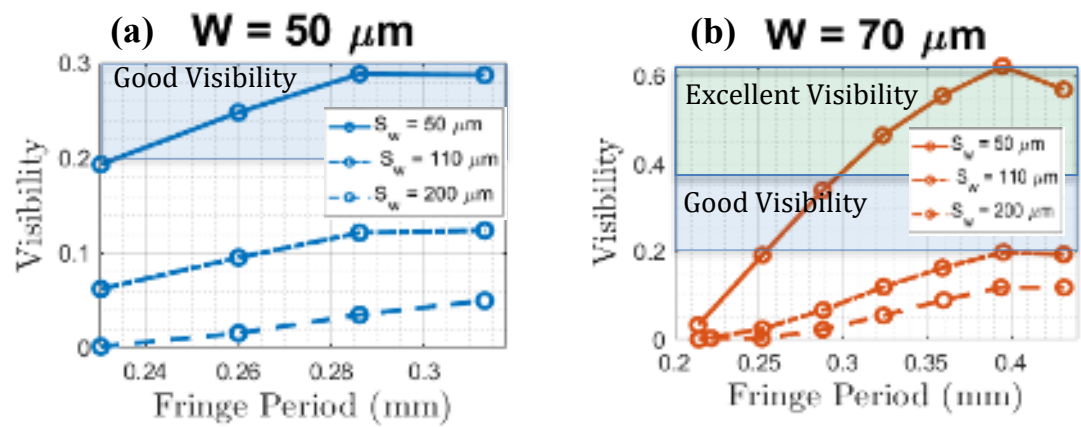

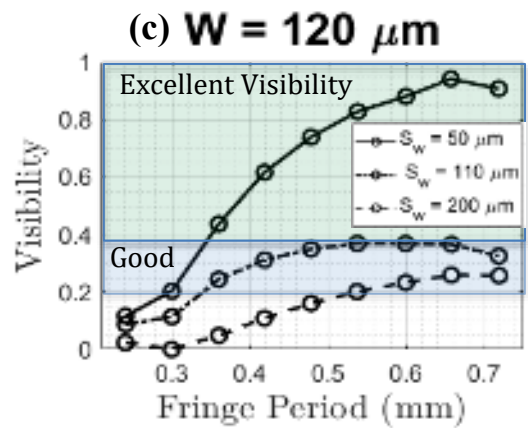

Figure 18. (a-c) Fringe visibility for slit sizes $S_w$ (or Ls) of 50 µm, 110 µm, and 200 µm, respectively. The source-to-MPG distance is 1.0 m and the pitch is 2.0 µm. The MPG-to-detector distances range from 1.0 m to 5.0 m. The pixel size is 100 µm. The slit convolution kernel is calculated as $K_s = S_w \times L_2/L_1$.

Lastly, we calculated the maximum phase sensitivities of our MPG system. The maximum phase sensitivity is given by $S_{max} = D_{gd} / W'$, (where the object is closest to the MPG, i.e. $D_{od} = D_{gd}$). This is analyzed for different MPG-to-detector distances as shown in Figure 19(a) for different grating modulation periods with source-to-MPG distance kept at 100cm. There is also an increase in phase sensitivity with lower $W'$ (lower fringe period) as shown in Figure 19(b) for a grating-to-detector distance of 5.0 m. Phase sensitivity information is sparse for the standard-dual-grating system but available in the literature for the conventional neutron Talbot-Lau interferometer (TLI). For MPG modulation periods W between 50 µm to 120 µm, we note that the theoretical maximum phase sensitivities are between 4.1 x 10$^3$ to 10.0 x 10$^3$ (0.41-1 x 10$^4$) when using SDDs up to 3.5 m (MPG-to-detector distance up to 2.5 m). These sensitivities are comparable or greater than that of the conventional neutron TLI of 4.5 x 10$^3$ with a SDD of 3.5 m and a beam wavelength of 0.44 nm [7,23].

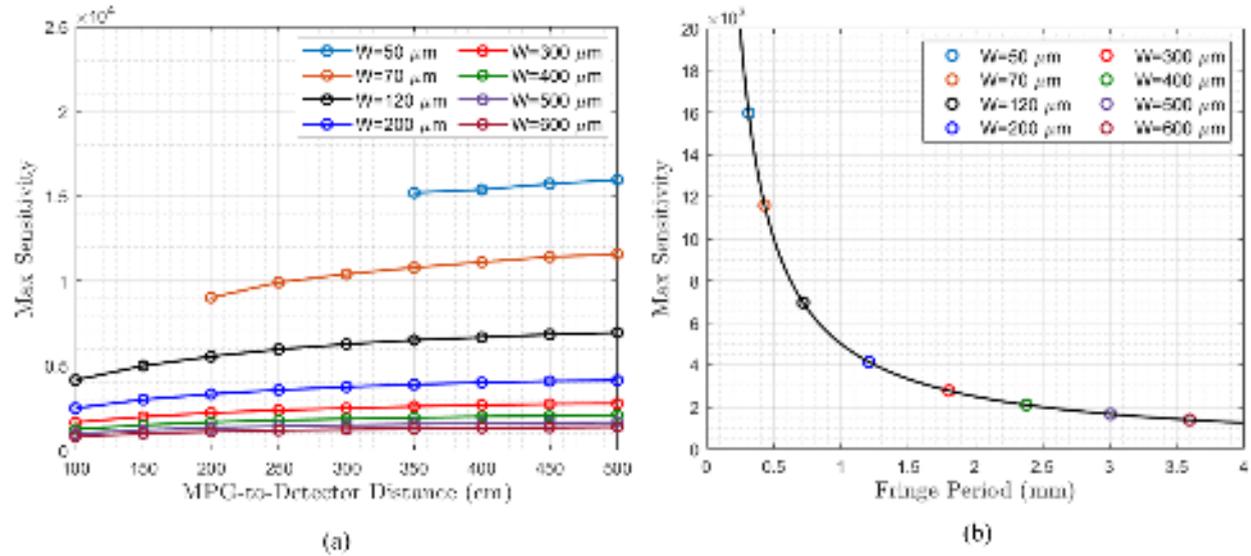

Figure 19. (a) Maximum phase sensitivity for different MPG-to-detector distances (b) Phase sensitivity versus fringe period for the MPG-to-detector distance of 5 m. The pixel size is 100 μm and the slit width is 200 μm. Fringe sampling frequency is ensured to be greater than the Nyquist rate (pixel size $< W'/2$).

## 3.3 N-SRDI Simulations for a Modulated Phase Grating (MPG) Neutron Interferometer Illuminated by a Polychromatic Beam

We used a polychromatic beam for the setup which produced the best visibility of 94.2% in the monochromatic simulations ($\lambda = 0.44$ nm) to see how the fringe visibility would degrade. The polychromatic beam is approximately described by Maxwell-Boltzmann spectrum with a peak wavelength $\lambda_c = 0.5$ nm. The MPG used in both monochromatic and polychromatic simulations were the same and had fixed phases of $(0, \pi)$ at $\lambda = 0.44$ nm. The system parameters include a grating with modulation period W of 300 μm, pitch of 2 μm, source-to-MPG distance of 1 m, and MPG-to-detector distance of 2.0 m. The pixel size is 100 μm and the slit width is 200 μm. While we note that the visibility dropped from 94.2% to 68.9% (as calculated from intensities shown in system with a best-case monochromatic visibility of around 39% in Ref [9].

Figure 20), this polychromatic visibility is still significantly higher than the standard-dual-grating

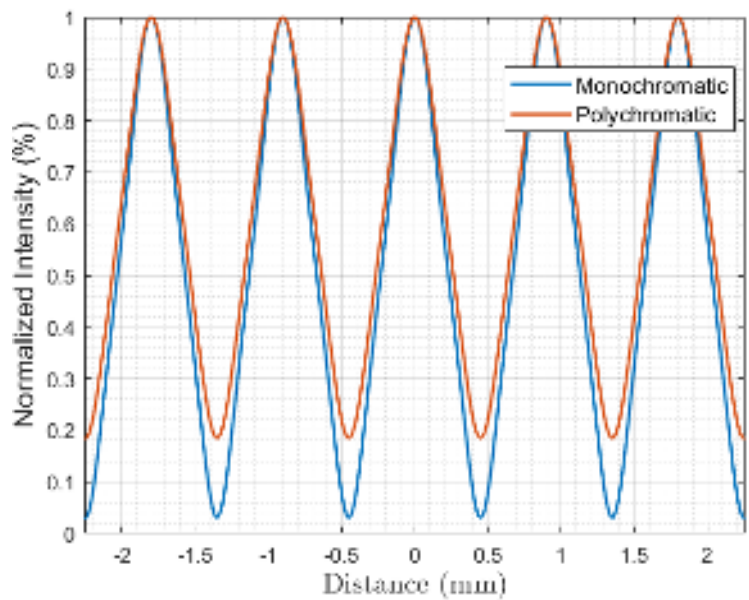

Figure 20. Fringe visibility degradation due to polychromatic beam for our MPG system. We consider the configuration that produced the best visibility of **94.2%** using a monochromatic beam of $\lambda = 0.44$ nm (blue trace). The polychromatic beam described by a Maxwell-Boltzmann spectrum with peak wavelength $\lambda_c$ =0.5 nm produces a fringe pattern with visibility of **68.9**% (orange trace). The system parameters include a grating with modulation period W of 300 µm, pitch of 2 µm, source-to-MPG distance of 1 m, and MPG-to-detector distance of 2.0 m.

### 3.4 Single-shot Phase Contrast Recovery with MPG

Figure 21 shows examples of linear, trapezoid and quadratic phase objects recovered using the

single-shot method described in Section 2.6. The recovery is shown for noiseless case and with

realistic Poisson noise. The noise addition step is explained as follows. First have added Poisson

statistics with a realistic flux level to our detector intensity. To determine a realistic flux we have

taken the average intensity for experiments shown in Ref [9] shown to be about 10,000. Taking

into account the scintillation gain camera settings, the actual neutron counts are about a factor of

10 smaller for the case shown in Ref [9]. This makes the neutron counts realistically 1000. We

add the noise to each pixel *after* pixel and slit convolutions and then perform the subsampling

and interpolation. Instead of phase stepping acquisitions, we take several single shot acquisitions

and add them. In Ref [9], 8 phase steps and 3 frames were taken for each step. For our case about

16 lines noisy acquisitions were added together – this would be similar to 8 steps at 2 frames. We compared a typical noisy recovery with noiseless case. The parameters of MPG system used are W= 50 µm or 120 µm, Pitch = 2.0 µm , $D_{sg} = 1$ m, and $D_{sd} = 3$ m. The pixel size is 50 µm or 100 µm and the slit width is 37.5 µm or 45 µm (convolution kernel of $L_s$ x $L_2/L_1$ ). A 50% duty cycle G0 grating may be used instead of a single slit (see Discussion). The main detriments to the recovery were at the transitions due to the phase-sensitivity effects. For each phase object the noisy recovery process described above was repeated 100 individual times (100 different seeds) and the root-mean-square-error (RMSE) and the standard deviation are reported. For system parameters of W=120 µm, pixel size 50 µm or 100 µm, and slit width of 45 µm, the average RMSE error was higher as expected from the smaller phase sensitivity (larger fringe period 360 µm at W=120 µm as opposed to 150 µm for W=50 µm).

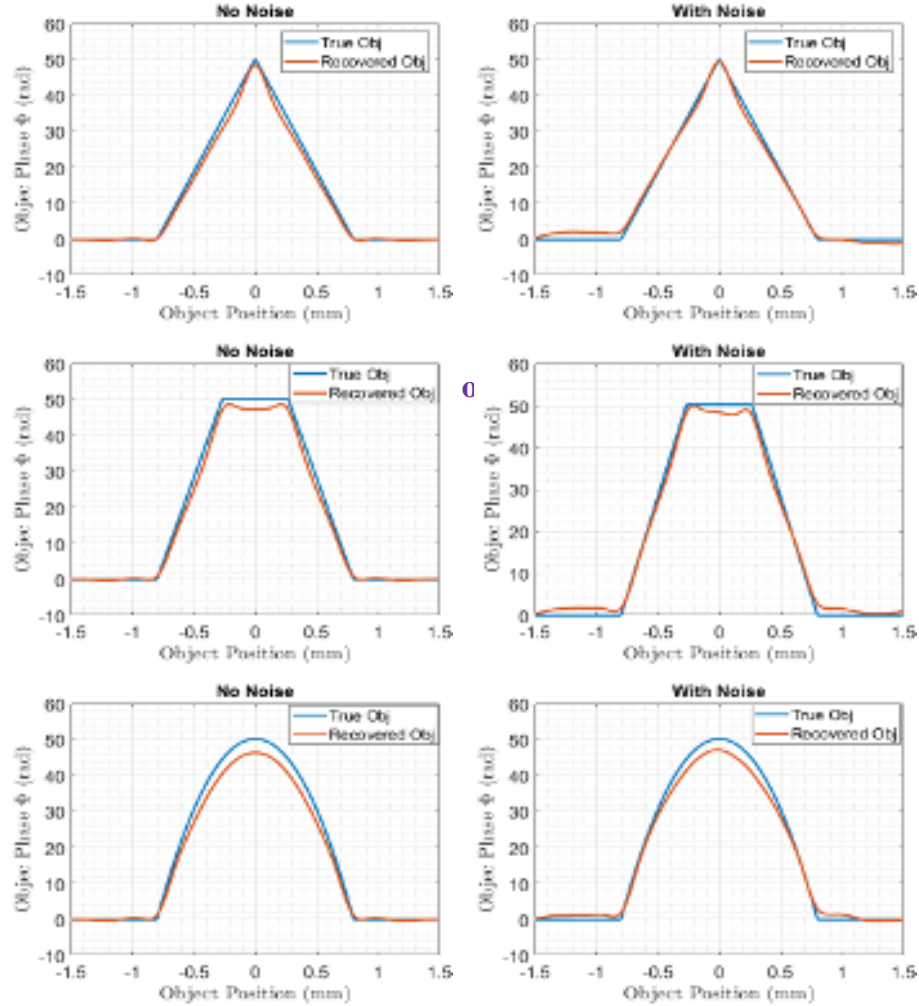

Figure 21. Example Phase recovery for a ramp, trapezoid and a quadratic shaped object, for noiseless and a typical with-noise case. The acquisition parameters are MPG W= 50 µm, MPG Pitch = 2.0 µm, $D_{sd}$ = 3m, and $D_{sg}$ = 1 m. The object distance to detector $D_{od}$ = 1.7m. The pixel size is 50 µm and slit size 37.5 µm (slit Kernel Ks = 37.5µm x $L_2/L_1$=75µm). A G0 with 37.5 µm opening and 50% duty cycle may be used instead of a single slit. Realistic Poisson noise was added to the detector simulation. 16 noisy lines were added before phase recovery.

**Table3. MPG phase recovery errors for Triangular, Trapezoid and Paraboloid phase objects**

| MPG | | Slit Size* Ls (µm) | Pixel Size (µm) | Triangular RMSE (+/-std) Over 100 seeds | | Trapezoid RMSE (+/-std) Over 100 seeds | | Paraboloid RMSE (+/-std) Over 100 seeds | |
|---|---|---|---|---|---|---|---|---|---|
| W (µm) | P (µm) | | | No Noise | With Noise | No Noise | With Noise | No Noise | With Noise |
| 50 | 2 | 37.5 | 50 | 1.44 | 2.22 +/- 0.86 | 1.88 | 2.48+/- 0.91 | 2.55 | 2.91+/- 0.93 |
| 120 | 2 | 45 | 50 | 1.64 | 3.04 +/- 1.20 | 2.82 | 3.84 +/- 1.70 | 2.80 | 3.74 +/- 1.86 |
| 120 | 2 | 45 | 100 | 1.68 | 4.08 +/- 1.74 | 2.89 | 4.89+/- 2.30 | 2.66 | 4.69 +/- 2.37 |

* The convolution kernel used is Ks = $2L_s$ due to the pinhole magnification, $L_2/L_1$

## 4. Discussion

Our wave-propagation simulator for neutron PGMI applications can accept various designs (e.g. modulated phase gratings and standard-dual-grating). These analytical simulations use Sommerfeld-Rayleigh diffraction integrals (SRDI) to calculate the intensity of the diffracted neutron wave at the detector plane. We derived the theory behind the MPG and compared intensity and minima/maxima conditions with the SRDI. Fringe pattern blurring at the detector due to the spatial coherence loss of a slit source and pixel size resolution was accounted for by convolving the fringe pattern with independent appropriate rectangular window functions. Fringe patterns are also subsampled according to the pixel size of detector and interpolated to increase the sample rate of their signal. Simulations of the standard-dual-grating system with a monochromatic configuration produced visibility results in good agreement with experimental data from Ref [9] and fringe periods which also agreed well the theory given in Ref [8]. Note also that experiments include gratings with imperfect square profiles of gratings, slight misalignment between gratings, coherence loss from air scattering over long distances, and mechanical vibrations [10-11]. While our analytical SRDI simulations and fringe processing techniques do not account for these sources of noise, the fringe visibilities of our simulations still matched well to experimental values. The SRDI and fringe pattern processing techniques were used to investigate our MPG system, which demonstrated it to be a promising type of PGMI system by producing higher visibilities (50% to 94.2%) than the standard-dual-grating system for a range of possible modulation period W parameters (Figure 16). At the lower end of the modulation periods, a narrower slit can aid in yielding higher visibilities (Figures 18). The margin of improvement the MPG system displays is high for most cases and appreciable from what is currently reported in literature for the standard-dual-grating system. For example, W = 300µm and pitch = 2 µm produces a visibility of near 94.2% with a comparable SDD as the standard-dual-grating system operating in a monochromatic configuration. When that same MPG configuration was illuminated by a polychromatic beam, the visibility dropped to 68.9% which is still higher than the best visibility case in Pushin et al. [9], and it demonstrates that our MPG system is potentially robust to polychromatic neutron sources. Operating in the smaller W range also improves the phase sensitivity, yielding as much as 10.0 x $10^3$ for SDDs up to 3.5 m.

Our mathematical treatise of intensity and the equation for maximum visibility was borne out with N-SRDI simulations, as shown in Figure 6 and Table 1.

For small fringe period a smaller slit size is important due to the pinhole magnification amplifying the effects of slit blurring. For a smaller slit, a $G_0$ grating will be useful. Therefore, one aspect to consider is the design of the $G_0$ for the MPG system as shown in Ref [14]. If $P_0$ = actual $G_0$ pitch, the projected pitch at the detector will have the pinhole magnification $P_0' = D_{gd}/D_{sg}$ x $P_0$ (or in other notation $L_2/L_1$ x $P_0$). *This should equate the fringe period $W' = M$ x $W = D_{sd}/D_{sg}$ x $W$ for coherence [14]. So,* $D_{gd}/D_{sg}$ x $P_0 = D_{sd}/D_{sg}$ x $W$. *Thus, P0 can be calculated from geometry for a given W as* $\boldsymbol{P_0 = D_{sd}/D_{gd}}$ **x W**.  For clarity it is emphasized that, the magnification $M = D_{sd}/D_{sg}$ is a different factor than the pinhole magnification effect, which is $L_2/L_1$ (that is, $D_{gd}/D_{sg}$).

For a sample calculation, we consider the phase recovery example in Figure 21. For this case, W = 50 µm, P = 2 µm, slit size = 37.5 µm for $D_{sg}$ = 1m, $D_{sd}$ = 3m, (or $D_{gd}$ = 2m). The required $\boldsymbol{P_0 = D_{sd}/D_{gd}}$ **x W** = 3/2 x 50 µm = 75 µm. The derivation can also be shown step by step. For W=50 µm, the fringe period at the detector is $W' = M$ x $W = D_{sd}/D_{sg}$ x $W = 3$ x 50 µm = 150 µm. The projected $P_0'$ has to be therefore 150 µm. Therefore $P_0$ = 75 µm. This makes the duty cycle $L_s/P_0$ = 37.5/75 or 50%.

The performance improvement with MPG from the dual phase grating system comes with the added benefit of the MPG system being a single-phase-grating system which does not require precise alignment of two gratings.

A potential advantage of the MPG system compared to TLI (for X-rays or neutron interferometry) is that the fluence absorbing analyzer grating between object and detector that is necessary for the TLI is not necessary for MPG. The advantage is potentially more than just one less grating element and less alignment. For the same fluence at the detector for both systems (and for similar object positions relative to the source) the dose to the object will be less for the MPG system than for the TLI, where the analyzer absorbs half the X-rays or neutrons. This is particularly important when imaging bio-samples in X-ray or neutrons. The caveat is that, source brilliance, exposure time, and source grating $G_0$ has to be considered to get same fluence at the detector. If the duty-cycle or open-ratios are the same, say 50% for both systems, then exposure time may be kept the similar for similar fluence-level at the detector. But for example, if the $G_0$ has a 50% open ratio for the TLI and 40% for MPG system then the exposure time has to be increased about 25% to obtain the same fluence at the detector.

The gratings can be manufactured using existing manufacturing techniques according to the lead manufacturer Microworks, GmbH.

For X-ray imaging (energies around 25 keV) we have obtained from Microworks GmbH rectangular MPGs with W=120 µm, pitch P = 1 µm, as well as triangular MPGs with W=120 µm and P=1.8 µm. We have conducted successful experiments with both and are preparing a manuscript for submission of those results [26].

For neutrons, Si is the material of choice with deep UV grey tone lithography being one option to modulate etch depth in the silicon with locally different heights of the resist serving as etch stops.

Other modulation shapes such as triangular or sinusoidal MPG is possible [14]. We observed in simulation that the fringe pattern may have ringing and harmonics for the RECT modulated grating at some geometries. The quality of fringes may be better for an ideal sinusoidal modulation in the sense that there are less harmonics for sinusoidal MPG due to the smoother fall off. However, the RECT modulated grating is easier to manufacture and maintain quality control and of lower cost. For these reasons, in this work we limited our investigation to RECT MPG.

The 1D N-SRDI simulations were performed using the LSU LONI QB3 cluster. A single compute node was utilized which has two Intel Xeon Platinum 8260 Processors each with 24 cores (2.4 GHz). OpenMP, an application program interface (API) for the thread-based parallelism, is utilized to generate multiple threads that run in parallel in order to speed up calculation times when running simulations. On average, a single simulation can take about 30 core-hours to complete. We expect more complicated/2D geometries to be amenable to an MPI (Message Passing Interface) implementation for parallel computing since it would allow the use of multiple compute nodes.

## 5. Conclusions

We have proposed a novel phase-grating moiré interferometer system for cold neutrons that only requires a single modulated phase grating (MPG) for phase-contrast imaging, as opposed to the two or three phase gratings in previously employed PGMI systems. The MPG system promises to deliver a significantly better fringe visibility relative to previous experiments in the literature that use a two-phase-grating-based PGMI. When operating in the smaller W range and using SDDs of up to 3.5 m, the MPG system could provide comparable or greater theoretical maximum phase sensitivity when compared to the conventional Talbot-Lau interferometer. The single MPG reduces the precise alignment requirements needed for multi-grating systems. Like other PGMI systems, the MPG system does not require the high-aspect-ratio absorption gratings used in Talbot-Lau interferometers which are challenging to manufacture and reduce the neutron flux reaching the detector.

## 6. ACKNOWLEDGEMENT

The work presented here is in part from first-author Ivan Hidrovo's MSc thesis work [28] and in part from ongoing thesis work of Hunter Meyer [27] (both advised by co-author J. D), towards their Medical Physics MSc degree, Department of Physics and Astronomy, LSU, Baton Rouge, LA. This work was funded in part by NSF EPSCOR RII Track 4, Award # 1929150. The simulations were conducted with high-performance computational resources provided by Louisiana State University and the Louisiana Optical Network Infrastructure.

## Disclaimer

Certain commercial equipment, instruments, or materials (or suppliers, or software, ...) are identified in this paper to foster understanding. Such identification does not imply recommendation or endorsement by the National Institute of Standards and Technology, nor does it imply that the materials or equipment identified are necessarily the best available for the purpose.

## APPENDIX A

### Field Amplitude Derivation

See Figure 1 for a general set up of the system. The Fresnel condition is derived by Patorski et al [19] for a standard periodic grating with period d in terms of the discrete Fourier coefficients. Since our grating has W and P components, we expect to have complicated periodic frequencies, with harmonic combinations of $\frac{m}{W} + \frac{n}{P}$. Hence, we derived the Fresnel conditions from first principles [27]. First, we derive the amplitude of the field under Fresnel condition for MPG grating $T(y)$ (see Eq. 6, reproduced here for convenience).

$$T(y_1) = \left[ \left\{ g(y_1) \sum_{n=-\infty}^{\infty} \delta(y_1 - nP) \right\} + \sum_{n=-\infty}^{\infty} \delta\left(y_1 - \frac{nP}{2} - P/2\right) \right] \otimes rect\left(\frac{y_1}{P/2}\right)$$

where $g(y_1) = \left\{ \exp(i\phi_1) \, rect\left(\frac{y_1}{W/2}\right) + \exp(i\phi_2) \, rect\left(\frac{y_1 - W/2}{W/2}\right) \right\} \otimes \left\{ \frac{1}{W} \sum_{-\infty}^{\infty} \delta(y_1 - nW) \right\}$, the

$\phi_1$ and $\phi_2$ being the phase heights in regions $h_1$ and $h_2$, each of width W/2 for the MPG of interest.

The first curly bracket shows one period of the envelope which is repeated via the comp-convolution in the second curly bracket.

where the field in the Fresnel region is derived in Goodman (Eq. 4-10 in [16]),

$U(x, y, z)$

$$= \frac{\exp(ikz)}{i\lambda z} \exp\left(\frac{ik(x^2 + y^2)}{2z}\right) FT\left( U(x', y', z) \exp\left(\frac{ik(x^2 + y^2)}{2z}\right) \right) \Bigg|_{f_x = \frac{x}{\lambda z}, f_y = \frac{y}{\lambda z}} \qquad (A.1)$$

where

$U(x', y', z) = \frac{\exp(ikr_0)}{r_0} T(y') \approx \frac{\exp(ikL_1)}{L_1} \exp\left(\frac{ik(x'^2 + y'^2)}{2L_1}\right) T(y')$, making a parabolic

approximation of incident spherical wave from point source, as done in Patorski [19].

Substituting $U(x', y', z)$, the FT-term in Eq. A.1 becomes

$$FT\left( U(x', y', z) \exp\left(\frac{ik(x^2 + y^2)}{2z}\right) \right) \Bigg| = FT(T(y')) \otimes FT\left( \exp\left(\frac{ik(x'^2 + y'^2)}{2}\left(\frac{1}{L_1} + \frac{1}{z}\right)\right) \right) \quad (A.2)$$

The $FT(T(y'))$ of the particular MPG $T(y')$ Eq. 6, is given by

$$FT(T(y')) = \frac{P}{2} sinc\left(\frac{Pf_y}{2}\right) \left[ \left\{ F(g(y)) \otimes \frac{1}{P} \sum_{n=-\infty}^{\infty} \delta\left(f_y - \frac{n}{P}\right) \right\} + \left\{ \frac{1}{P} \exp\left(\frac{-j2\pi Pf_y}{2}\right) \sum_{n=-\infty}^{\infty} \delta\left(f_y - \frac{n}{P}\right) \right\} \right]$$

The P and $\frac{1}{P}$ scaling terms cancel. Note the grating is assumed to be constant lines along the x-direction, which just introduces an overall $\delta(f_x)$ and leaves the function to convolute intact in the x-direction.

Then, carrying out the convolution in Eq. A.2 and, plugging in $f_y = \frac{y}{\lambda z}$ and $f_x = \frac{x}{\lambda z}$ and simplifying, we obtain the amplitude in Eq. A.1 as

$$U(y,z) = \text{A}exp(iB)\frac{1}{L_1+z}\left[\sum_{n=-\infty}^{+\infty}\sum_{m=-\infty}^{+\infty}g_m\,sinc\left\{\frac{1}{2}\left(\frac{mP}{W}+n\right)\right\}exp\left\{-\frac{j\pi\lambda L_1 z}{L_1+z}\left(\frac{m}{W}+\frac{n}{P}\right)^2\right\}exp\left\{\frac{j2\pi L_1 y}{L_1+z}\left(\frac{m}{W}+\frac{n}{P}\right)\right\}\right.$$

$$\left.+\sum_{n=-\infty}^{+\infty}exp\{-j\pi n\}sinc\left\{\frac{n}{2}\right\}exp\left\{-\frac{j\pi\lambda L_1 z}{(L_1+z)}\left(\frac{n}{P}\right)^2\right\}exp\left\{\frac{j2\pi L_1 y}{(L_1+z)}\left(\frac{n}{P}\right)\right\}\right] \qquad (A.3)$$

$$= Aexp(iB)\frac{1}{L_1+z}[U_1(y,z,W,P)+U_2(y,z,P)] \qquad (A.4)$$

Here $U_1(y,z,W,P)$ is the first dual sum (depends on W and P) and $U_2(y,z,P)$ is the second single sum. Also, we lumped some real (and z-independent) and exponential phasor terms in $Aexp(iB)$ as they will result in z-independent scaling term $A^2$ in intensity $I(y,z) = U(y,z)U^*(y,z) = |U(y,z)|^2$. Note that the $g_m$ are the Fourier coefficients of the envelope function, with period W,

$$g(y) = \sum_{m=-\infty}^{\infty}g_m exp\left(\frac{j2\pi my}{W}\right) \quad or, \quad F(g(y)) = \sum_{m=-\infty}^{\infty}g_m\,\delta\left(f_y-\frac{m}{W}\right)$$

where $g_m$ is found by Fourier transform integral of $g(y)$. Eq. A.3 is reminiscent of Patorski [19] or Guigay [18], except with a double summation due to the dual-harmonics. We are interested in $I(y,z) = |U(y,z)|^2$ which is what we observe in detector.

**Field Intensity and Visibility**

From the Eq. A.4, we can write

$$I(y,z) = \left(\frac{A}{L_1+z}\right)^2[U_1U_1^*+U_2U_1^*+U_1U_2^*+U_2U_2^*]$$

Writing each of $U_1$ and $U_2$ in terms of z-dependent and y-dependent terms,

$$U_1(y,z) = \sum_{n=-\infty}^{+\infty}\sum_{m=-\infty}^{+\infty}b_1(m,n,z)\,exp\left\{\frac{j2\pi L_1 y}{L_1+z}\left(\frac{m}{W}+\frac{n}{P}\right)\right\}$$

$$U_2(y,z) = \sum_{n=-\infty}^{+\infty} b_2(m,n,z) exp\left\{\frac{j2\pi L_1 y}{(L_1+z)}\left(\frac{n}{P}\right)\right\}$$

where $b_1(m,n,z) = g_m sinc\left\{\frac{1}{2}\left(\frac{mP}{W}+n\right)\right\} exp\left\{-\frac{j\pi\lambda L_1 z}{L_1+z}\left(\frac{m}{W}+\frac{n}{P}\right)^2\right\}$ and

$b_2(m,n,z) = exp\{-j\pi n\} sinc\left\{\frac{n}{2}\right\} exp\left\{-\frac{j\pi\lambda L_1 z}{(L_1+z)}\left(\frac{n}{P}\right)^2\right\}$

We can show

$$U_1 U_1^* = \sum_{n=-\infty}^{+\infty}\sum_{m=-\infty}^{+\infty} C_{11}(m,n,z) \, exp\left\{\frac{j2\pi L_1 y}{L_1+z}\left(\frac{m}{W}+\frac{n}{P}\right)\right\} \quad A.5$$

where $C_{11}(m,n,z) = b_1(m,n,z) \otimes b_1^*(-m,-n,z)$, which is the **auto-correlation** of the

$b_1(m,n,z)$. *Note immediately that this component of the intensity (and in fact for the others) has*

$exp\left\{\frac{j2\pi L_1 y}{L_1+z}\left(\frac{n}{P}\right)\right\}$ *which is a modulation of the harmonics of P and cannot really be detected by*

*the practical detector of 50-100μm without an analyzer grating as is typically used in Talbot-Lau*

*Interferometers.*

Similarly, $U_1 U_2^*$, $U_2 U_1^*$ and $U_2 U_2^*$ can be written in equivalent forms, to get

$$I(y,z) = \left(\frac{A}{L_1+z}\right)^2 \sum_{n=-\infty}^{+\infty}\sum_{m=-\infty}^{+\infty}[C_{11}(z,m,n)+C_{12}(z,m,n)$$

$$+\, C_{12}^*(z,-m,-n)]\, exp\left\{\frac{j2\pi L_1 y}{L_1+z}\left(\frac{m}{W}+\frac{n}{P}\right)\right\} + C_{22}(z,n) exp\left\{\frac{j2\pi L_1 y}{L_1+z}\left(\frac{n}{P}\right)\right\}$$

where $C_{12}(z,m,n)$ is the cross-correlation of $b_1(m,n,z)$ and $b_2(n,z)$ and $C_{22}(z,n)$ is the auto-correlation of $b_2(n,z)$.

It is noted again that since this is the final *intensity* at the detector, the P-harmonics, i.e., $\left(\frac{n}{P}\right)$ will be blurred by pixel size. This is the reason why for a Talbot-Lau system with standard phase grating with pitch P (of the order of $1 - 4\ \mu m$), a second grating is needed (the G2, absorption grating) to observe the fringes with a practical detector of pixel size $50\text{-}100\ \mu m$.

Therefore, we can consider only $n=0$ and the intensity can be shown to be

$$I(y,z) \approx \left(\frac{A}{L_1 + z}\right)^2 \left[C_{22}(z,0) + \sum_{m=-\infty}^{+\infty} [C_{11}(z,m,0) + C_{12}(z,m,0) + C_{12}^*(z,-m,0)]\ exp\left\{\frac{j2\pi L_1 y}{L_1 + z}\left(\frac{m}{W}\right)\right\}\right]$$

$$= \left(\frac{A}{L_1 + z}\right)^2 \left[I_0 + \sum_{m=1}^{+\infty} I_m(z)\ cos\left\{\frac{2\pi L_1 y}{L_1 + z}\left(\frac{m}{W}\right) + \theta_m\right\}\right] \qquad A.6$$

*Therefore, the intensity produces fringe patterns of period $\frac{m}{MW}$ where magnification $M = \frac{L_1 + z}{L_1}$.* In our case of an even grating, $\theta_m = 0$. The $I_m = \kappa(m) = 2|C_{11}(z,m,0) + C_{12}(z,m,0) + C_{12}^*(z,-m,0)|$. The average is contributed by $I_0 = C_{22}(0) + \kappa(0)$.

The auto-correlations and cross-correlations are given by (for $n = 0$, i.e., not considering standalone $n/P$ harmonic modulations, as explained earlier)

$$C_{11}(z,m,0) = \sum_{m'}\sum_{n'} b_1(m - m', -n')\ b_1^*(-m', -n')$$

$$= \sum_{m'}\sum_{n'} g_{m-m'} g_{-m'}^*\ sinc\left\{\frac{\pi}{2}\left(\frac{(m - m')P}{W} - n'\right)\right\} sinc\left\{\frac{\pi}{2}\left(\frac{(-m')P}{2W} - n'\right)\right\} exp\left\{\frac{j\pi L_1 z}{L_1 + z}\left(\frac{2mn'}{WP} + \frac{2mm' - m^2}{W^2}\right)\right\}$$

And,

$$C_{12}(z,m,0) = g_m \sum_{n'} sinc\left\{\frac{\pi}{2}\left(\frac{mP}{W} - n'\right)\right\}\ sinc\left\{\frac{\pi n'}{2}\right\} exp\{-j\pi n'\} exp\left\{\frac{j\pi \lambda L_1 z}{L_1 + z}\left(\frac{2mn'}{WP} - \frac{m^2}{W^2}\right)\right\}$$

The visibility is guided by the first harmonic component that is $\frac{1}{MW}$ that is for m = 1.

In other words, visibility is given by

$$\text{visibility} = \frac{A_1}{A_0} = \frac{2|C_{11}(z,1,0) + C_{12}(z,1,0) + C_{12}^*(z,-1,0)|}{C_{22}(z,0) + |C_{11}(z,0,0) + C_{12}(z,0,0) + C_{12}^*(z,0,0)|}$$

The visibility will maximize and minimize according to the exponentials in $C_{11}(z, 1, 0)$ and $C_{12}(z, 1, 0)$. This is difficult to pin down in general, however, setting m=1 in the equations for $C_{11}(z, m, 0)$ and $C_{12}(z, m, 0)$ ignoring $\frac{1}{W^2}$ in comparison to $\frac{1}{WP}$ terms, we can see both terms have $exp\left\{\frac{j\pi\lambda L_1 z}{L_1 + z}\left(\frac{2n'}{WP}\right)\right\}$.

This may lead to a condition, $\frac{2\pi\lambda L_1 z}{L_1 + z}\left(\frac{1}{WP}\right) = \pi k'$

$$z = \frac{L_1 WP k'}{2\lambda L_1 - WP k'} \qquad (A.7)$$

Where $k'$ is an integer.

We also observe that the location of Z is independent of the grating envelope function (though the visibility is).

**APPENDIX B**

**Source Shift Derivation**

Here the effect of applying a shift to a parabolic wavefront source will be shown to follow pinhole-type magnification under the Fresnel Approximation. First, the diffracted field amplitude as derived by Goodman (ignoring constant multipliers) [16]:

$$U(x, y, z) \propto \iint_{-\infty}^{\infty} U(x', y', 0)\exp\left[\frac{jk}{2z}(x'^2 + y'^2)\right]\exp\left[\frac{-jk}{z}(xx' + yy')\right]dx'dy' \qquad B.1$$

For a source located on the optical axis, the field at the grating under parabolic wavefront illumination is:

$$U(x', y', 0) = T(y')\frac{e^{jkL_1}}{L_1}\exp\left[\frac{jk}{2L_1}(x'^2 + y'^2)\right]$$

where T(y') is the grating transmission function. Shifting the source by $y_0$ yields:

$$U(x', y', 0) = T(y') \frac{e^{jkL_1}}{L_1} \exp\left[\frac{jk}{2L_1}\left(x'^2 + (y' - y_0)^2\right)\right]$$

Plugging this into B.1 and extracting the constants yields:

$$U(x, y, z) \propto \iint_{-\infty}^{\infty} T(y') \exp\left[\frac{jk}{2L_1}\left(x'^2 + (y' - y_0)^2\right)\right] \exp\left[\frac{jk}{2z}(x'^2 + y'^2)\right] \exp\left[\frac{-jk}{z}(xx' + yy')\right] dx' dy'$$

Expanding and combining the exponentials yields:

$$U(x, y, z) \propto \exp\left[-\frac{jky_0^2}{2L_1}\right] \iint_{-\infty}^{\infty} T(y') \exp\left[\frac{jk}{2L_1}(x'^2 + y'^2) + \frac{jk}{2z}(x'^2 + y'^2) - \frac{jky'y_0}{L_1} - \frac{jk}{z}(xx' + yy')\right] dx' dy'$$

Which can then be arranged to:

$$U(x, y, z) \propto \exp\left[-\frac{jky_0^2}{2L_1}\right] \iint_{-\infty}^{\infty} T(y') \exp\left[\frac{jk}{2L_1}(x'^2 + y'^2) + \frac{jk}{2z}(x'^2 + y'^2) - \frac{jk}{z}\left(xx' + y'\left(y + \frac{y_0 z}{L_1}\right)\right)\right] dx' dy'$$

It is immediately evident that (besides multiplicative factors), the field for a grating illuminated by a source shifted by $y_0$ will result in a field that is equivalent but shifted (in the opposite direction) by $\frac{y_0 z}{L_1}$. This is corroborated by Figure 5. It should be noted that Figure 5 is generated using a true spherical wave (point source) for illumination, and the corroboration further validates the parabolic wave approximation used in Appendix A.

# 7. REFERENCES

[1] F. Pfeiffer, C. Grünzweig, O. Bunk, G. Frei, E. Lehmann, and C. David, Phys. Rev. Lett. **96**, 215505 (2006).

[2] K.W. Herwig, *Introduction to the Neutron*, in *Neutron Imaging and Applications. A Reference for the Imaging Community*, H.Z. Bilheux, R. McGreevy, and I. Anderson (eds.), (Springer, Boston, M.A., 2009), pg.3-12.

[3] H. Rauch and S. A. Werner, *Neutron Interferometry: Lessons in Experimental Quantum Mechanics*, (Oxford University Press, Oxford, 2000).

[4] C. Grünzweig, C. David, O. Bunk, M. Dierolf, G. Frei, G.Kühne, J. Kohlbrecher, R. Schäfer, P. Lejcek, H. M. R. Rønnow et al. Phys. Rev. Lett. **101**, 025504 (2008).

[5] B. Betz, R. P. Harti, M. Strobl, J. Hovind, A. Kaestner, E. Lehmann, H. Van Swygenhoven, and C. Grünzweig, Rev. Sci. Instrum. **86,** 123704 (2015).

[6] T. Neuwirth, A. Backs, A. Gustschin, S. Vogt, F. Pfeiffer, P. Böni, and M. Schulz, Sci. Rep. **10,** 1764 (2020).

[7] Y. Kim, J. Kim, D. Kim, D.S. Hussey, and S. W. Lee, Rev. Sci. Instrum. **90**, 073704 (2019).

[8] H. Miao, A. Panna, A. A. Gomella, E. E. Bennett, S. Znati, L. Chen, and H. Wen, Nat. Phys. **12**, 830 (2016).

[9] D. Pushin, D. Sarenac, D. Hussey, H. Miao, M. Arif, D. Cory, M. Huber, D. Jacobson, J. LaManna, J. Parker et al., Phys. Rev. A **95**, 043637 (2017).

[10] D. Sarenac, D. Pushin, M. Huber, D. Hussey, H. Miao, M. Arif, D. Cory, A. Cronin, B. Heacock, D. Jacobson et al., Phys. Rev. Lett. **120**, 113201 (2018).

[11] B. Heacock, D. Sarenac, D. G. Cory, M. G. Huber, D. S. Hussey, C. Kapahi, H. Miao, H. Wen, and D. A. Pushin, AIP Adv. **9**, 085115 (2019).

[12] M. Strobl, J. Valsecchi, R.P. Harti, P.Trtik, A. Kaestner, C. Gruenzweig, E. Polatidis, J. Capek, Sci Rep. **12,** 3461 (2019).

[13] J. Dey, N. Bhusal, L. Butler, J. P. Dowling, K. Ham, V. Singh, "Phase Contrast X-ray Interferometry" US Patent No., 10872708 , Dec 22, 2020

[14] J. Xu, K. Ham, and J. Dey, "X-ray interferometry without analyzer for breast CT application: a simulation study," *J. Med. Imag.*, vol. 7, no. 2, 2020, doi: 10.1117/ 1.JMI.7.2.023503

[15] A. Pandeshwar, M. Kagias, Z. Shi, *and* M. Stampanoni "Envelope modulated x-ray grating interferometry", Appl. Phys. Lett. 120, 193701 (2022) https://doi.org/10.1063/5.0087940

[16] J. W. Goodman, *Introduction to Fourier Optics*, 2nd ed. (McGraw-Hill, New York, 1996).

[17] J. P. Wilde and L. Hesselink, "Statistical optics modeling of dark-field scattering in X-ray grating interferometers: Part 1. Theory," Opt. Express 29, 40891-40916 (2021)

[18] J.P. Guigay (1971) On Fresnel Diffraction by One-dimensional Periodic Objects, with Application to Structure Determination of Phase Objects, Optica Acta: International Journal of Optics, 18:9, 677-682, DOI: 10.1080/713818491

[19] Patorski, K. (1989). I The Self-Imaging Phenomenon and its Applications. *Progress in Optics, 27*, 1-108.

[20] T. Weitkamp, C. David, C. Kottler, O. Bunk, F. Pfeiffer, "Tomography with grating interferometers at low-brilliance sources," Proc. SPIE 6318, Developments in X-Ray Tomography V, 63180S (7 September 2006); https://doi.org/10.1117/12.683851

[21] A. Macovski, *Medical Imaging Systems*, Prentice Hall, Inc, NJ, 1983

[22] N. Bevins, J. Zambelli, K. Li, Z. Qi, and G. H. Chen, Med. Phys. **39**, 424–428 (2012).

[23] J. Kim, S. W. Lee, G. Cho, Nucl. Instrum. Meth. A **746**, 26–32 (2014).

[24] Y. Kim, J. Valsecchi, J. Kim, S. W. Lee, M. Strobl, Sci. Rep. **9,** 18973 (2019).

[25] T. Donath, M. Chabior,F. Pfeiffer, O. Bunk, E. Reznikova, J. Mohr, E. Hempel, S. Popescu, M. Hoheisel, M. Schuster et al., J. Appl. Phys. **106,** 054703 (2009).

[26] **J. Dey,** H. Meyer, S. Carr, K. Ham, L. Butler, I. Hidrovo, J. Xu, T. Varga, J. Schulz, T. Beckenbach, K. Kaiser, "Imaging Experiments with Modulated Phase Grating Interferometry", manuscript in preparation.

[27] H. Meyer, MSc Thesis in prep, Louisiana State University (expected May 2023).

[28] I. J. Hidrovo Giler., "Neutron Interferometry Using a Single Modulated Phase Grating" (2022). LSU Master's Theses. 5630.

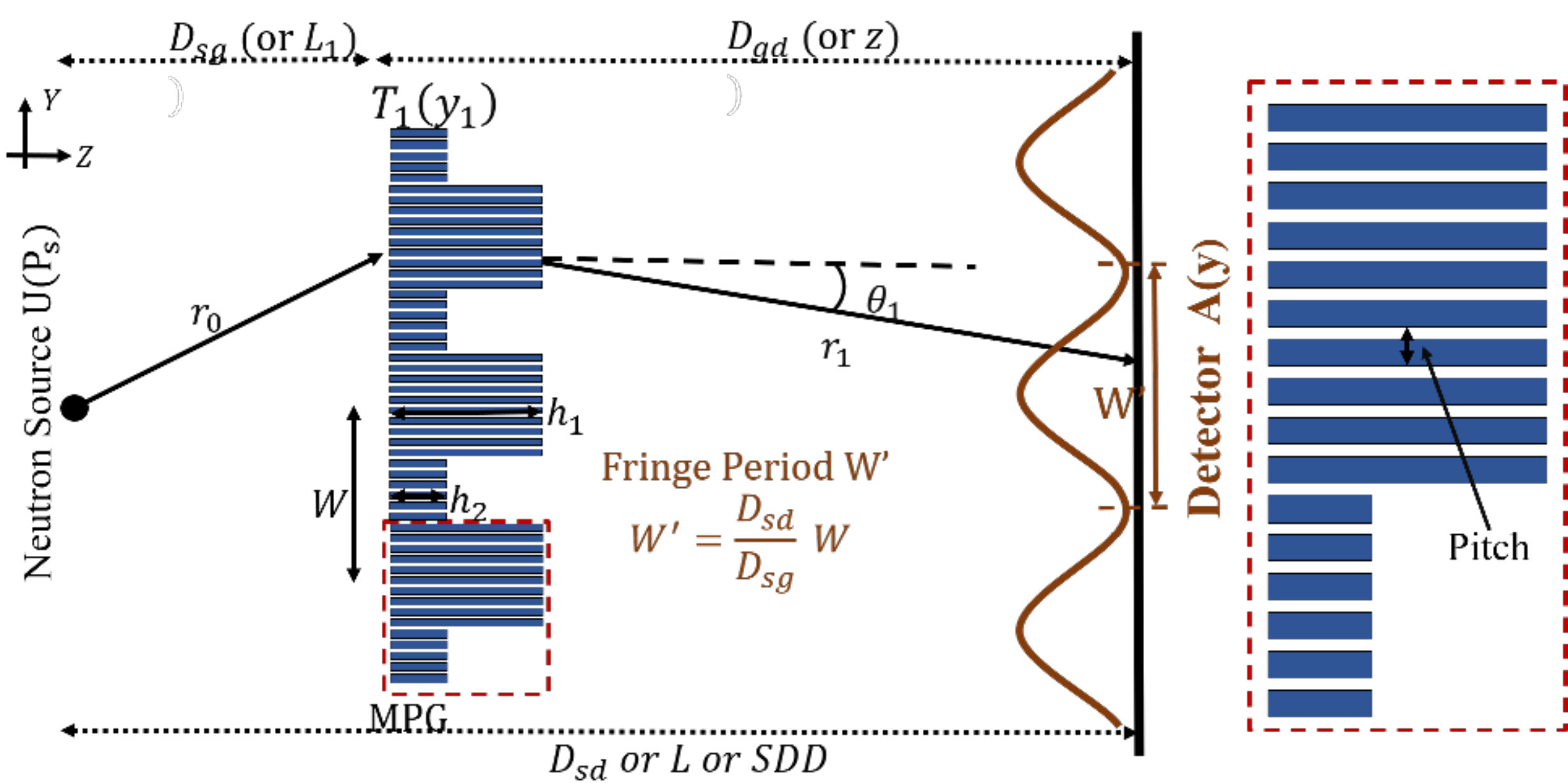

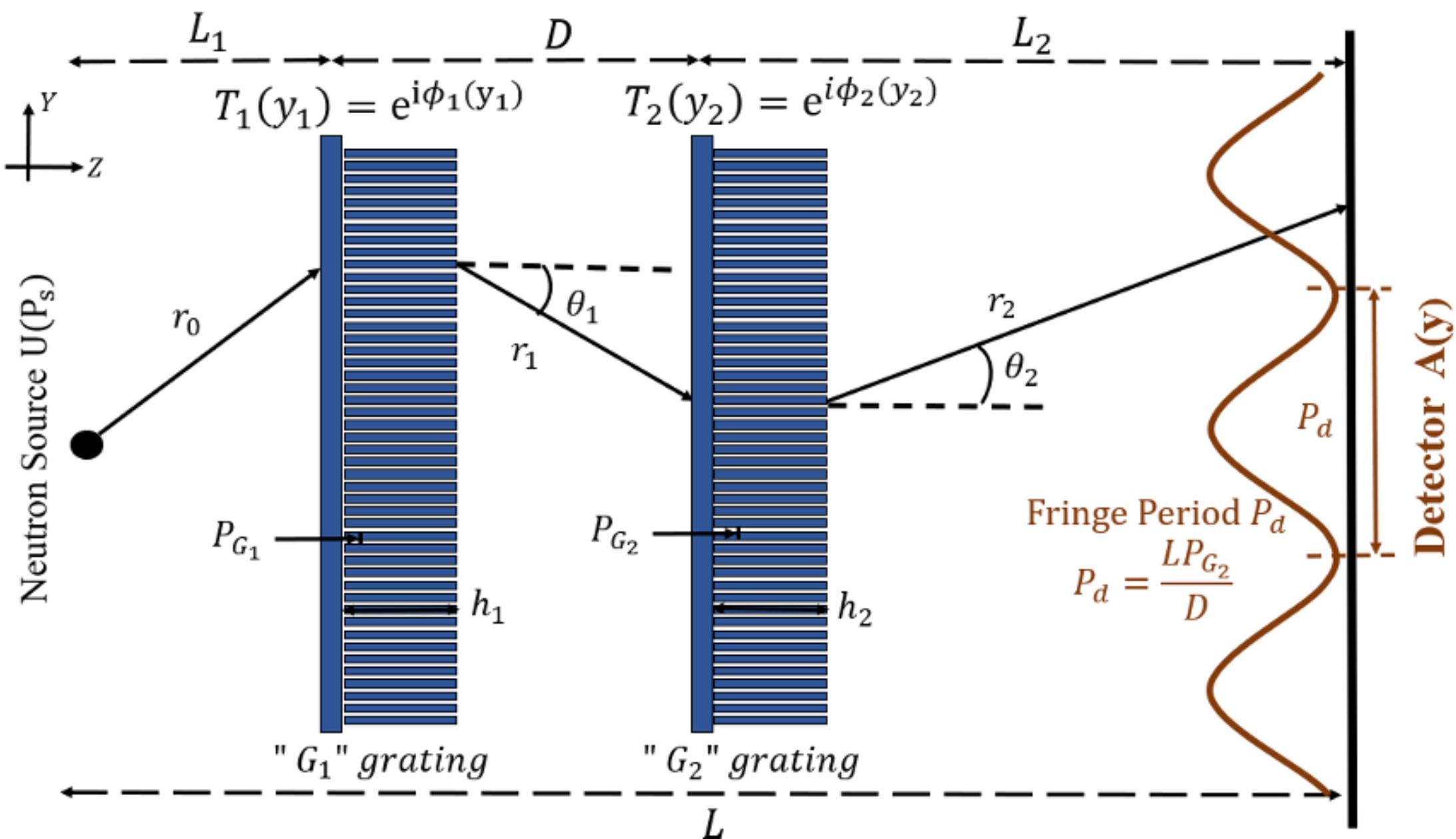

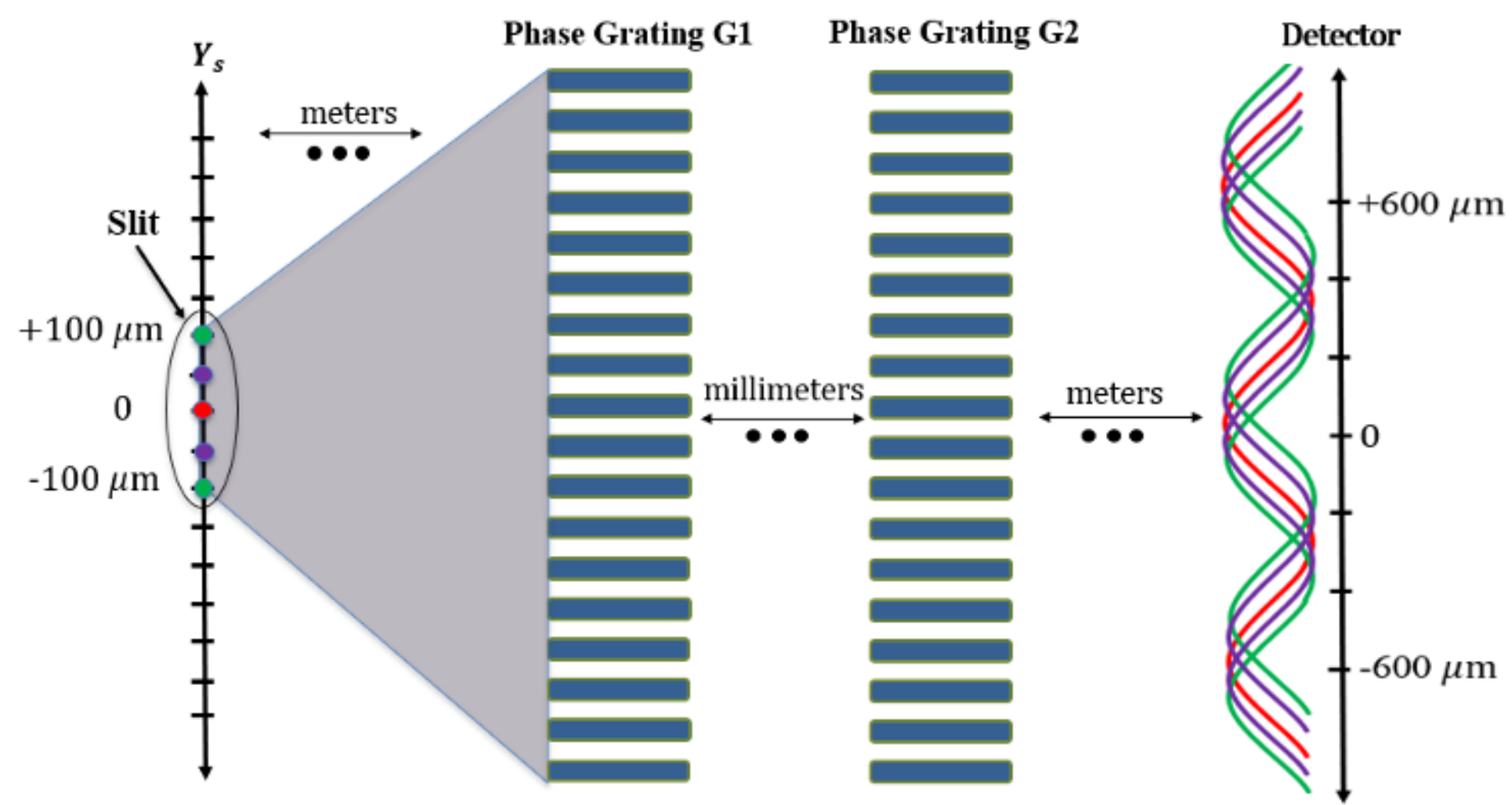

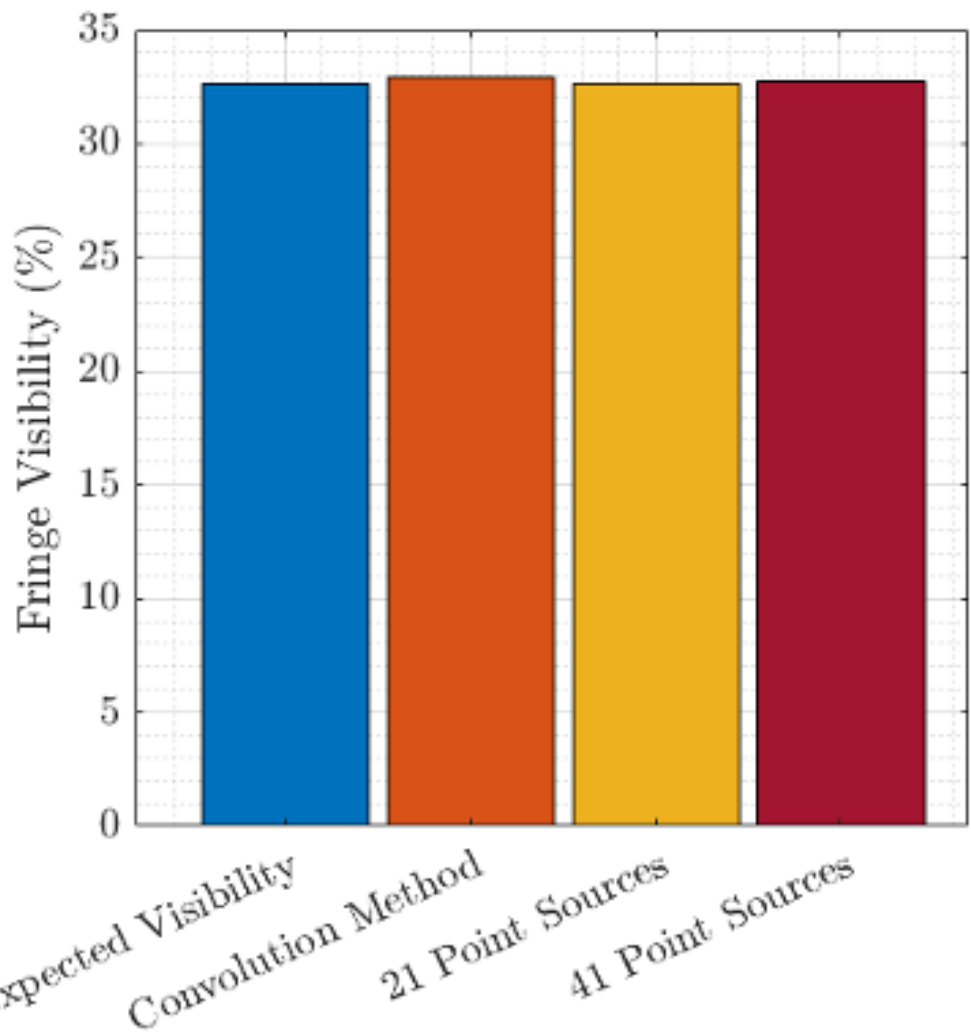

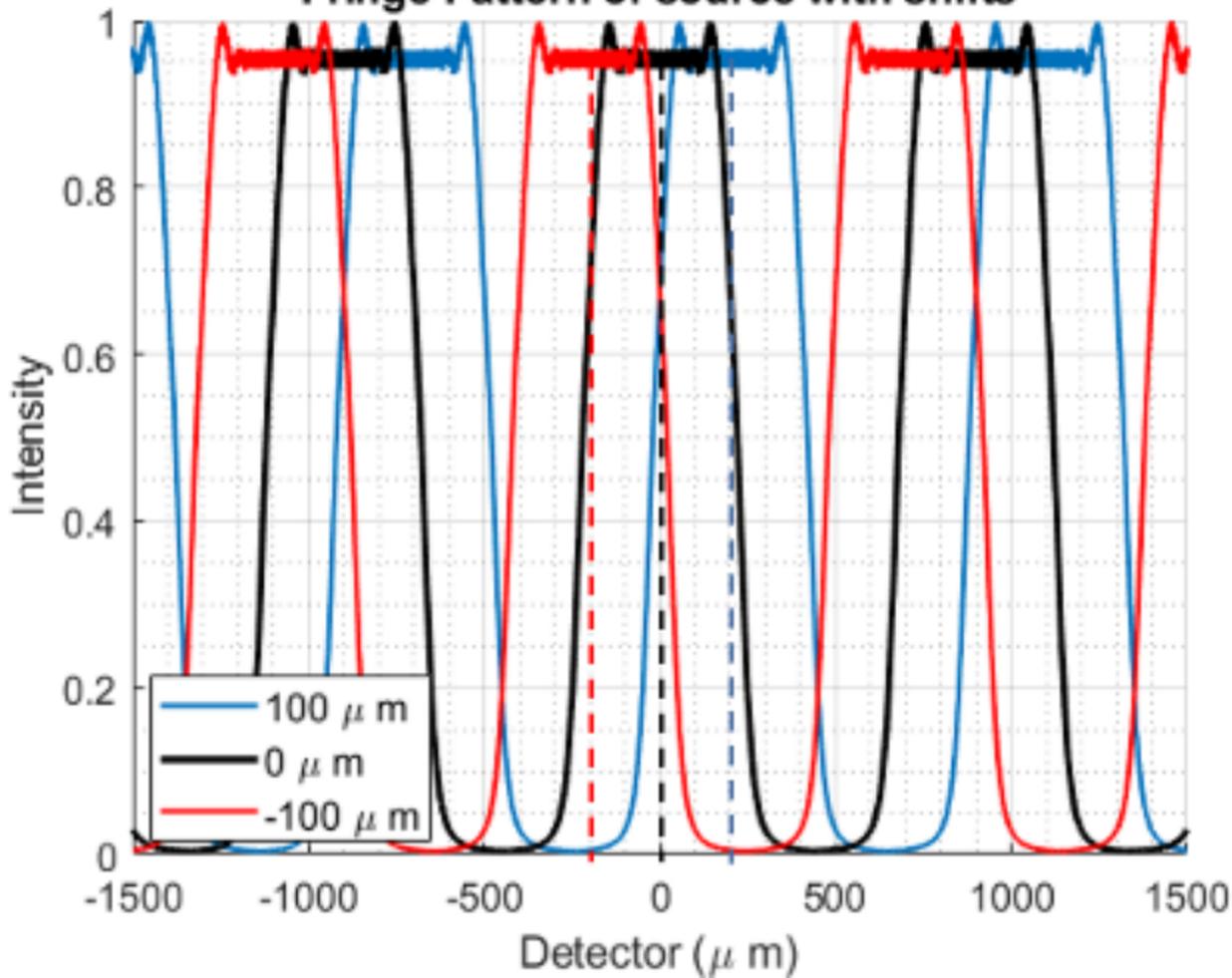

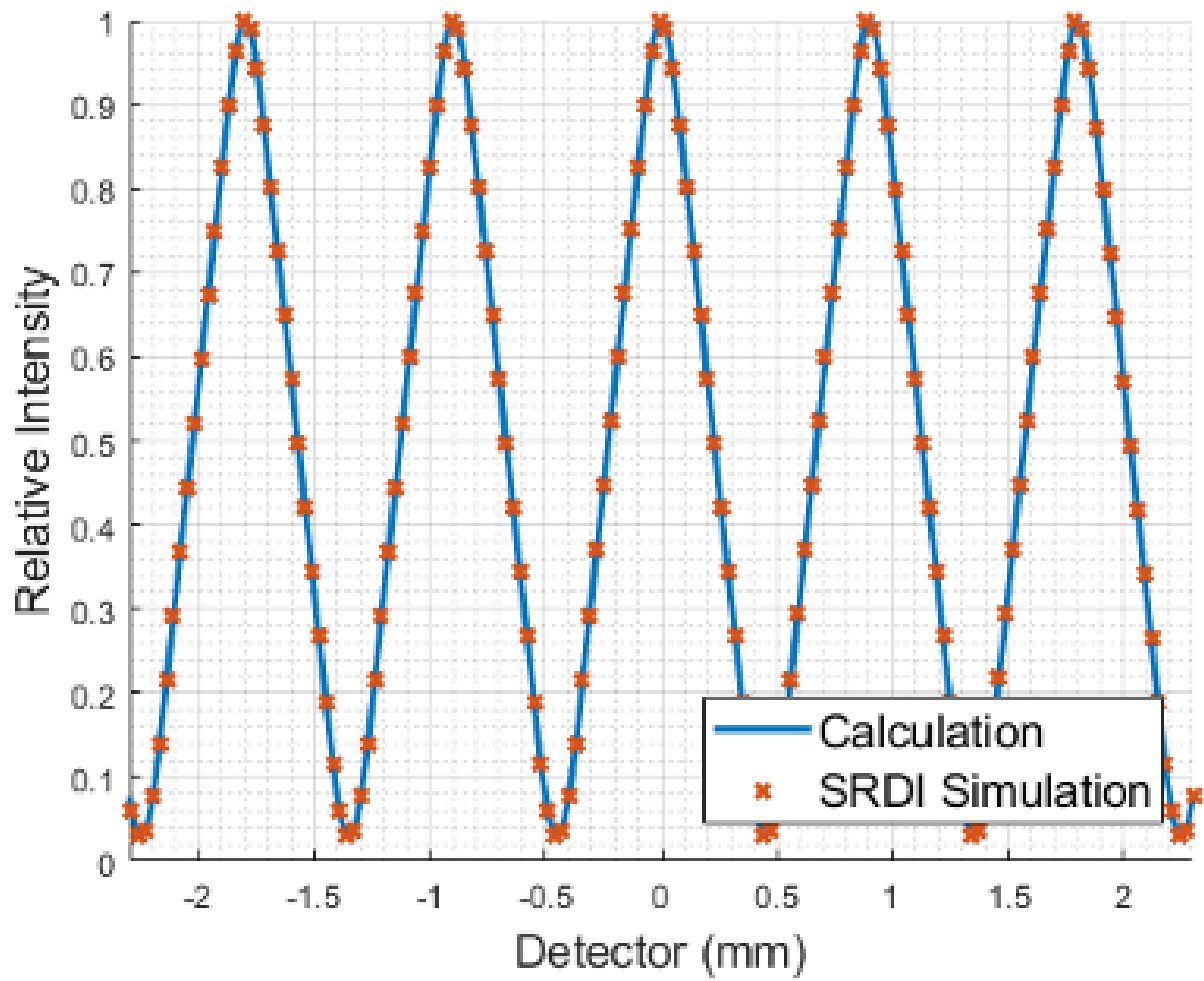

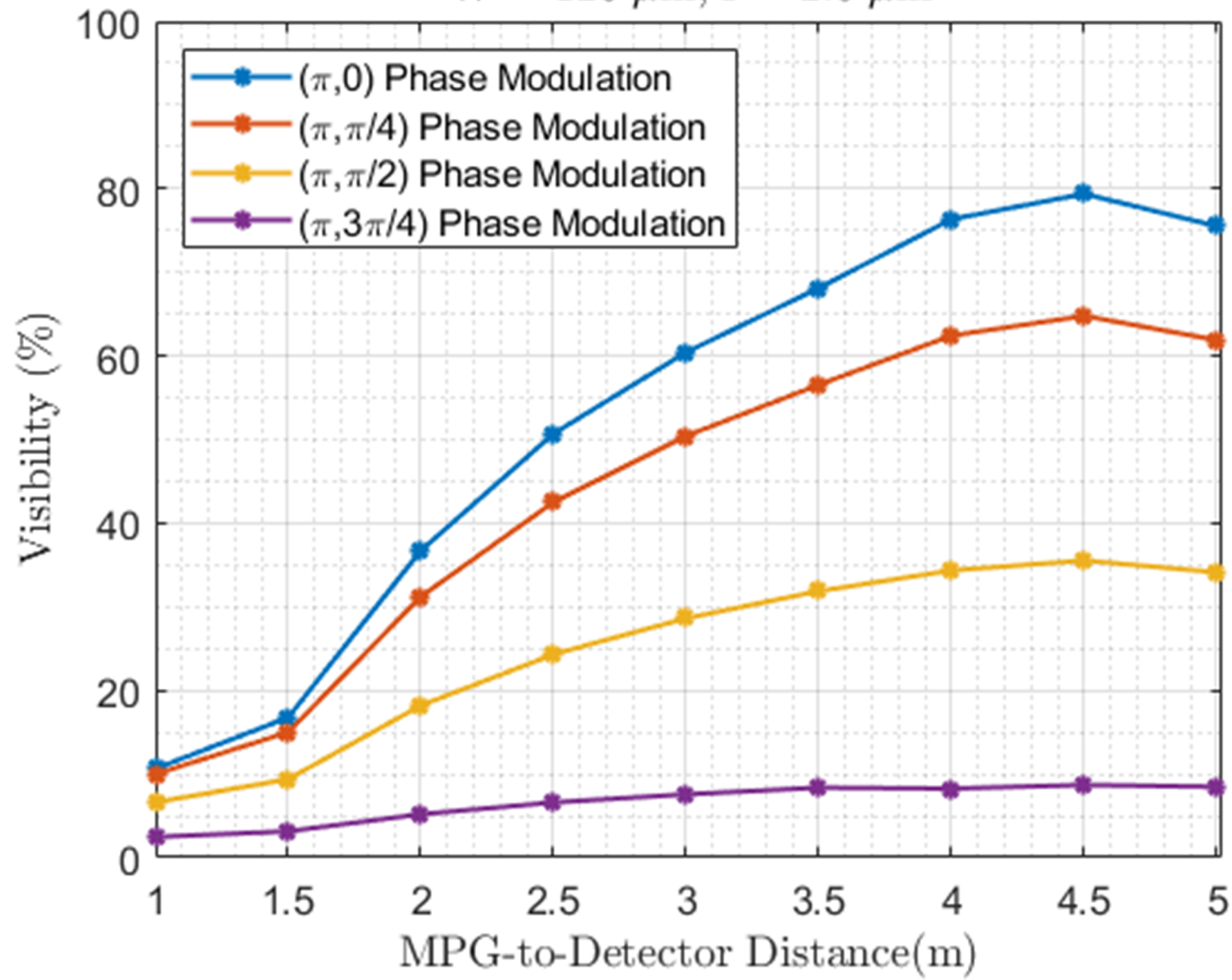
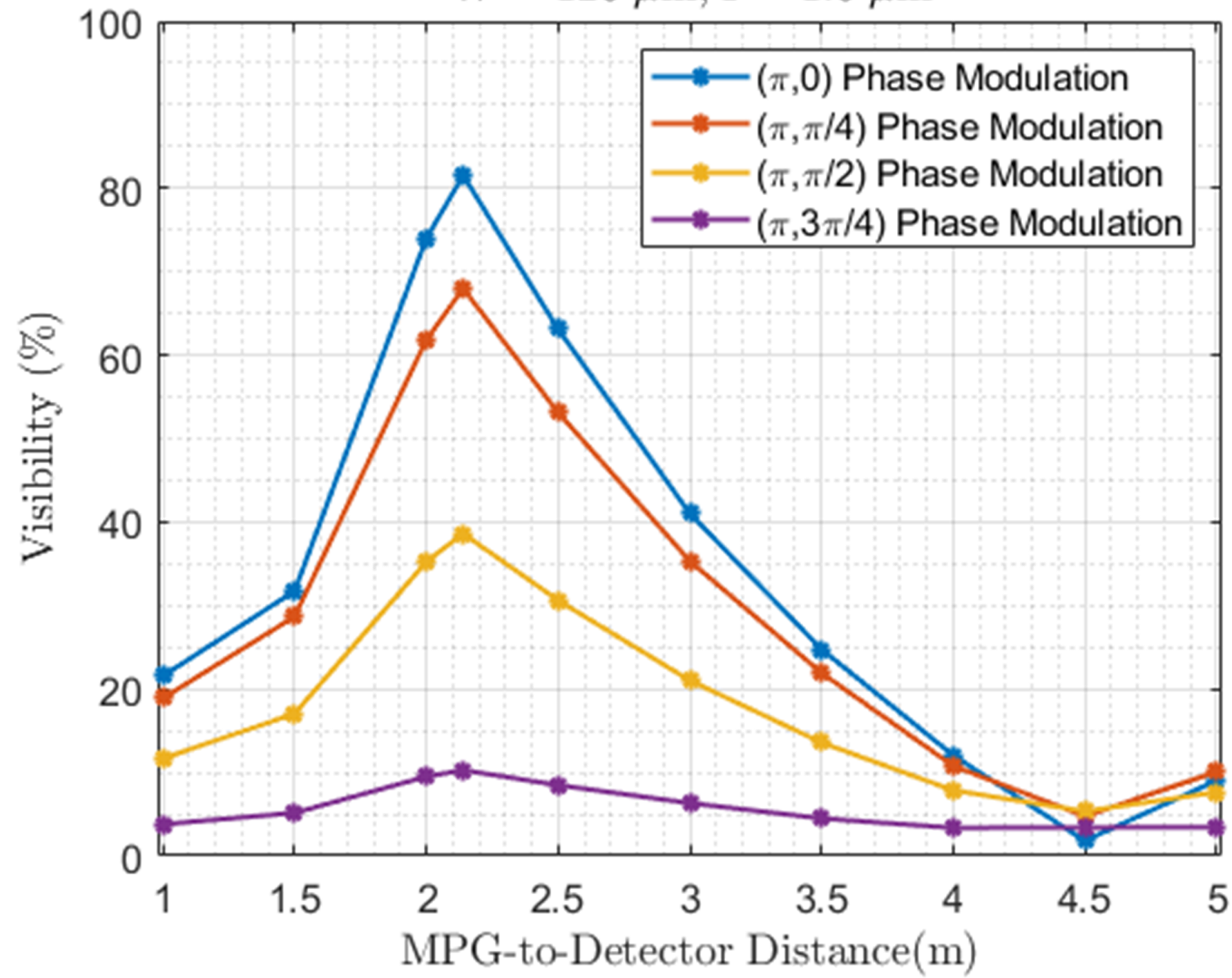

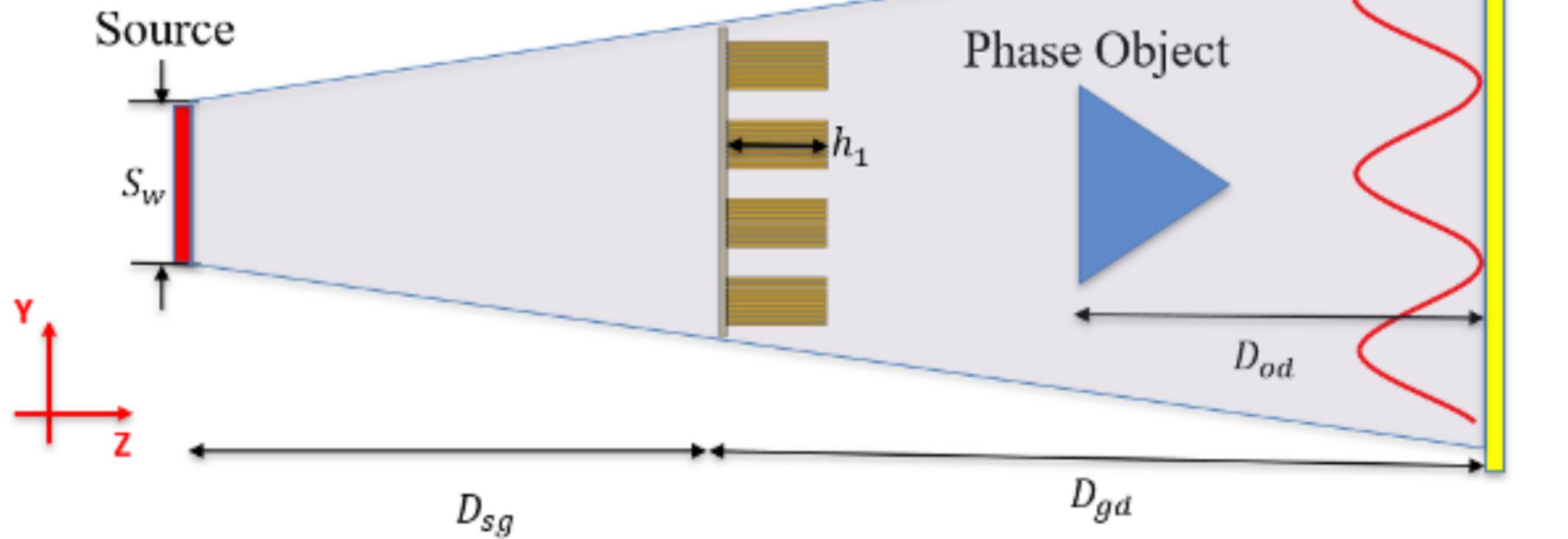

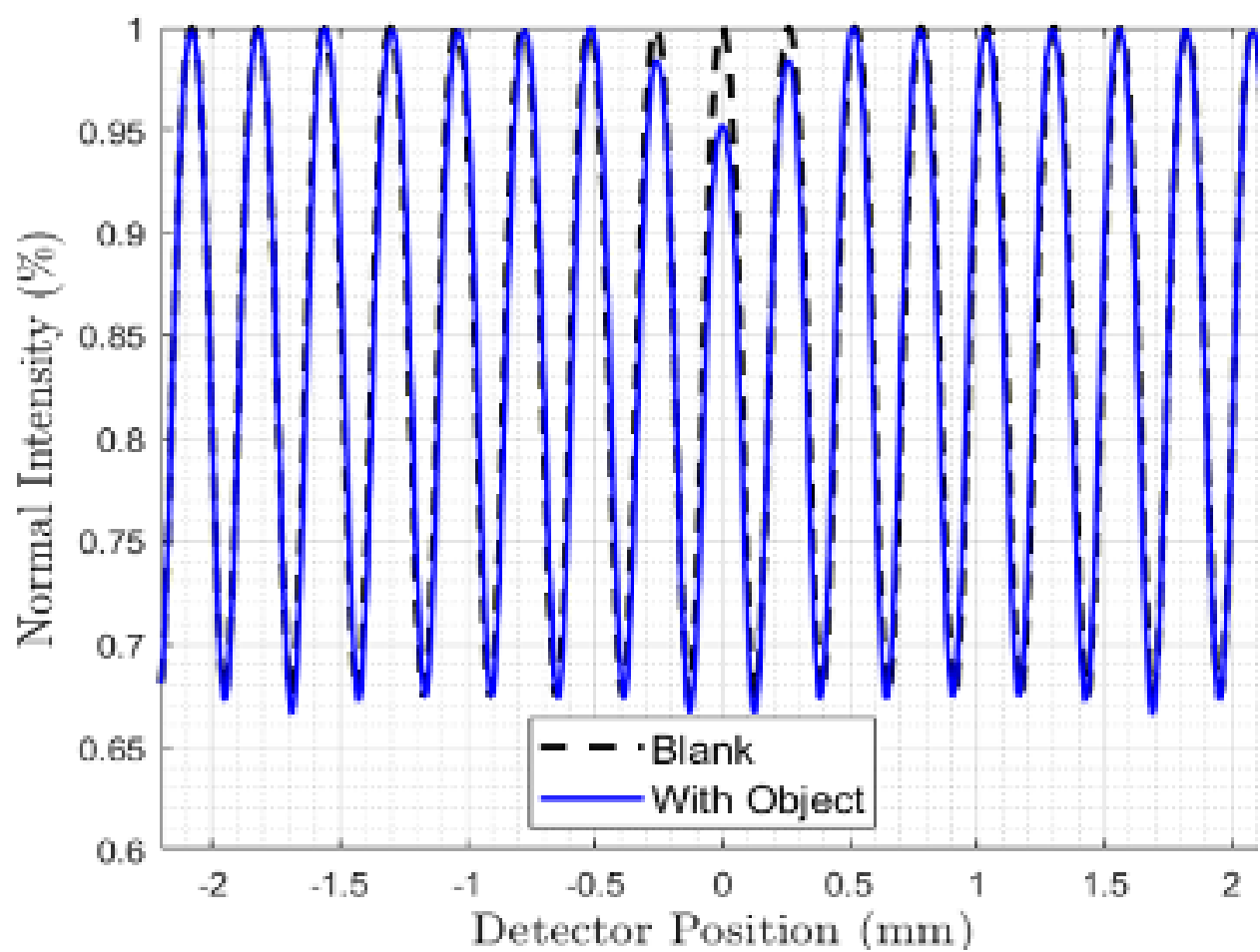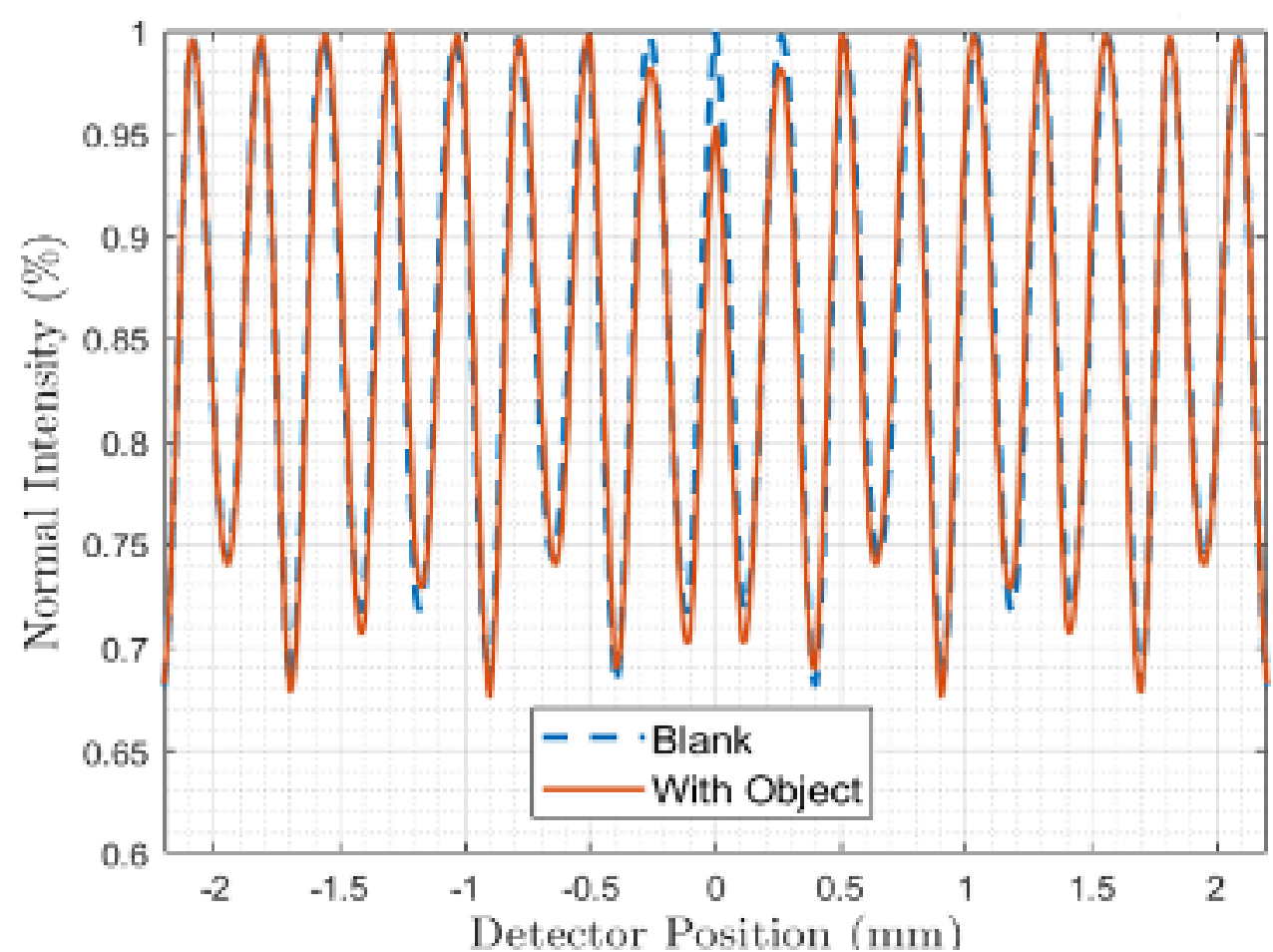

**(a)**                                   **(b)**

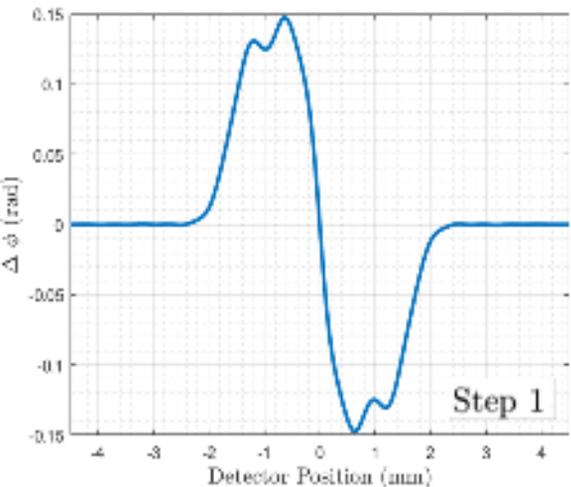

(a)

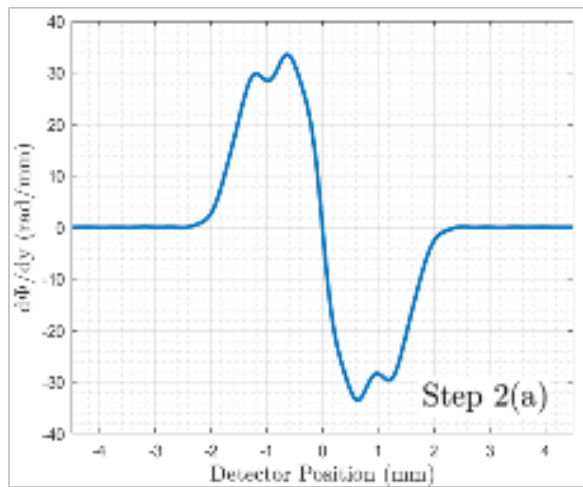

(b)

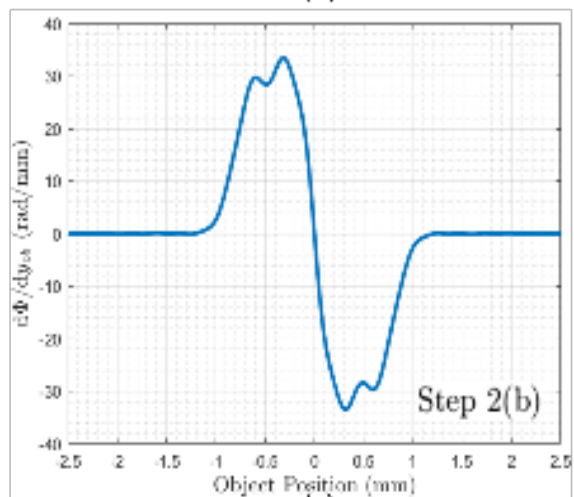

(c)

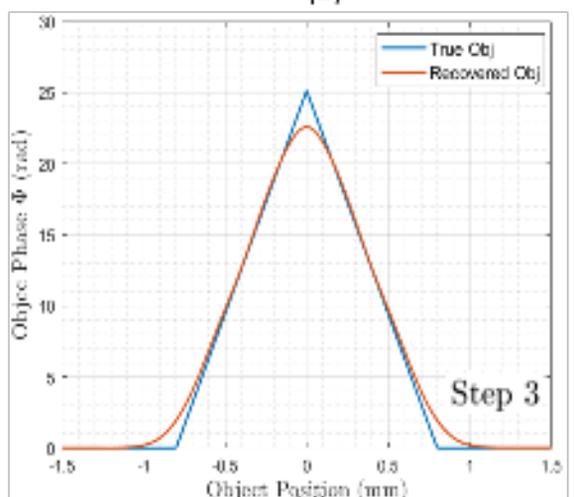

(d)

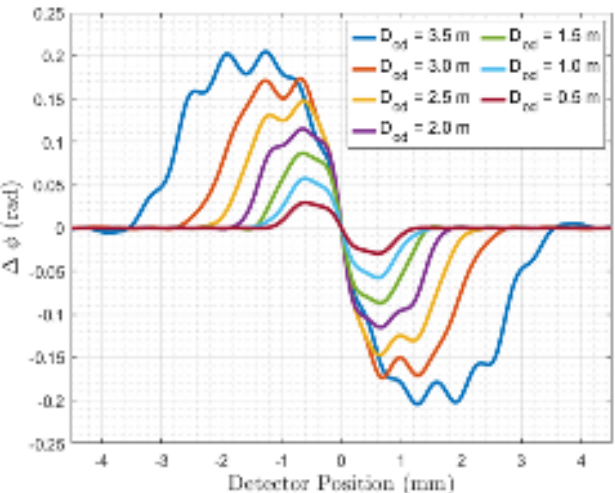

(a)

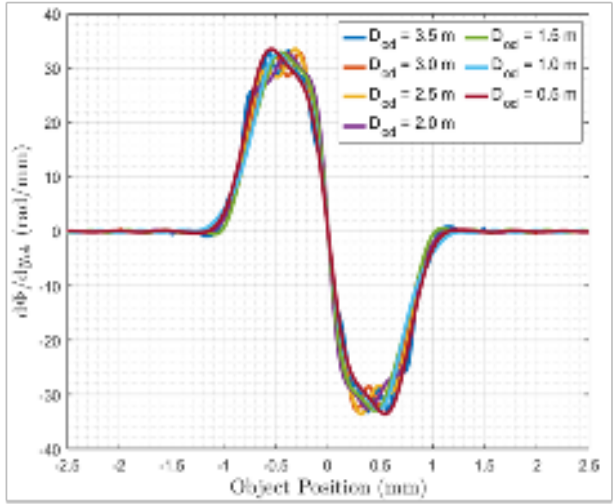

(b)

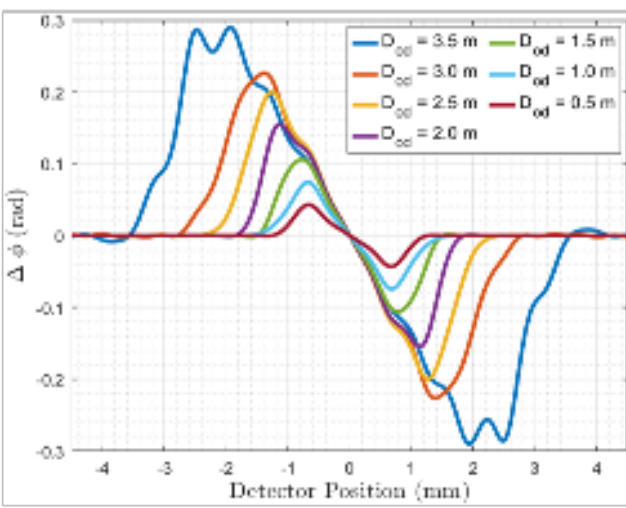

(c)

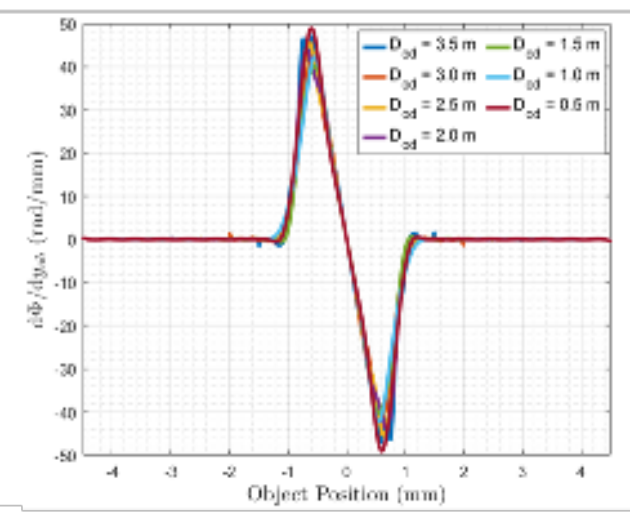

(d)

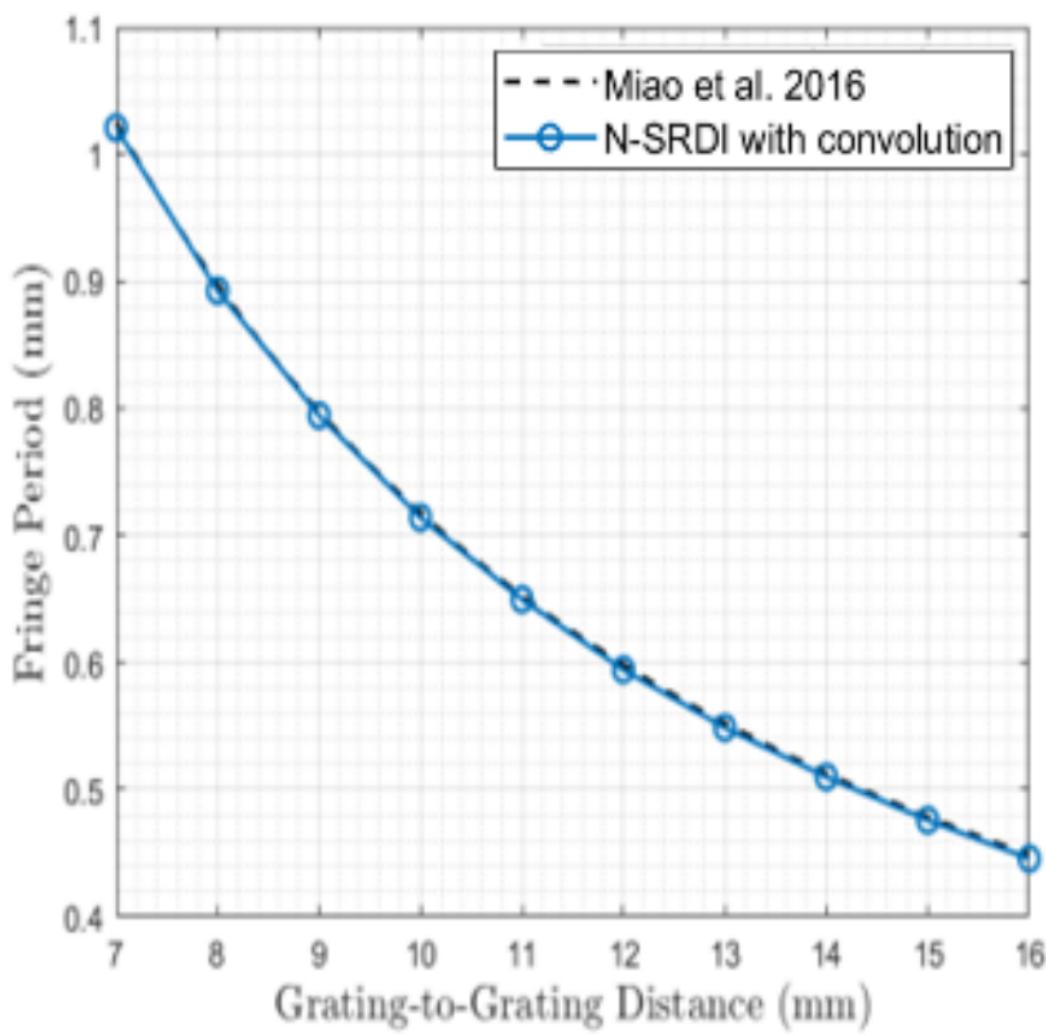

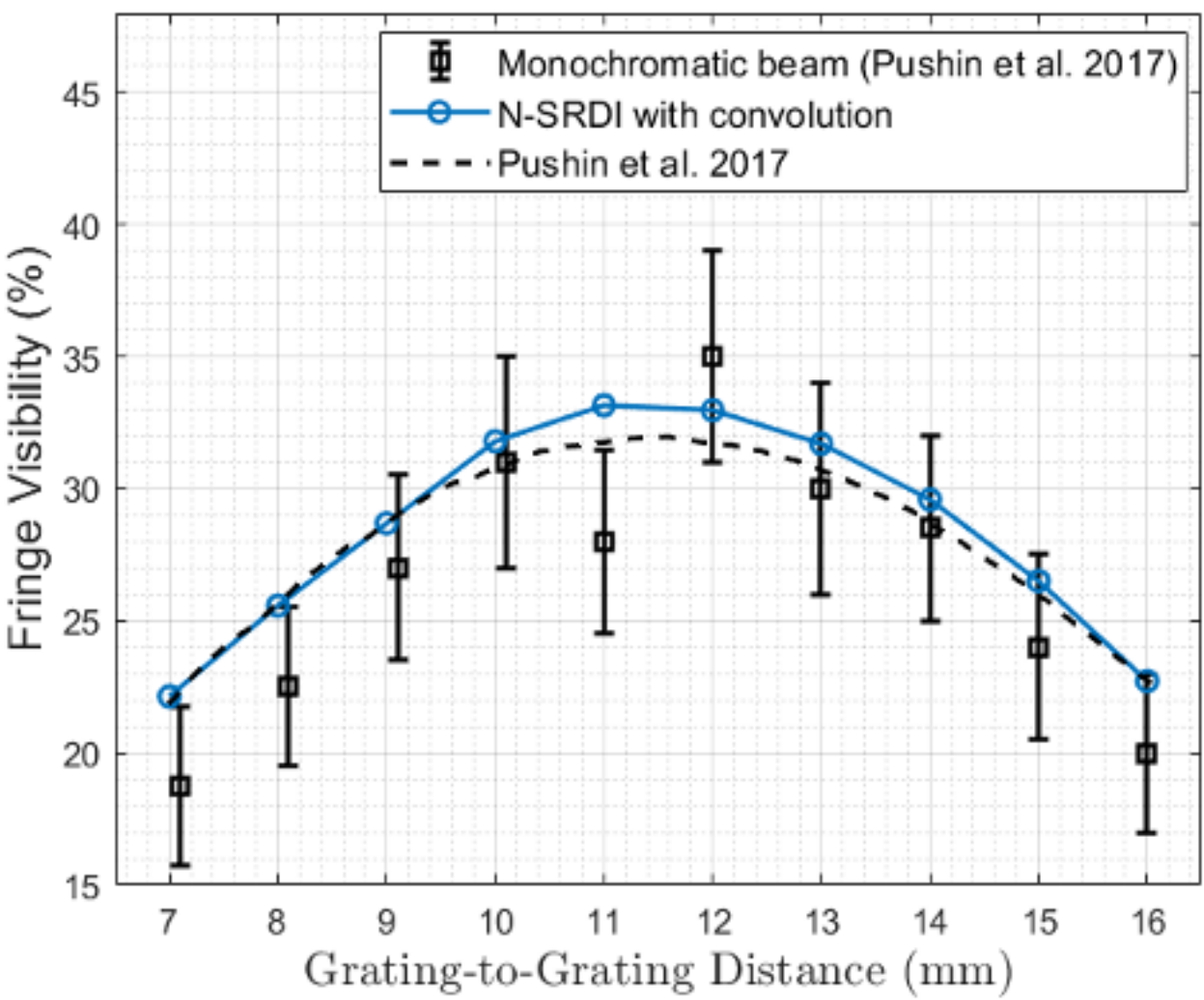

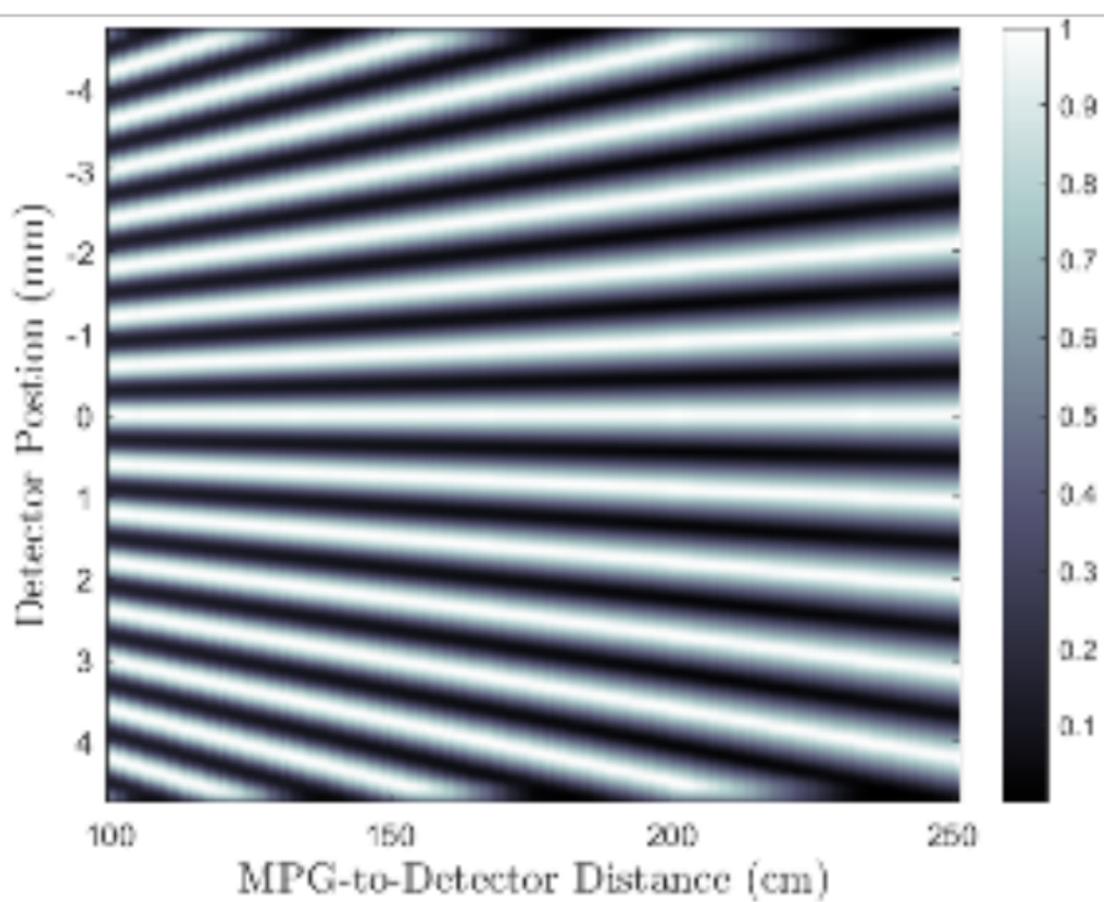

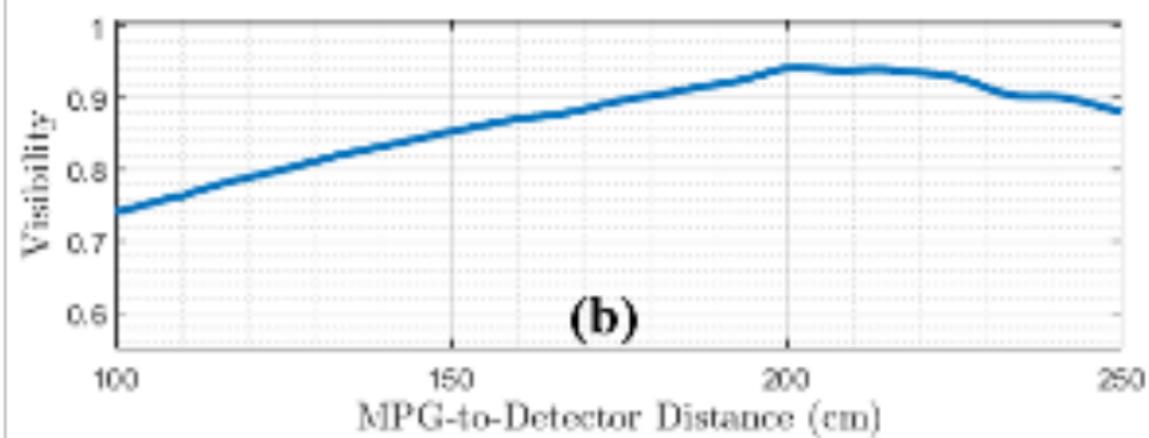

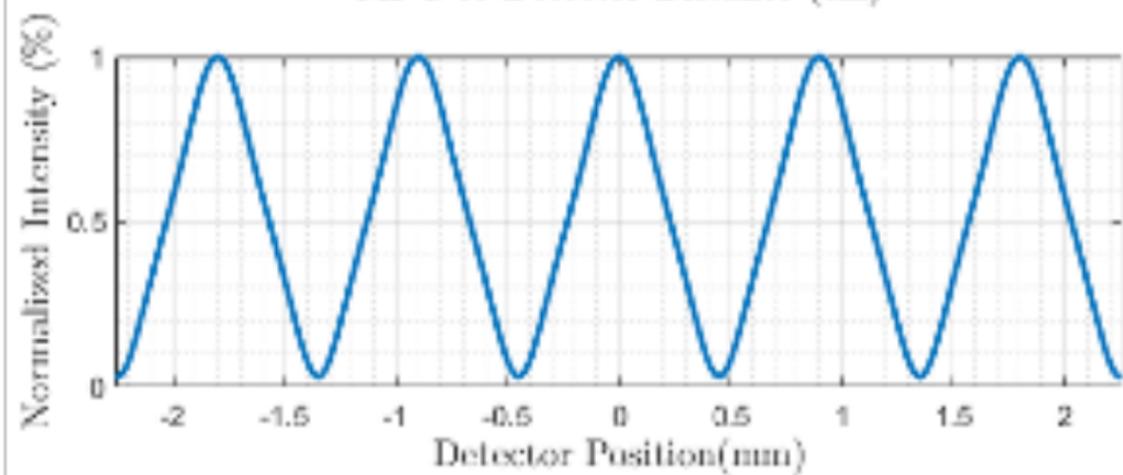

**(a)**

**(b)**

**(c)**

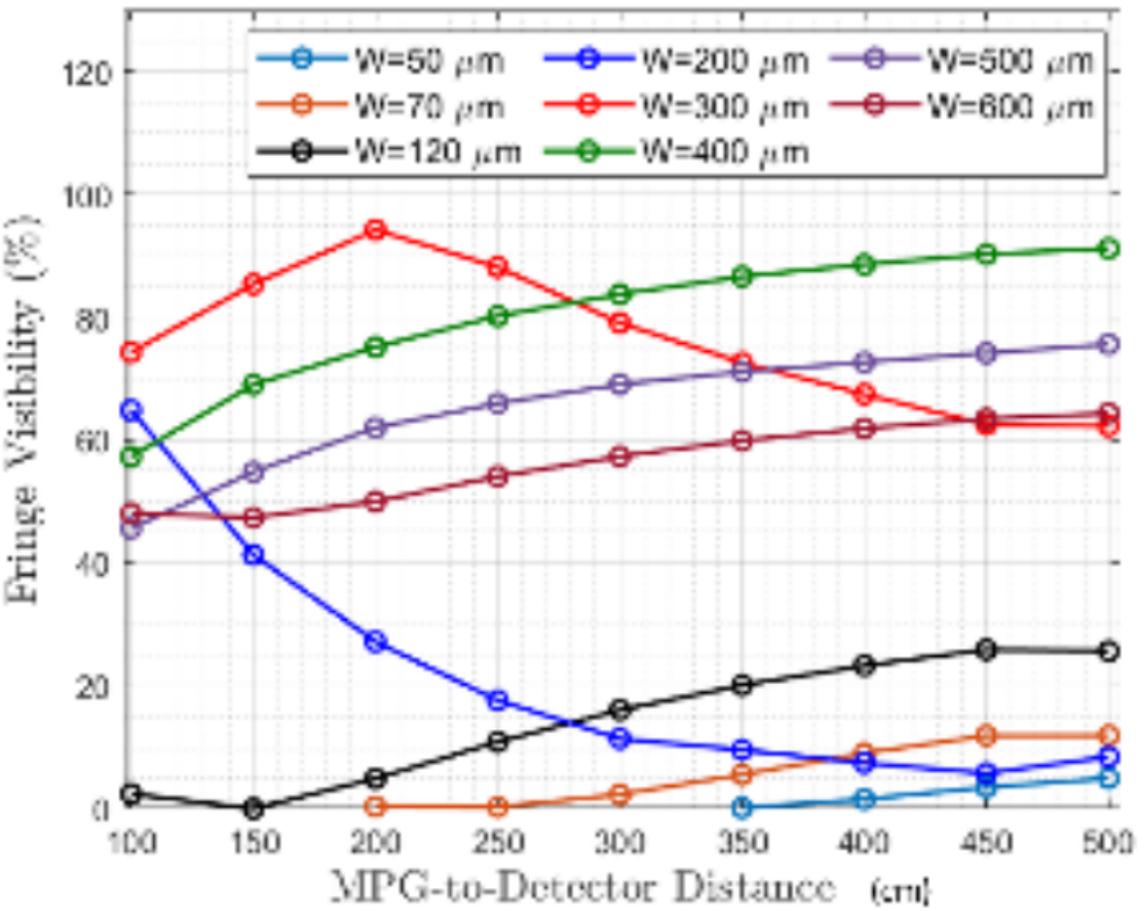

**(a)**

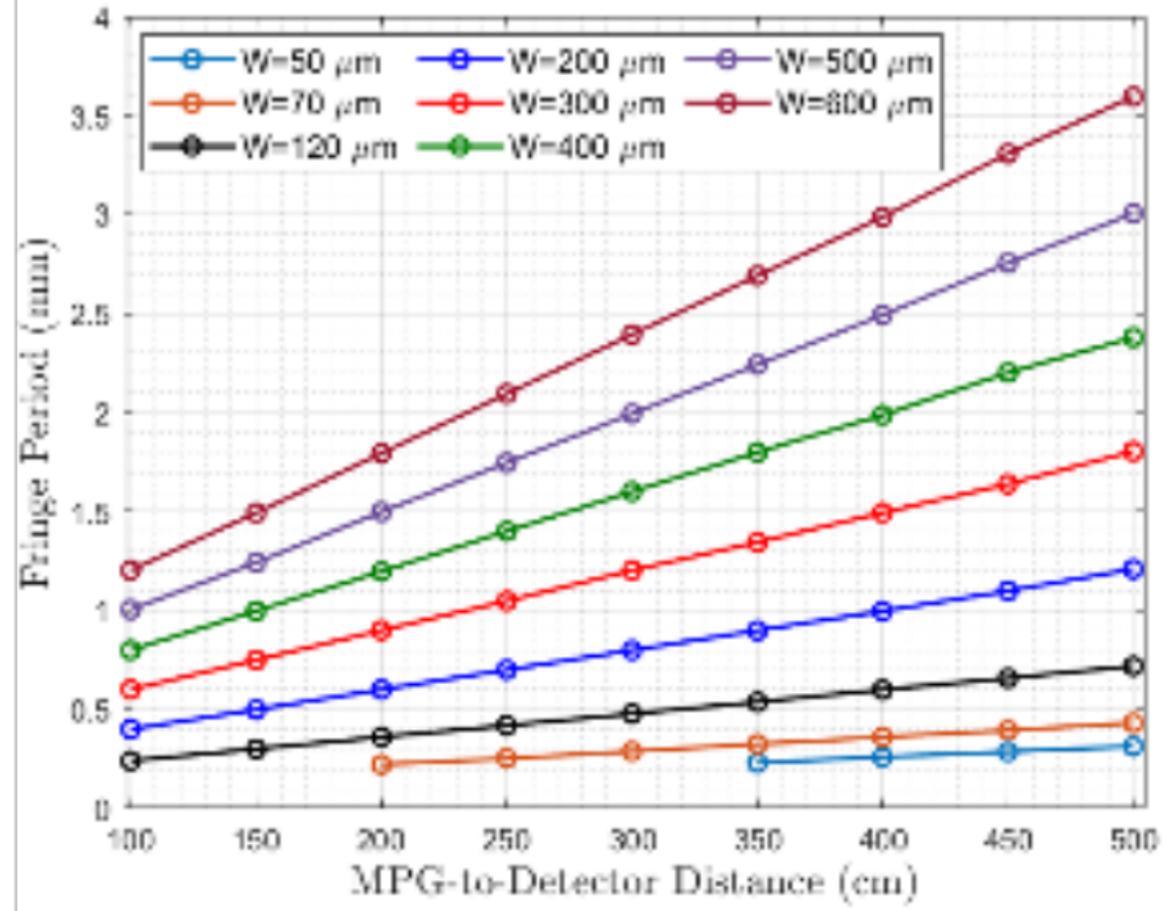

**(b)**

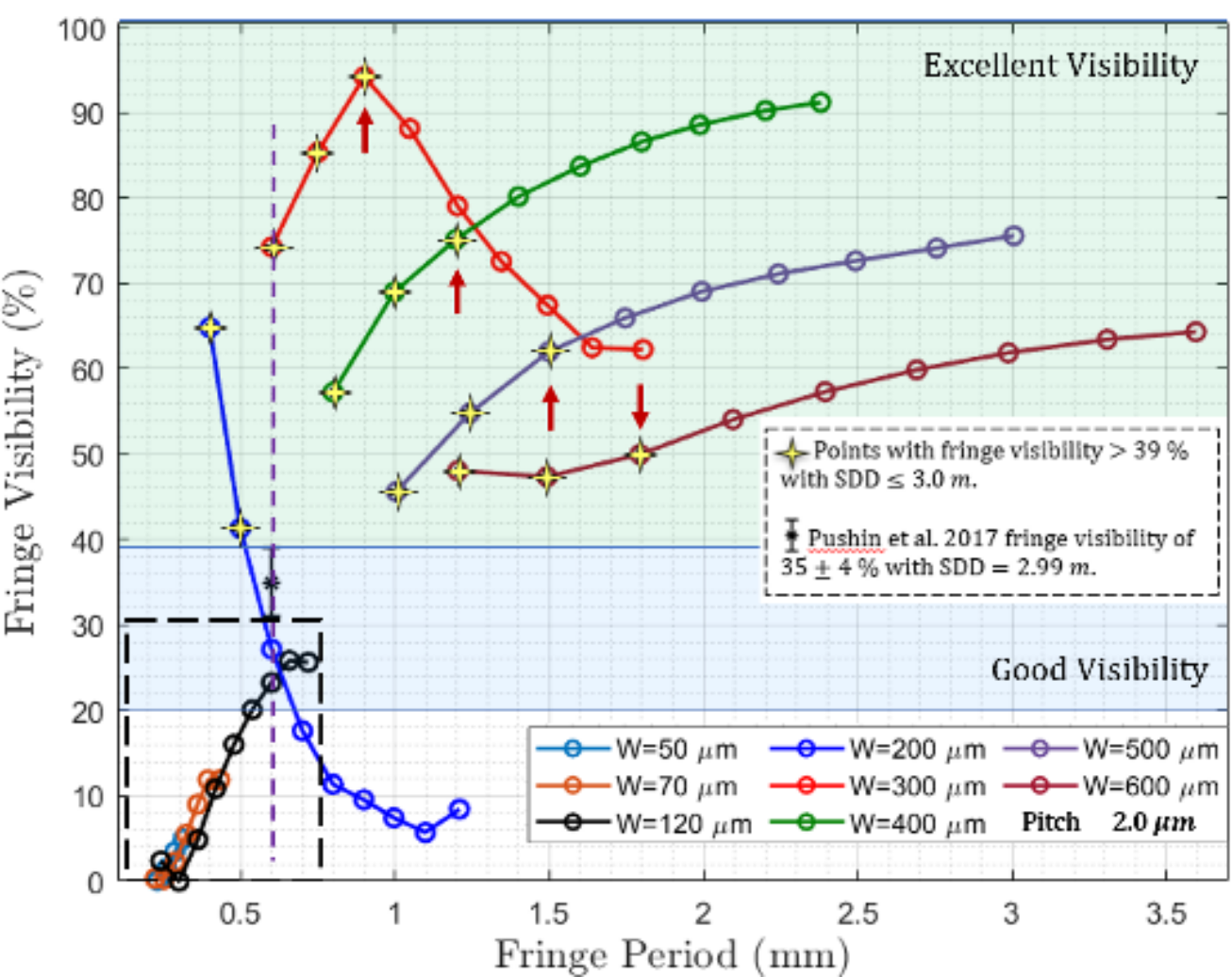

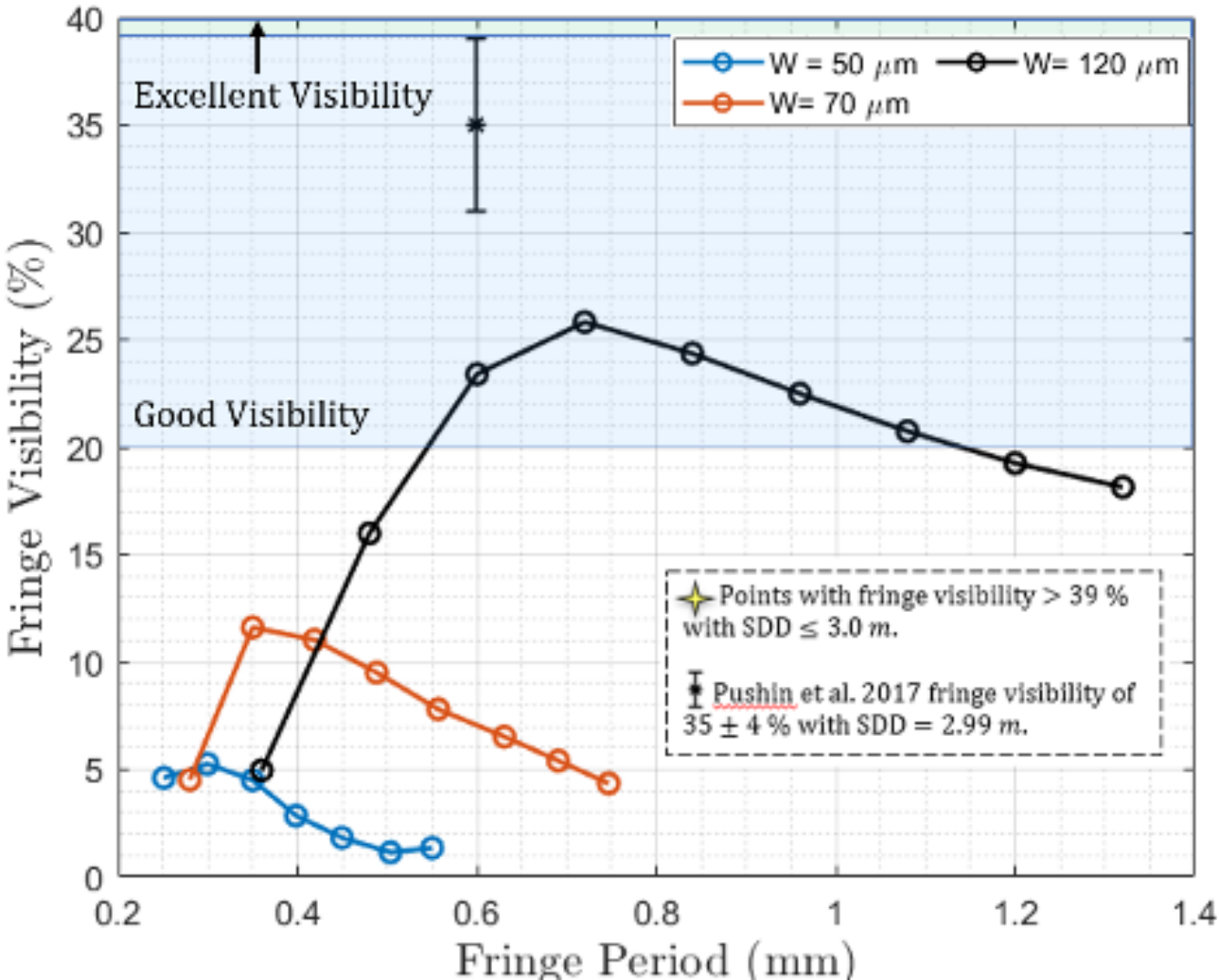

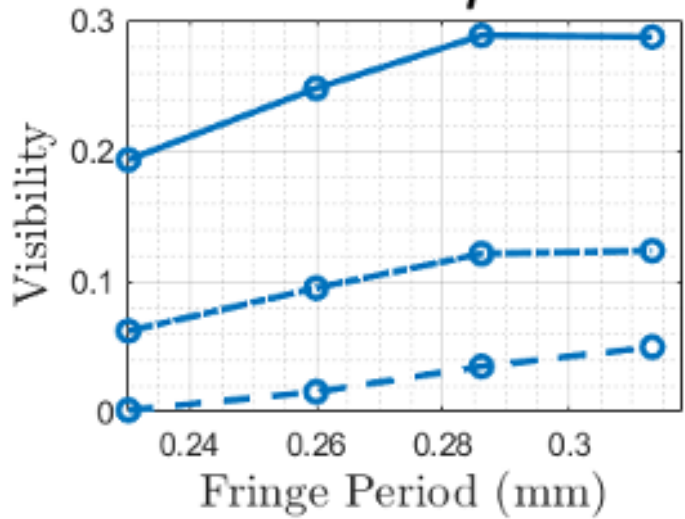

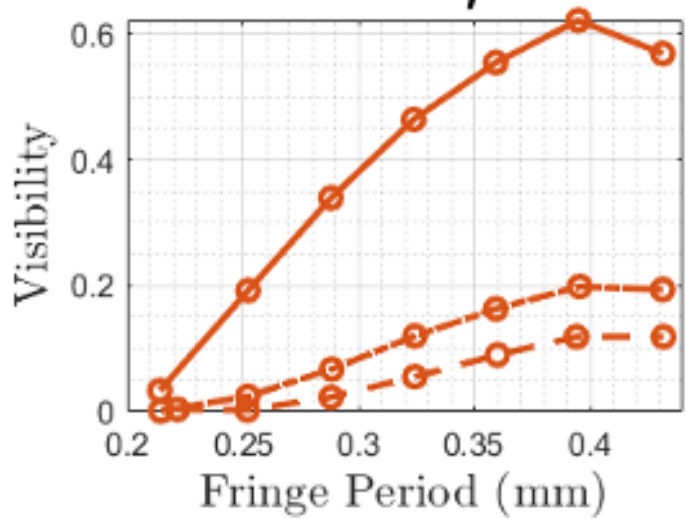

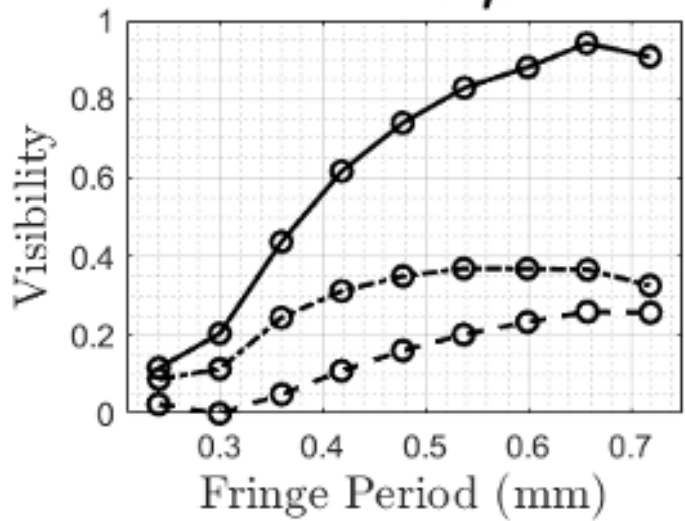

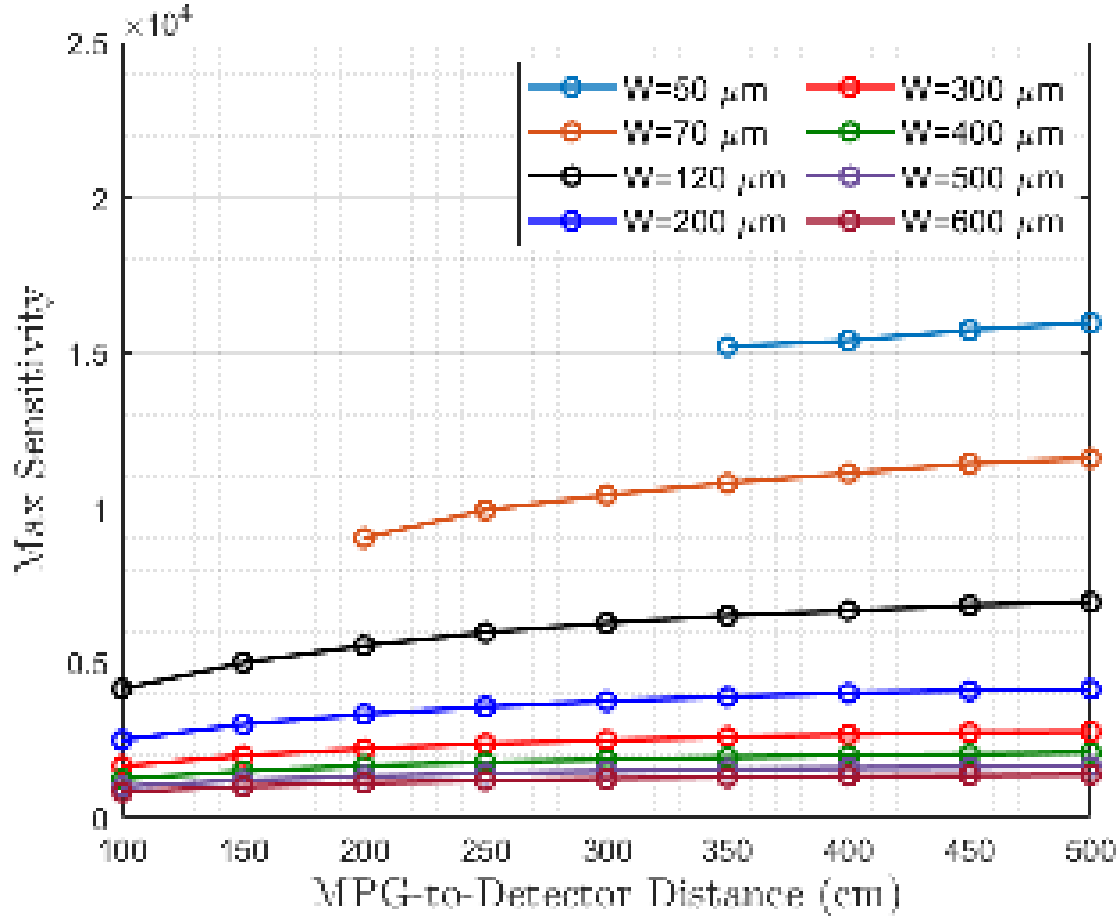

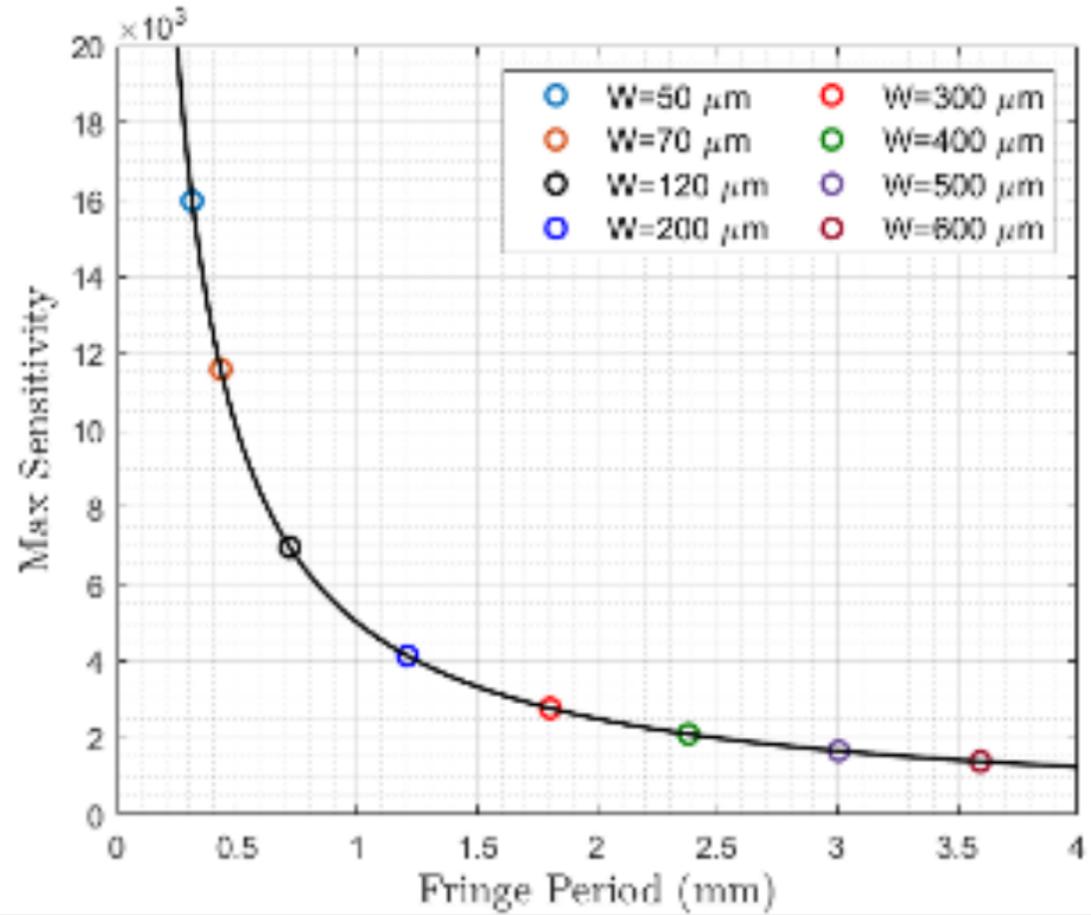

(a)

(b)

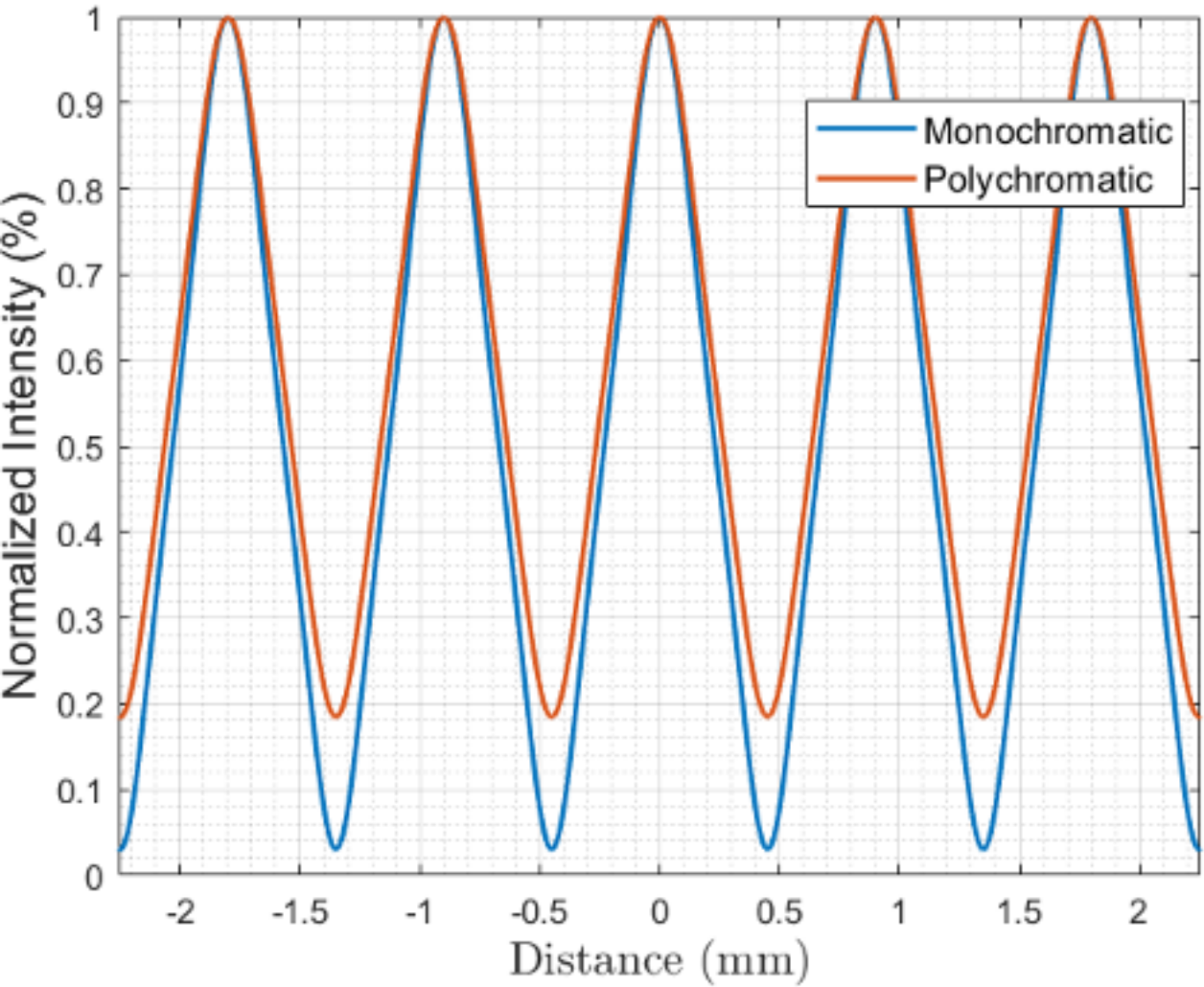

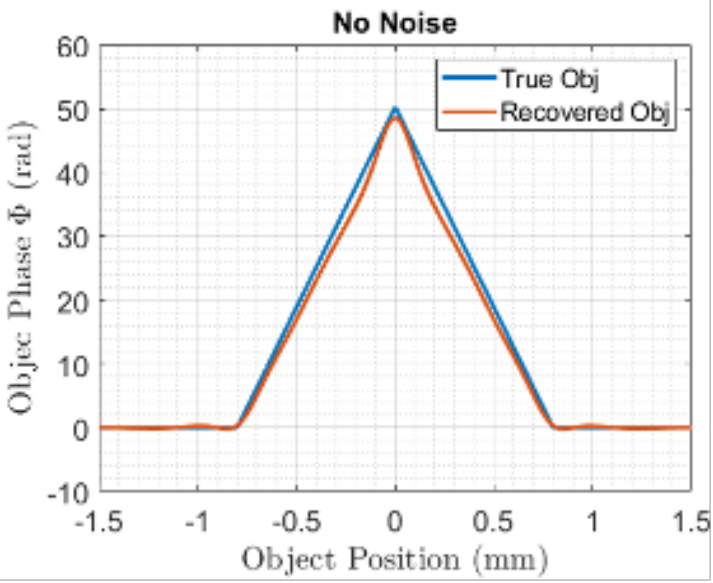
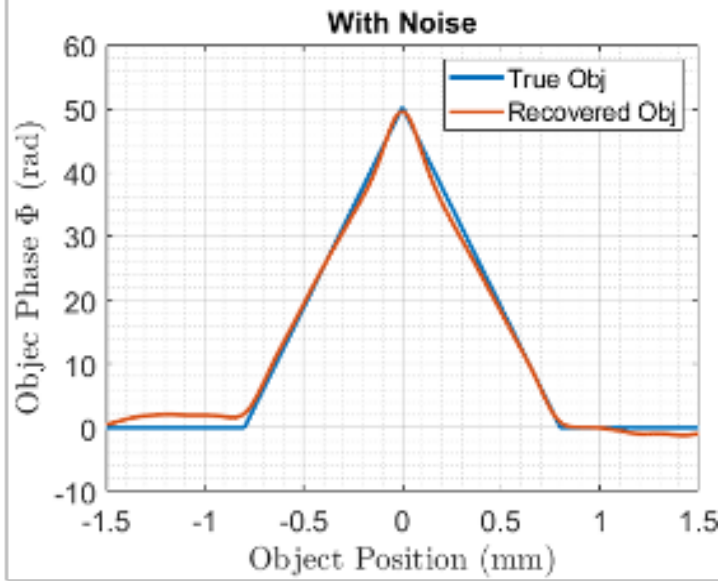
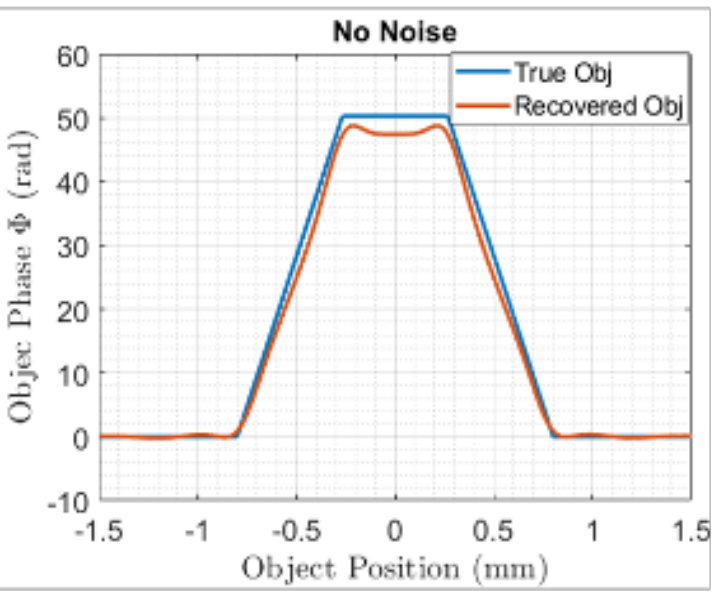
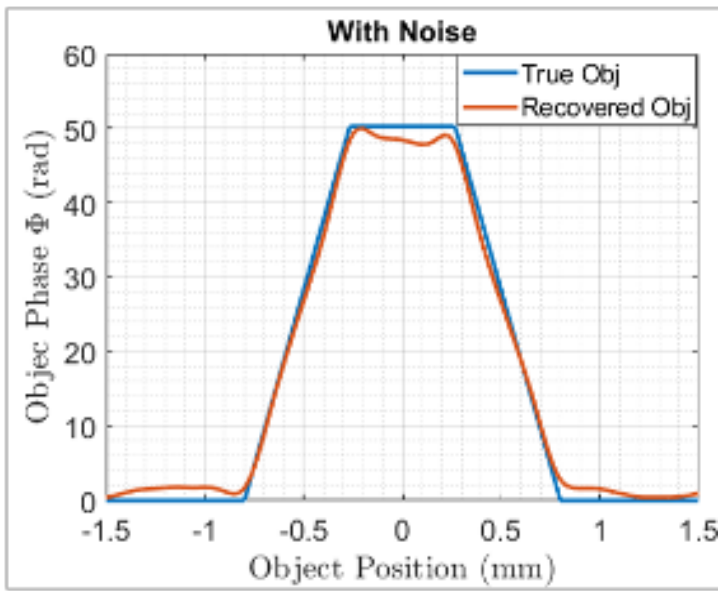
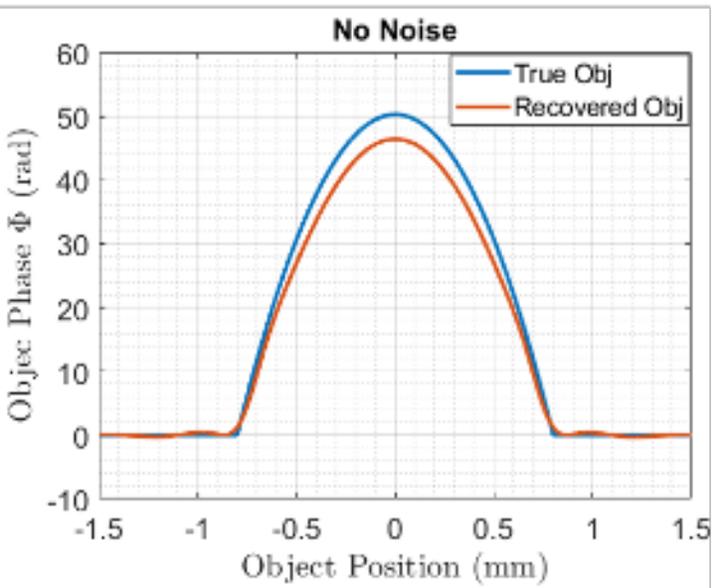
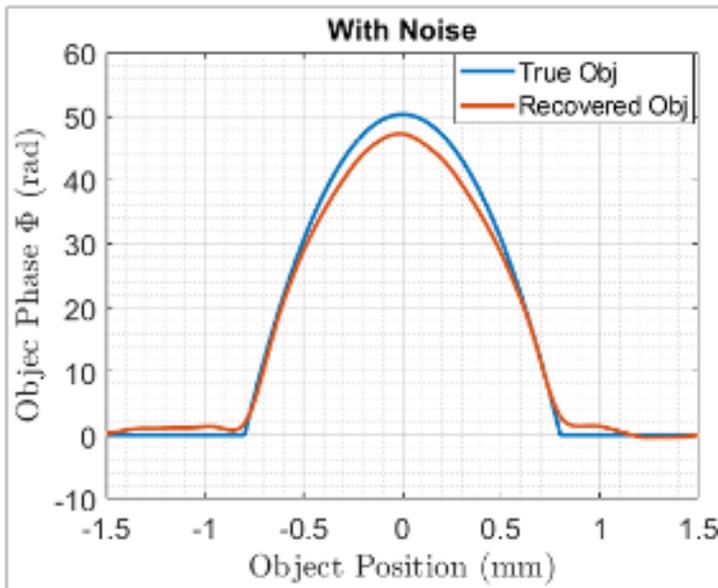